\documentclass[a4paper,11pt]{article}
\usepackage{heppub}
\pdfoutput=1
\usepackage{xspace}

\newcommand{\WidthDiagonalSubfig}{0.66\textwidth}
\newcommand{\WidthTwoSubfigs}{0.5\textwidth}
\newcommand{\WidthThreeSubfigs}{0.33\textwidth}

\newcommand{\HeightTwoSubfigs}{0.350694\textwidth} % = \WidthTwoSubfigs for 2/3 aspect ratio and default padding

\newcommand{\abs}[1]{\lvert#1\rvert}
\newcommand{\ord}[1]{\mathcal{O}(#1)}
\newcommand{\ORd}[1]{\mathcal{O}\Bigl(#1\Bigr)}

\newcommand{\df}{\mathrm{d}}
\newcommand{\img}{\mathrm{i}}

\newcommand{\nn}{\nonumber}

\let\oldvec\vec
\renewcommand*\vec[1]{\oldvec{\kern0pt #1}}

\newcommand{\sdot}{\!\cdot\!}

\renewcommand{\Re}{\operatorname{Re}}

\newcommand{\cI}{\mathcal{I}}
\newcommand{\cL}{\mathcal{L}}
\newcommand{\cS}{\mathcal{S}}
\newcommand{\cV}{\mathcal{V}}

\newcommand{\tB}{\tilde{B}}
\newcommand{\tS}{\tilde{S}}
\newcommand{\tf}{\tilde{f}}
\newcommand{\tgamma}{\tilde{\gamma}}
\newcommand{\tcS}{\tilde{\cS}}
\newcommand{\bn}{{\bar{n}}}
\newcommand{\bq}{{\bar{q}}}

\newcommand{\bulk}{\mathrm{bulk}}
\newcommand{\cusp}{\mathrm{cusp}}
\newcommand{\cumul}{\mathrm{cumul}}
\newcommand{\cut}{\mathrm{cut}}
\newcommand{\diff}{\mathrm{diff}}
\newcommand{\FO}{\mathrm{FO}}
\newcommand{\I}{\mathrm{I}}
\newcommand{\II}{\mathrm{II}}
\newcommand{\match}{\mathrm{match}}
\newcommand{\np}{\mathrm{np}}
\newcommand{\QCD}{\mathrm{QCD}}
\newcommand{\total}{\mathrm{total}}
\newcommand{\run}{\mathrm{run}}
\newcommand{\vary}{\mathrm{vary}}

\newcommand{\as}{\alpha_s}
\newcommand{\bmax}{b_\mathrm{max}}
\newcommand{\Ecm}{E_\mathrm{cm}}
\newcommand{\MSbar}{$\overline{\text{MS}}$\xspace}
\newcommand{\muFO}{{\mu_\FO}}
\newcommand{\qTcut}{{q_T^\cut}}
\newcommand{\Tau}{{\mathcal T}}
\newcommand{\TauCut}{{\mathcal{T}_\cut}}

\newcommand{\GeV}{\,\mathrm{GeV}}
\newcommand{\TeV}{\,\mathrm{TeV}}

\allowdisplaybreaks[0]

\title{\boldmath Joint Two-Dimensional Resummation in $q_T$ and $0$-Jettiness at NNLL}

\author[a,b]{Gillian Lustermans,}
\author[c]{Johannes K.~L.~Michel,}
\author[c]{Frank J.~Tackmann,}
\author[a,b]{\\and Wouter J.~Waalewijn}

\affiliation[a]{Institute for Theoretical Physics Amsterdam and Delta Institute for Theoretical Physics, \\University of Amsterdam,\\Science Park 904, 1098 XH Amsterdam, The Netherlands}
\affiliation[b]{Nikhef, Theory Group,\\Science Park 105, 1098 XG, Amsterdam, The Netherlands}
\affiliation[c]{Theory Group, Deutsches Elektronen-Synchrotron (DESY),\\D-22607 Hamburg, Germany}

\emailAdd{g.h.h.lustermans@uva.nl}
\emailAdd{johannes.michel@desy.de}
\emailAdd{frank.tackmann@desy.de}
\emailAdd{w.j.waalewijn@uva.nl}

\abstract{%
We consider Drell-Yan production $pp \to Z/\gamma^* \to \ell^+\ell^-$ with the
simultaneous measurement of the $Z$-boson transverse momentum $q_T$ and
$0$-jettiness $\mathcal{T}_0$. Since both observables resolve the initial-state QCD
radiation, the double-differential cross section in $q_T$ and $\mathcal{T}_0$ contains
Sudakov double logarithms of both $q_T/Q$ and $\mathcal{T}_0/Q$, where $Q \sim m_Z$ is
the dilepton invariant mass. We simultaneously resum the logarithms in $q_T$ and
$\mathcal{T}_0$ to next-to-next-to-leading logarithmic order (NNLL) matched to
next-to-leading fixed order (NLO). Our results provide the first genuinely
two-dimensional analytic Sudakov resummation for initial-state radiation.
Integrating the resummed double-differential spectrum with an appropriate scale choice over either $\mathcal{T}_0$ or
$q_T$ recovers the corresponding single-differential resummation for the
remaining variable. We discuss in detail the required effective field theory
setups and their combination using two-dimensional resummation profile scales.
We also introduce a new method to perform the $q_T$ resummation
where the underlying resummation is carried out in impact-parameter space,
but is consistently turned off depending on the momentum-space target value for
$q_T$. Our methods apply at any order and for any color-singlet production process,
such that our results can be systematically extended when the relevant perturbative
ingredients become available.
}
\date{January 10, 2019}

\preprint{\vbox{%
\hbox{DESY 19-004}
\hbox{NIKHEF 2018-053}
}}


\arxivnumber{1901.03331}

\journalref[https://doi.org/10.1007/JHEP03(2019)124]{JHEP 03 (2019) 124}

\begin{document}

\maketitle

%%%%%%%%%%%%%%%%%%%%%%%%%%%%%%%%%%%%%%%%%%%%%%%%%%%%%%%%%%%%%%%%%%%%%%%%%%%%%%%%
\section{Introduction}
\label{sec:intro}
%%%%%%%%%%%%%%%%%%%%%%%%%%%%%%%%%%%%%%%%%%%%%%%%%%%%%%%%%%%%%%%%%%%%%%%%%%%%%%%%

The increasing accuracy of measurements at the LHC
places high demands on the precision and versatility of theoretical predictions.
Fixed-order perturbation theory has proven to be a powerful tool
to describe a large number of LHC processes, provided the measurement is sufficiently inclusive.
With increasing data sets, however, more fine-grained measurements
become possible and increasingly differential quantities come into focus.
These more exclusive cross sections often involve several physical scales
set by the hard interaction and the differential measurements or
cuts applied on the final state. When these scales are widely separated,
the perturbative series at each order is dominated by logarithms of their ratios.
The resummation of these logarithms to all orders is crucial to arrive at the
best possible predictions.

The resummation for measurements sensitive to infrared (soft and/or collinear) physics
can, in part, be achieved through the use of parton-shower Monte Carlo event generators;
popular examples include \textsc{Pythia}~\cite{Sjostrand:2006za,Sjostrand:2014zea}, \textsc{Herwig}~\cite{Bahr:2008pv,Bellm:2015jjp}, or \textsc{Sherpa}~\cite{Gleisberg:2008ta}.
Parton showers provide fully exclusive final states so that in principle, any desired measurements
or cuts can be imposed on the generated events.
Existing implementations of parton showers are only formally accurate at about
leading-logarithmic (LL) level, depending on the shower's evolution variable (and other
implementation details) and the observable in question. (A recent
detailed analysis can be found in \refcite{Dasgupta:2018nvj}.)
Furthermore, estimating the perturbative uncertainties of parton showers is challenging,
which is in part due to their limited perturbative accuracy.

Analytic methods for the higher-order resummation of infrared-sensitive observables are available.
These include the CSS formalism~\cite{Collins:1985ue,Collins:1988ig,Collins:1989gx},
seminumerical methods based on the coherent-branching formalism~\cite{Banfi:2001bz, Banfi:2004yd, Banfi:2014sua, Bizon:2017rah},
and methods using renormalization group evolution (RGE) in
effective field theories (EFTs) of QCD such as soft-collinear effective theory (SCET)~\cite{Bauer:2000ew,Bauer:2000yr,Bauer:2001ct,Bauer:2001yt,Bauer:2002nz,Beneke:2002ph}.
The common drawback of analytic resummation methods is that they only apply
after a sufficient amount of emissions have been integrated over,
which is why they have been primarily used for the resummation of single-differential observables.
Their crucial advantage is that they can be systematically extended
to higher orders, and theoretical uncertainties can be addressed in a more reliable way.

There has been much progress in extending analytic resummation
methods to cases involving multiple resummation variables. Examples include
the joint resummation of transverse momentum $q_T$
and threshold (large $x$) logarithms~\cite{Li:1998is, Laenen:2000ij, Kulesza:2002rh, Kulesza:2003wn, Lustermans:2016nvk, Marzani:2016smx, Muselli:2017bad}, $q_T$ and small $x$~\cite{Marzani:2015oyb},
$N$-jettiness (or jet mass) together with dijet invariant masses~\cite{Bauer:2011uc, Pietrulewicz:2016nwo},
two angularities~\cite{Larkoski:2014tva, Procura:2018zpn},
jet mass and jet radius~\cite{Kolodrubetz:2016dzb}, jet vetoes and jet rapidity~\cite{Hornig:2017pud, Michel:2018hui},
or threshold and jet radius in inclusive jet production~\cite{Liu:2017pbb, Liu:2018ktv}.
Most of these examples either involve different variables that effectively resolve
different subsequent emissions, or involve a primary resummation variable
that is modified by an auxiliary measurement or constraint.
Another well-understood case is when an infrared-sensitive measurement
is separated into its contributions from mutually exclusive regions of phase
space~\cite{Fleming:2007qr, Jouttenus:2011wh, Bertolini:2017efs}.%
\footnote{Yet another case, which will not be relevant here, arises when different infrared-sensitive
measurements are performed in different regions of phase space, which may require the resummation of nonglobal
logarithms~\cite{Dasgupta:2001sh, Banfi:2002hw, Hatta:2013iba, Larkoski:2015zka, Caron-Huot:2015bja, Becher:2015hka}.}

In contrast, here we are interested in resolving emissions at the same
level by simultaneously measuring two independent
infrared-sensitive observables.
Extending analytic resummation to such genuinely multi-dimensional
resolution variables is of key theoretical concern, as it allows for a more complete description
of the emission pattern beyond LL, effectively filling a gap between
analytic resummations and parton showers.
So far, this has been achieved at NNLL for the case of simultaneously measuring two
angularities in $e^+e^-$ collisions~\cite{Procura:2018zpn}.

In this paper, we consider Drell-Yan, $pp \to Z/\gamma^\ast \to \ell^+ \ell^-$,
with a simultaneous measurement of (1)~the transverse momentum $q_T$ of the Drell-Yan lepton pair
and (2)~the hadronic resolution variable 0-jettiness $\Tau\equiv \Tau_0$~\cite{Stewart:2009yx, Stewart:2010tn}.
Achieving their combined resummation is important conceptually
because $q_T$ and $\Tau$ are prototypes
for two large classes of infrared-sensitive observables:
$q_T$ constrains the transverse momentum of initial-state radiation,
while $\Tau$ constrains its virtuality.
These different behaviors lead to very different logarithmic structures already at LL,
which in SCET is reflected in the RGE structure
of two distinct effective theories, SCET$_\I$ and SCET$_\II$.
(For parton showers, these correspond to evolution variables based on
either transverse momentum or virtuality, respectively.)

Beyond providing a prototype for combining SCET$_\I$ and SCET$_\II$
resummations, the joint resummation of $q_T$ and $\Tau$ is also of direct
phenomenological interest. First, they are important variables
individually.
The measurement of $\Tau$ in bins of $q_T$~\cite{Aad:2016ria}
can probe the so-called underlying event in hadronic collisions.
Furthermore, the \textsc{Geneva}~Monte Carlo event generator~\cite{Alioli:2012fc, Alioli:2015toa}
uses $\Tau$ as the underlying resolution variable for the event generation,
achieving NNLL$'+$NNLO accuracy in $\Tau$ in conjunction with fully showered and hadronized
events. While other observables, such as $q_T$, benefit from the underlying high resummation order, they do not enjoy the same level of formal accuracy in \textsc{Geneva} as $\Tau$ itself.
The joint resummation of $\Tau$ and $q_T$ to a given order enables extending
the event generation in \textsc{Geneva} to also be accurate in $q_T$ to the same order.

\begin{figure*}
\centering
\includegraphics[width=\WidthDiagonalSubfig]{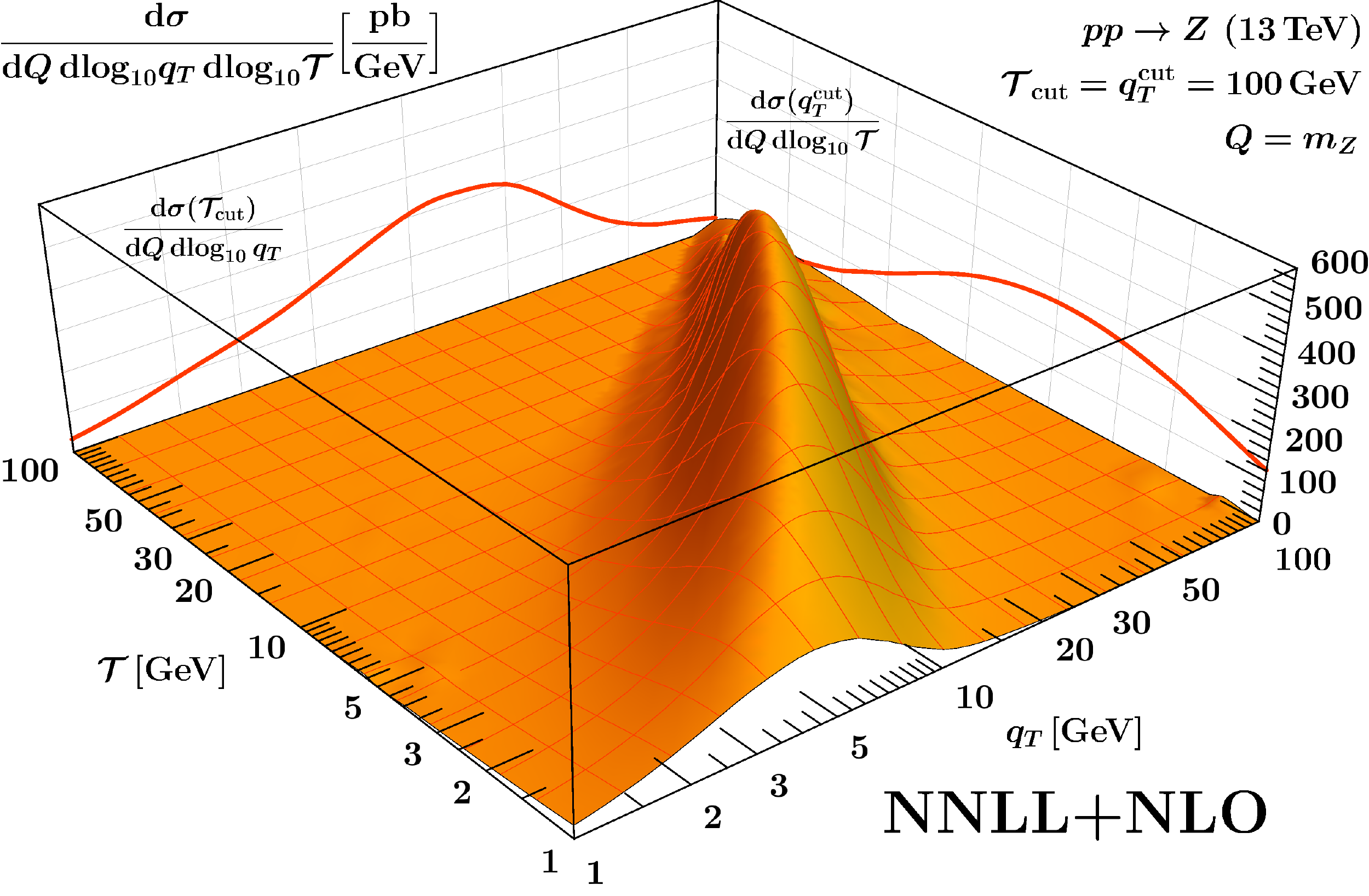}%
\caption{The Drell-Yan cross section
double-differential in the transverse momentum $q_T$ of the $Z$ boson
and the 0-jettiness event shape $\Tau$ at NNLL$+$NLO.
For better visibility, the spectrum is plotted with respect to $\log_{10} q_T$ and $\log_{10} \Tau$.
On the two side walls we show the corresponding single-differential spectra in $q_T$ and $\Tau$
obtained by integrating the double-differential spectrum up to $\TauCut = 100 \GeV$ and
$\qTcut = 100 \GeV$, respectively.}
\label{fig:3d_plot_intro}
\end{figure*}

The double-differential factorization for $q_T$ and $\Tau$ was first considered in \refcite{Procura:2014cba}. There, the regions of phase space where $q_T$ (SCET$_\II$)
and $\Tau$ (SCET$_\I$) determine the resummation structure were identified, together with the appropriate intermediate effective theory SCET$_+$~\cite{Bauer:2011uc,Procura:2014cba} that
connects them.
Here, we develop an explicit matching procedure that combines the three different theories,
SCET$_\I$, SCET$_+$, and SCET$_\II$,
such that the resummation structure of each is recovered in its respective region of phase space.
In particular, our method ensures that the single-differential resummation in one variable
is recovered upon integration over the other. We discuss in detail the technical challenges involved.
These include the construction of appropriate two-dimensional profile scales to combine the SCET$_\II$ resummation for $q_T$, which is performed in position
(impact-parameter) space, with the SCET$_\I$ resummation for $\Tau$, which is performed in momentum space, the estimation of perturbative uncertainties, and the matching to full QCD at large $q_T$ and/or $\Tau$ in a flexible way and consistent with the corresponding single-differential cases.
We obtain explicit numerical predictions for the double-differential $(q_T,\Tau)$ spectrum,
achieving its complete and fully two-dimensional Sudakov resummation at NNLL$+$NLO.
Our main result is shown in \fig{3d_plot_intro}, featuring a nice two-dimensional Sudakov peak
structure.

We like to stress that our methods are completely general and can be applied
to any color-singlet production process and at any order for which the relevant
perturbative ingredients are available. (Some of the double-differential
ingredients required at NNLL$'$ and N$^3$LL are already known~\cite{Gaunt:2014xxa}.)
Furthermore, our matching procedure is
generic and can be applied to any type of two-dimensional resummation
for which the relevant EFTs on the boundaries and in the bulk are known.

The remainder of the paper is organized as follows.
In \sec{framework}, we discuss the three different parametric regimes
and the factorization and resummation for each individually. In \sec{matching},
we  then discuss in detail our method for consistently combining them to obtain
a complete description of the two-dimensional $(q_T,\Tau)$ plane. Our numerical
results for the double-differential spectrum at NNLL$+$NLO are presented in \sec{results}.
We conclude in \sec{conclusions}.
In \app{plus_distributions_and_fourier_transform} we summarize our conventions
for plus distributions and Fourier transforms. All required perturbative
ingredients are collected in \app{perturbative_ingredients}.

%%%%%%%%%%%%%%%%%%%%%%%%%%%%%%%%%%%%%%%%%%%%%%%%%%%%%%%%%%%%%%%%%%%%%%%%%%%%%%%%
\section{Resummation framework}
\label{sec:framework}
%%%%%%%%%%%%%%%%%%%%%%%%%%%%%%%%%%%%%%%%%%%%%%%%%%%%%%%%%%%%%%%%%%%%%%%%%%%%%%%%

%===============================================================================
\subsection{Overview of parametric regimes}
%===============================================================================

\begin{figure*}
\centering
\includegraphics[width=\WidthTwoSubfigs]{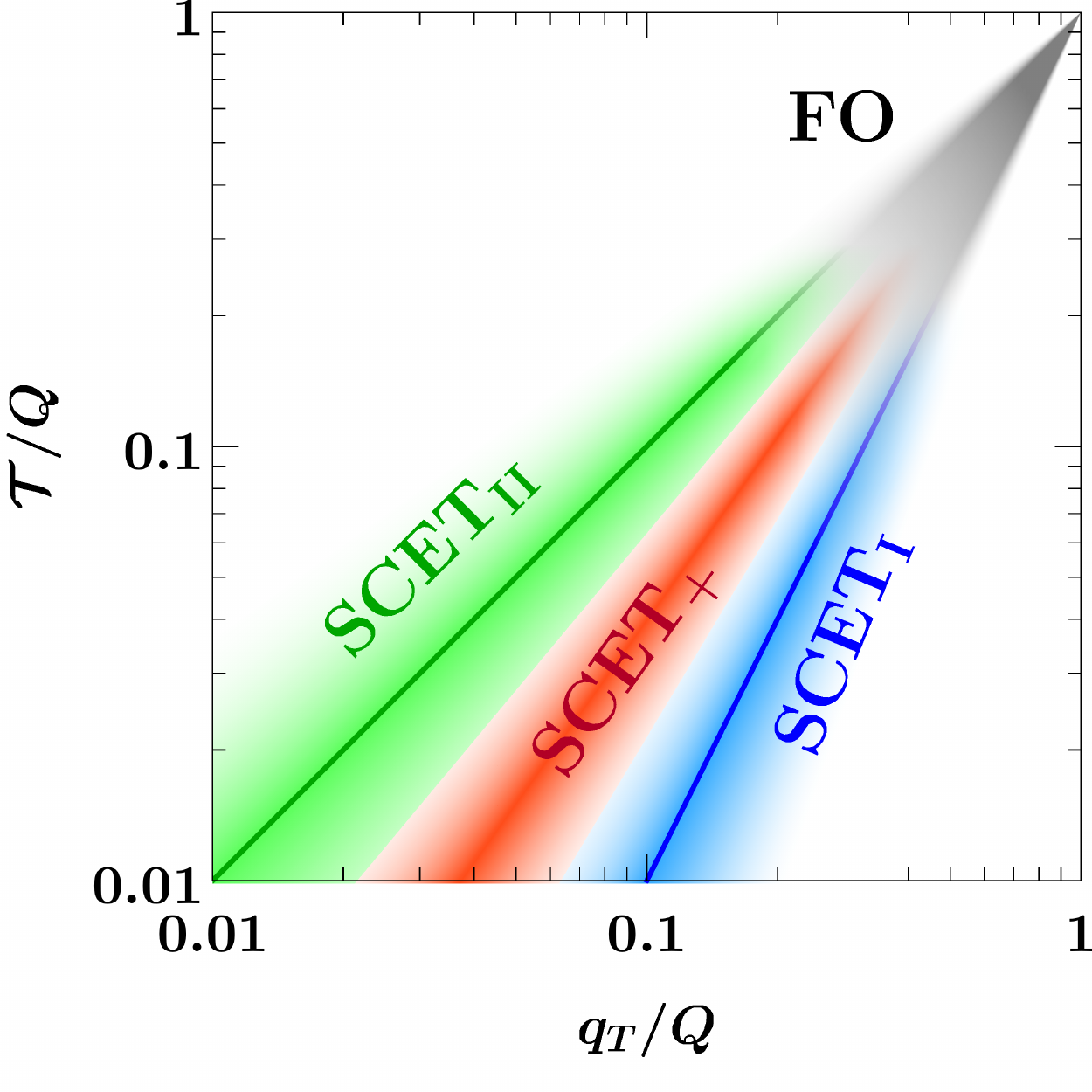}%
\caption{Parametric regimes in the $(q_T,\Tau)$ plane and their SCET description.
The solid lines correspond to the phase-space boundaries $q_T = \Tau$ (green) and $q_T = \sqrt{Q \Tau}$ (blue).}
\label{fig:overview_regimes}
\end{figure*}

We consider color-singlet production at hadron colliders. Although the process dependence
is not important for our discussion, we consider the example of
Drell-Yan production, $pp \to Z/\gamma^\ast (\to \ell^+\ell^-)$, for concreteness.
We measure the total invariant mass $Q$ and rapidity $Y$ of the color-singlet
final state (the lepton pair). The two resolution variables we measure are the
transverse momentum $q_T$ of the color-singlet final state and
the 0-jettiness $\Tau$ (aka beam thrust)~\cite{Stewart:2009yx, Stewart:2010tn, Stewart:2010pd, Jouttenus:2011wh}, defined as
%%%
\begin{align} \label{eq:def_Tau_generic}
\Tau \equiv \Tau_0
= \sum_i\, \min\Bigl\{ \frac{2q_a \cdot k_i}{Q_a}, \frac{2q_b \cdot k_i}{Q_b} \Bigr\}
\,.\end{align}
%%%
The sum runs over all particles $i$ with momentum $k_i$ in the final state,
excluding the color-singlet final state.
We choose the massless reference momenta $q_a$ and $q_b$ as
%%%
\begin{align} \label{eq:def_Tau_reference_momenta}
q_a^\mu = \frac{Q e^{+Y}}{2} \, n_a^\mu
\,, \qquad
q_b^\mu = \frac{Q e^{-Y}}{2} \, n_b^\mu
\,,\end{align}
%%%
where $n_a^\mu = (1,+\hat{z})$ and $n_b^\mu = (1,-\hat{z})$ are lightlike vectors along the beam axis $\hat z$.
For definiteness we use the leptonic definition of 0-jettiness for all numerical results in this paper,
for which the measure factors $Q_{a,b}$ are simply given by
%%%
\begin{equation}
Q_a = Q_b = Q
\,.\end{equation}
%%%
Our setup applies equally well to other definitions of $\Tau$,
so we keep $Q_a$ and $Q_b$ (with $Q_a Q_b = Q^2$) generic for the rest of this section.

We are interested in the contribution of initial-state radiation (ISR)
to the simultaneous measurement of $q_T, \Tau \ll Q$, where $Q \gg \Lambda_\QCD$ sets the scale of
the hard interaction.
The dynamics of perturbative ISR is then governed by three distinct momentum
scales set by the measurement of $q_T$ and $\Tau$.
First, the typical transverse momentum of emissions that recoil against the lepton pair is set by $q_T$.
Second, isotropic (soft) emissions at central rapidities can contribute to $\Tau$ via either
of the projections onto $q_a^\mu$ and $q_b^\mu$ in \eq{def_Tau_generic}.
This implies that their characteristic transverse momentum is $\sim \Tau$.
Third, ISR with typical energy $\sim Q$ can contribute to $\Tau$
as long as it is collinear to either of the incoming beams,
such that its contribution to $\Tau$ in \eq{def_Tau_generic} is small.
These collinear emissions then have a typical transverse momentum $\sim \sqrt{Q\Tau}$.
The factorization and resummation structure of the cross section for $q_T, \Tau \ll Q$
depends on the parametric hierarchy between these scales.
There are three relevant parametric regimes~\cite{Procura:2014cba}, which are illustrated in \fig{overview_regimes}
and are discussed in the following.

In the first (blue) regime, $\Tau \ll q_T \sim \sqrt{Q \Tau}$,
soft emissions with transverse momentum $\sim \Tau$ and collinear emissions with transverse momentum $\sim \sqrt{Q \Tau}$
both contribute to the $\Tau$ measurement.
Due to the separation in transverse momentum, the $q_T$ measurement is determined
by collinear emissions, while soft emissions do not contribute to it.
The appropriate EFT description for this regime is SCET$_\I$. It
has the same RG structure as the single-differential $\Tau$ spectrum,
with $q_T$ acting as an auxiliary variable.
The SCET$_\I$ regime is discussed in more detail in \sec{scet1}.

In the opposite (green) regime, $\Tau \sim q_T \ll \sqrt{Q \Tau}$,
both soft and collinear emissions have transverse momentum $\sim q_T$ and thus
contribute to $q_T$.
On the other hand, only soft radiation at central rapidities contributes to $\Tau$,
while the contribution from collinear radiation is suppressed.
This regime is described by SCET$_\II$, whose RG structure is analogous to that of the
single-differential $q_T$ spectrum, with $\Tau$ as the auxiliary variable.
The SCET$_\II$ regime is discussed in more detail in \sec{scet2}.

Third, the intermediate (orange) regime in the bulk, $\Tau \ll q_T \ll \sqrt{Q \Tau}$,
shares features with both boundary cases.
As in the SCET$_\I$ regime, central soft radiation contributes to $\Tau$,
while as in the SCET$_\II$ regime, collinear radiation contributes to $q_T$.
In addition, this regime requires a distinct collinear-soft mode at an intermediate
rapidity scale that can contribute to both measurements~\cite{Procura:2014cba}. The
relevant EFT description is provided by SCET$_+$, which in this case shares elements of both
SCET$_\I$ and SCET$_\II$. The SCET$_+$ regime, as well as its relation to the regimes
on the two boundaries, is discussed in \sec{scetp}.
We briefly comment on the regions beyond the phase-space boundaries (left blank in \fig{overview_regimes}) in \sec{outer}.

All numerical results for the SCET predictions in the following are obtained from our
implementation in \texttt{SCETlib}~\cite{scetlib}. All fixed NLO results in full QCD
are obtained from \texttt{MCFM~8.0}~\cite{Campbell:1999ah,Campbell:2011bn,Campbell:2015qma}.
Throughout this paper we use \texttt{MMHT2014nnlo68cl}~\cite{Harland-Lang:2014zoa} NNLO PDFs with $\as(m_Z)= 0.118$
and five active quark flavors.

%===============================================================================
\subsection[\texorpdfstring{SCET$_\I$: $\Tau \ll q_T \sim \sqrt{Q \Tau}$}{SCET1: Tau << qT sim sqrt(Q*Tau)}]
                 {\boldmath SCET$_\I$: $\Tau \ll q_T \sim \sqrt{Q \Tau}$}
\label{sec:scet1}
%===============================================================================

In this regime, both soft and collinear modes are constrained by $\Tau$,
while only collinear modes can contribute to $q_T$,
whose characteristic transverse momentum $\sqrt{Q\Tau}$ coincides
parametrically with $q_T$. The scaling of the relevant EFT modes reads
%%%
\begin{align}
n_a \text{-collinear:}
&\quad
p^\mu \sim \Bigl(\Tau, Q, \sqrt{Q \Tau}\Bigr)
\sim \Bigl(\tfrac{q_T^2}{Q}, Q, q_T\Bigr)
\,, \nn \\
n_b \text{-collinear:}
&\quad
p^\mu \sim \Bigl(Q, \Tau, \sqrt{Q \Tau}\Bigr)
\sim \Bigl(Q, \tfrac{q_T^2}{Q}, q_T\Bigr)
\,, \nn \\
\text{soft:}
&\quad
p^\mu \sim \Bigl(\Tau, \Tau, \Tau\Bigr)
\,,\end{align}
%%%
in terms of lightcone coordinates defined by (with $n \equiv n_a$, $\bn \equiv n_b$)
%%%
\begin{equation} \label{eq:lightcone_coordinates}
p^\mu = n \sdot p \, \frac{\bn^\mu}{2} + \bn \sdot p \, \frac{n^\mu}{2} + p_\perp^\mu
\equiv (n \sdot p, \bn \sdot p, p_\perp)
\equiv (p^+, p^-, p_\perp)
\,.\end{equation}
%%%
This leads to the following factorization formula for the cross section~\cite{Stewart:2009yx,Jain:2011iu},
%%%
\begin{align} \label{eq:factorization_scet1}
\frac{\df \sigma_\I}{\df Q\, \df Y\, \df q_T\, \df \Tau}
&= H_\kappa(Q, \mu)
\int\!\df t_a \int\! \df^2 \vec k_a \, B_a(t_a, x_a, \vec k_a, \mu)
\int\!\df t_b \int\! \df^2 \vec k_b \, B_b(t_b, x_b, \vec k_b, \mu)
 \nn\\ &\quad \times
\int\! \df k \, S_\kappa(k, \mu) \,
\delta \bigl( q_T - \abs{\vec k_{a} + \vec k_{b}} \bigr) \,
\delta \Bigl( \Tau - \frac{t_a}{Q_a} - \frac{t_b}{Q_b} - k \Bigr)
\,,\end{align}
%%%
which holds up to power corrections of the form%
\footnote{Lorentz invariance suggests that power corrections in $q_T$ always appear in terms of $q_T^2$.
This distinction is irrelevant for our discussion.}
%%%
\begin{align} \label{eq:power_corrections_scet1}
\frac{\df \sigma}{\df Q\, \df Y\, \df q_T\, \df \Tau}
&= \frac{\df \sigma_\I}{\df Q\, \df Y\, \df q_T\, \df \Tau}\,
\Big[1+ \ORd{\frac{\Tau}{Q}, \frac{q_T^2}{Q^2}, \frac{\Tau^2}{q_T^2}}\Big]
\,.\end{align}
%%%

The hard function $H_\kappa(Q,\mu)$ describes the short-distance scattering that produces the lepton pair through the off-shell $\gamma^\ast$ or $Z$.
In addition to $Q$, it depends on the partonic channel $\kappa \equiv \{a, b\}$,
which is implicitly summed over all relevant combinations of quark and antiquark flavors $a, b$
on the right-hand side of \eq{factorization_scet1}.
The beam functions $B_q(t, x, \vec{k}_T, \mu)$ describe extracting a quark (or antiquark)
from the proton with momentum fraction $x$, virtuality $t$, and transverse momentum $\vec{k}_T$.
The momentum fractions are directly related to $Q$ and $Y$,
%%%
\begin{equation} \label{eq:def_x_ab}
x_a = \frac{Q}{\Ecm}\,e^{+Y}
\,,\qquad
x_b = \frac{Q}{\Ecm}\,e^{-Y}
\,.\end{equation}
%%%
The $t$ and $\vec{k}_T$ encode the contribution of the collinear radiation to the $\Tau$ and $q_T$ measurement,
as captured by the measurement $\delta$ functions on the last line of \eq{factorization_scet1}.
For $t \sim k_T^2 \gg \Lambda_\QCD^2$, these beam functions can be matched onto PDFs~\cite{Stewart:2009yx,Stewart:2010qs,Jain:2011iu},
%%%
\begin{align} \label{eq:matching_beam_function_scet1}
B_q(t,x,\vec{k}_T,\mu)
&= \sum_{j} \int_x^1 \frac{\df z}{z}\,
\cI_{qj}(t, z, \vec{k}_T, \mu) \, f_j\Bigl(\frac{x}{z},\mu\Bigr) \bigg[1 + \ORd{\frac{\Lambda_\QCD^2}{t}, \frac{\Lambda_\QCD^2}{k_T^2}}\bigg]
\,.\end{align}
%%%
The soft function $S_\kappa(k,\mu)$ encodes the contribution from soft radiation to the 0-jettiness measurement, and depends on the color charge of the colliding partons.

The factorization in \eq{factorization_scet1} separates the physics at the canonical
SCET$_\I$ scales
%%%
\begin{equation} \label{eq:canonical_scales_scet1}
\mu_H^\I \sim Q
\,, \qquad
\mu_B^\I \sim \sqrt{Q \Tau}
\,, \qquad
\mu_S^\I \sim \Tau
\,.\end{equation}
% %%%
By evaluating the ingredients at their natural scale and evolving them to a common scale,
all logarithms of $\Tau/Q \sim \mu^\I_S/\mu^\I_H \sim (\mu^\I_B / \mu^\I_H)^2 \sim (\mu^\I_S/\mu^\I_B)^2$ are resummed.

The hard and soft function in \eq{factorization_scet1} are the same as in the
single-differential $\Tau$ spectrum and do not depend on $q_T$.
The RG consistency of the cross section then implies that the RGE of
the double-differential beam functions cannot depend on $q_T$, such that
the overall RG structure of the cross section is equivalent to the single-differential case,
i.e., $q_T$ takes the role of an auxiliary measurement in the SCET$_\I$ resummation,
with no large logarithms of $q_T$ appearing in the cross section
as long as $q_T \sim \sqrt{Q \Tau}$ is satisfied.
We stress that \eq{factorization_scet1} nevertheless provides a nontrivial
and genuinely double-differential extension of the single-differential case. This
is already visible from the structure of power corrections in \eq{power_corrections_scet1}.
Furthermore, the $q_T$ dependence does affect and is affected by the $\Tau$ resummation
because the double-differential beam functions enter in a convolution
with the beam and soft renormalization group kernels. Physically, they
account for the total $q_T$ recoil from all collinear emissions
that are being resummed in $\Tau$.

The factorization of the double-differential spectrum in \eq{factorization_scet1}
(and in the following sections) does not account for effects from Glauber gluon exchange.
For active-parton scattering, they are expected to
enter at $\ord{\alpha_s^4}$ (N$^4$LL$'$)~\cite{Gaunt:2014ska, Zeng:2015iba}, which is well beyond the order we are
interested in. They can be included using the Glauber operator framework of
\refcite{Rothstein:2016bsq}. For proton initial states the factorization formula also does not
account for spectator forward scattering effects. Their complete treatment for
the single-differential $\Tau$ spectrum is not yet available, but we expect that their
treatment for the double-differential case would follow in a similar way.

\paragraph{Scale setting and fixed-order matching.}

\begin{figure*}
\centering
\includegraphics[width=\WidthTwoSubfigs]{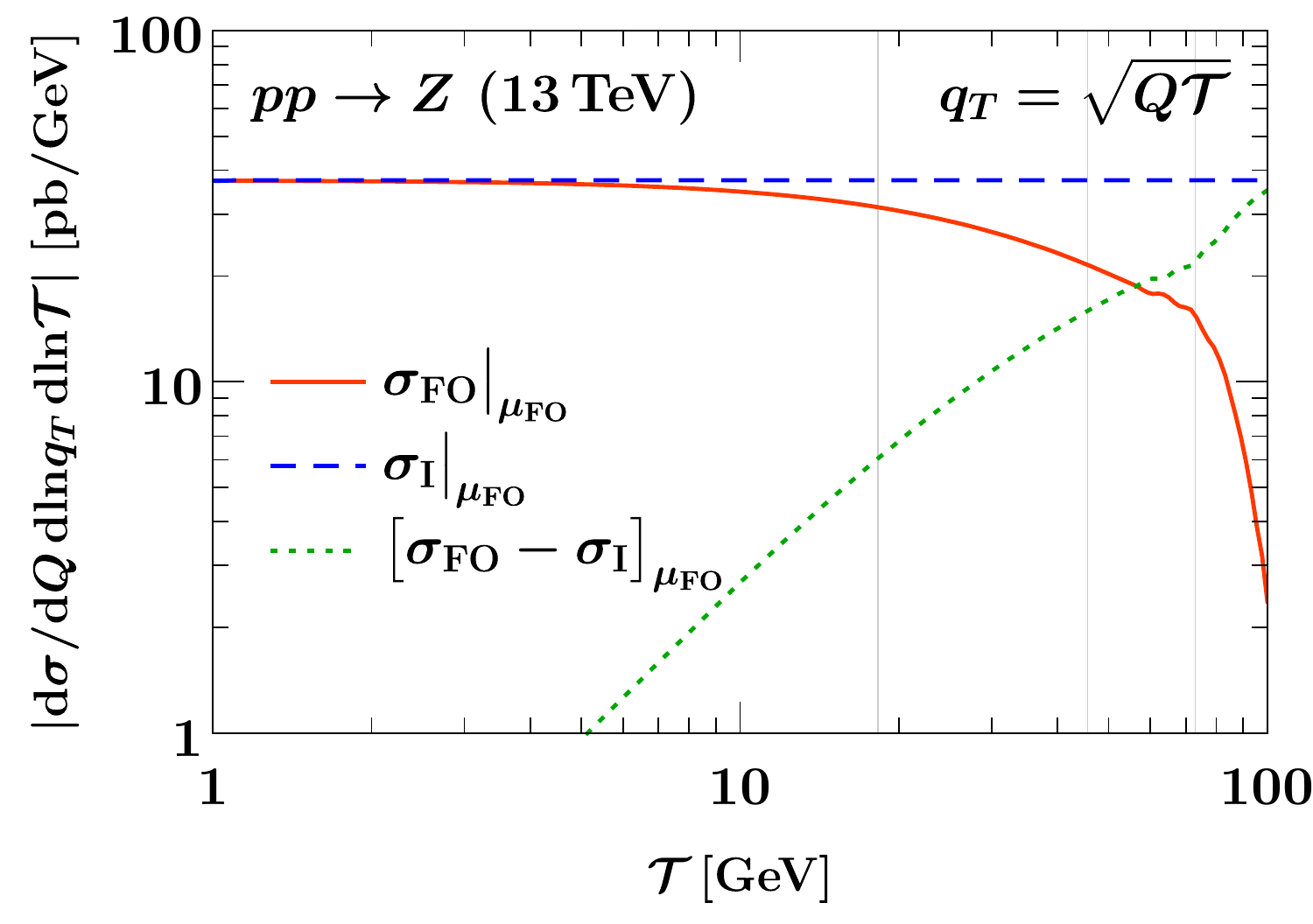}%
\hfill%
\includegraphics[width=\WidthTwoSubfigs]{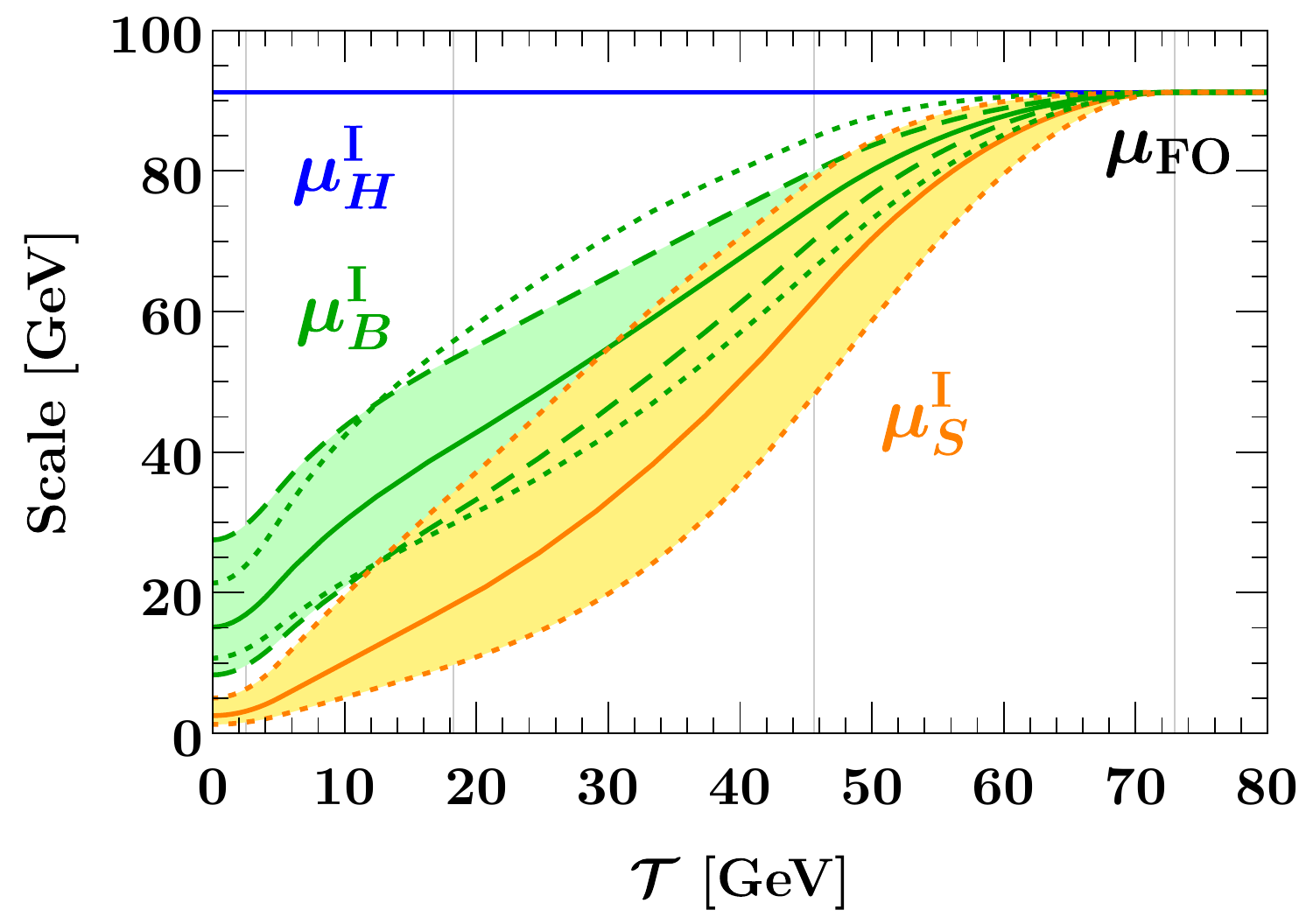}%
\caption{Left: Comparison of singular and nonsingular contributions to the fixed $\ord{\as}$ double spectrum
as a function of $\Tau$, with $q_T = \sqrt{Q \Tau}$ kept fixed.
The orange solid line shows the full QCD result
and the dashed blue line the singular contributions contained in the SCET$_\I$ result
\eq{factorization_scet1}. The dotted green line shows their difference, which
corresponds to the power corrections indicated in \eq{power_corrections_scet1}.
Right: SCET$_\I$ profile scales and their associated variations.
The dotted lines (and the yellow band) indicate common up/down variations of $\mu_S^\I$ and $\mu_B^\I$ from varying $\alpha$.
The dashed lines (and the green band) are variations of $\beta$ that only act on $\mu_B^\I$.
In both plots, the thin vertical lines correspond to the transition points $(x_0, x_1, x_2, x_3)$ given in the text.}
\label{fig:scet1_qcd_singnons}
\end{figure*}

To extend the description of the cross section to large $\Tau \sim q_T^2/Q \lesssim Q$,
we have to reinstate the power corrections dropped in \eq{power_corrections_scet1}.
This is achieved by matching to the full fixed-order result, for which we use the standard
additive matching,
%%%
\begin{equation} \label{eq:matching_scet1}
\df \sigma^\match_\I = \df \sigma_\I \bigr|_{\mu^\I} + \bigl[ \df \sigma_\FO - \df \sigma_\I \bigr]_{\mu_\FO}
\,.\end{equation}
%%%
Here we abbreviated $\df \sigma \equiv \df \sigma / (\df Q \, \df Y \, \df q_T \, \df \Tau)$,
and $\df\sigma_\FO$ denotes the fixed-order cross section in full QCD.
The scale subscripts on the right-hand side indicate
whether $\df\sigma_\I$ is RG evolved using the SCET$_\I$ resummation
scales $\mu^\I$, with their precise choices given below, or whether it is evaluated
with all scales set to a common fixed-order scale $\mu_\FO$.

By construction, $\df\sigma_\I$ evaluated at common scales $\mu_\FO$ exactly reproduces
the singular limit of $\df\sigma_\FO$, such that the term in square brackets in \eq{matching_scet1} is a pure
nonsingular power correction at small $\Tau$, which we can simply add to the resummed cross section.
In the left panel of \fig{scet1_qcd_singnons}, we explicitly check that this is satisfied at fixed $\ord{\as}$,
and numerically assess the size of the power corrections.
We compare the full QCD result (solid orange) to the SCET$_\I$ singular limit (dashed blue)
as a function of $\Tau$, while keeping $q_T = \sqrt{Q \Tau}$ fixed to ensure that all classes of power corrections
in \eq{power_corrections_scet1} uniformly vanish as $\Tau \to 0$. This is
indeed satisfied, as the difference (dotted green) vanishes like a power.

For $\Tau \sim Q$, the SCET$_\I$ singular contribution and the power corrections are of the same size,
implying that the resummation must be turned off to not upset the $\ord{1}$ cancellation between them
and correctly reproduce the fixed-order result.
This is commonly achieved by using profile scales~\cite{Ligeti:2008ac, Abbate:2010xh},
i.e., by having $\mu_B^\I \equiv \mu_B^\I(\Tau)$ and $\mu_S^\I \equiv \mu_S^\I(\Tau)$
transition from their canonical values \eq{canonical_scales_scet1} at small $\Tau$
to a common high scale for large $\Tau$, schematically,
%%%
\begin{equation}
\mu^\I_B(\Tau) \,, \mu^\I_S(\Tau) \to \mu_H^\I = \mu_\FO
\quad \text{for} \quad \Tau \to Q
\,.\end{equation}
%%%
As a result, the first and third term in \eq{matching_scet1} exactly cancel in this limit,
so the matched result reproduces $\df\sigma_\FO$ as desired.

For the concrete choices of $\mu^\I_B$, $\mu^\I_S$ we can rely on those
used for the single-differential spectrum due to the equivalent RG structure.
We use the profile scale setup developed
for the closely related case of SCET$_\I$-like jet vetoes in \refcite{Gangal:2014qda} and
used for the $\Tau$ resummation in \textsc{Geneva}~\cite{Alioli:2015toa}.
The profile scales are chosen as
%%%
\begin{align} \label{eq:SCET1_central}
\mu_S^\I = \muFO\, f_\run^\I\Bigl(\frac{\Tau}{Q}\Bigr)
\,, \qquad
\mu_B^\I = \muFO\, \Bigl[f_\run^\I\Bigl(\frac{\Tau}{Q}\Bigr) \Bigr]^{1/2}
\,, \qquad
\mu_H^\I = \mu_\FO
\,,\end{align}
%%%
with the profile function $f_\run^\I$ given by~\cite{Stewart:2013faa}
%%%
\begin{align} \label{eq:f_run_scet1}
f_\run^\I(x)
&= \begin{cases}
x_0 \Bigl(  1 + \frac{x^2}{4 x_0^2} \Bigr) &x \leq 2x_0\,, \\
x & 2x_0 < x \leq x_1\,, \\
x + \frac{(2-x_2-x_3)(x-x_1)^2}{2(x_2-x_1)(x_3-x_1)} & x_1 < x \leq x_2\,, \\
1 - \frac{(2-x_1-x_2)(x-x_3)^2}{2(x_3-x_1)(x_3-x_2)} & x_2 < x \leq x_3\,, \\
1 & x_3 < x\,.
\end{cases}
\end{align}
%%%
Based on \fig{scet1_qcd_singnons}, we take
$(x_1,x_2,x_3) = (0.2, 0.5, 0.8)$ for the transition points towards the fixed-order region $x \sim 1$.
In addition, \eq{f_run_scet1} turns off the resummation in the nonperturbative region
$x \lesssim 2x_0$, where we set $x_0 = 1 \GeV/Q$.
This cuts off the nonperturbative region and ensures that RG running induced by perturbative anomalous dimensions
always starts from a perturbative boundary condition.
For $\mu_\FO$ itself we use $\mu_\FO = Q$ as the central scale.
Our central scale choices are illustrated as solid lines in the right panel of \fig{scet1_qcd_singnons}.

\paragraph{Perturbative uncertainties.}

We estimate perturbative uncertainties in $\df \sigma_\I^\match$ by considering two different sources.
The first uncertainty contribution $\Delta_\I$ is inherent to the SCET$_\I$ resummation.
It is estimated by varying the individual SCET$_\I$ scales while keeping $\muFO$ fixed,
effectively probing the tower of higher-order logarithms that are being resummed.
For this we use the profile scale variations~\cite{Gangal:2014qda}
%%%
\begin{align} \label{eq:SCET1_vary}
\mu_S^\I
&= \muFO\, \Bigl[f_\vary\Bigl(\frac{\Tau}{Q}\Bigr)\Bigr]^{\alpha}\,
f_\run^\I\Bigl(\frac{\Tau}{Q}\Bigr)
\,, \nn \\
\mu_B^\I
&= \muFO\, \Bigl\{ \Bigl[f_\vary\Bigl(\frac{\Tau}{Q}\Bigr)\Bigr]^{\alpha}\, f_\run^\I\Bigl(\frac{\Tau}{Q}\Bigr) \Bigr\}^{1/2 - \beta}
\,,\end{align}
%%%
where $\alpha = \beta = 0$ corresponds to the central scale choice in \eq{SCET1_central},
and the variation factor is defined as
%%%
\begin{align} \label{eq:f_vary_def}
f_\vary(x) = \begin{cases}
2(1 - x^2/x_3^2) & 0 \leq x < x_3/2\,, \\
1 + 2(1 - x/x_3)^2 & x_3/2 \leq x < x_3\,, \\
1 & x_3 \leq x\,.
\end{cases}
\end{align}
%%%
It approaches a factor of two in the resummation region at small $x$
and reduces to unity toward the fixed-order regime at $x = x_3$, where the resummation
is turned off.
The estimate for $\Delta_\I$ is obtained by computing $\df \sigma^\I_\match$
for each of the four profile scale variations
%%%
\begin{equation}
(\alpha,\beta) = \{(+1,0),(-1,0),(0,+1/6),(0,-1/6)\}
\,,\end{equation}
%%%
and taking the maximum absolute deviation from the central result.
These variations are also indicated in the right panel of \fig{scet1_qcd_singnons}.
Note that for simplicity we do not perform explicit variations of the transition points
since they are known to have a subdominant effect, and
the uncertainty in the fixed-order matching is not essential to this paper.

For the second uncertainty contribution, $\Delta_\FO$,
we consider common variations of $\mu_\FO$ up and down by a factor of two in all pieces of \eq{matching_scet1}.
Since $\mu_\FO$ enters all $\mu^\I$ scales as a common overall factor, they
inherit the same variation, which keeps all resummed logarithms invariant. Hence,
the $\mu_\FO$ variation effectively probes the effect
of missing higher-order corrections in the fixed-order contributions.
The final uncertainty estimate for $\df \sigma_\I^\match$ is obtained by adding
both contributions in quadrature,
%%%
\begin{equation} \label{eq:Delta_I_total}
\Delta^\I_\total = \Delta_\I \oplus \Delta_\FO
\equiv \bigl( \Delta_\I^2 + \Delta_\FO^2 \bigr)^{1/2}
\,.\end{equation}
%%%

The matched result $\df \sigma_\I^\match$ in \eq{matching_scet1} on its own constitutes
a prediction for the double-differential spectrum that
covers the part of phase space where $q_T \sim \sqrt{Q\Tau}$.

%===============================================================================
\subsection[\texorpdfstring{SCET$_\II$: $\Tau \sim q_T \ll \sqrt{Q \Tau}$}{SCET2: Tau sim qT << sqrt(Q*Tau) }]
                 {\boldmath SCET$_\II$: $\Tau \sim q_T \ll \sqrt{Q \Tau}$}
\label{sec:scet2}
%===============================================================================

In this regime, both soft and collinear emissions are constrained by $q_T$.
Only soft radiation is constrained by the $\Tau$ measurement,
while collinear radiation at transverse momenta $\sim q_T \ll \sqrt{Q\Tau}$ is not affected by it.
The relevant EFT modes scale as
%%%
\begin{align}
n_a \text{-collinear:}
&\quad
p^\mu \sim \Bigl(\tfrac{q_T^2}{Q}, Q, q_T\Bigr)
\,, \nn \\
n_b \text{-collinear:}
&\quad
p^\mu \sim \Bigl(Q, \tfrac{q_T^2}{Q}, q_T\Bigr)
\,, \nn \\
\text{soft:}
&\quad
p^\mu \sim \Bigl(q_T, q_T, q_T\Bigr)
\sim \Bigl(\Tau, \Tau, \Tau\Bigr)
\,.\end{align}
%%%
In this case, the cross section factorizes as~\cite{Procura:2014cba}
%%%
\begin{align} \label{eq:factorization_scet2}
\frac{\df \sigma_\II}{\df Q\, \df Y\, \df q_T\, \df \Tau}
&= H_\kappa(Q, \mu)
\int\! \df^2 \vec{k}_a\, B_a(\omega_a, \vec{k}_a, \mu, \nu)
\int\! \df^2 \vec{k}_b\, B_b(\omega_b, \vec{k}_b, \mu, \nu)
\\ &\quad \times
\int\! \df k \int\! \df^2 \vec{k}_s\, S_\kappa(k, \vec{k}_s, \mu, \nu)\,
\delta \bigl(q_T - \abs{\vec k_{a} + \vec k_{b} + \vec{k}_s} \bigr)\,
\delta \bigl(\Tau - k \bigr)
\,.\nn \end{align}
%%%
The factorization receives power corrections of the form
%%%
\begin{align} \label{eq:power_corrections_scet2}
\frac{\df \sigma}{\df Q\, \df Y\, \df q_T\, \df \Tau}
&= \frac{\df \sigma_\II}{\df Q\, \df Y\, \df q_T\, \df \Tau}\,
\Big[1+ \ORd{\frac{\Tau}{Q}, \frac{q_T^2}{\Tau Q}}\Big]
\,.\end{align}
%%%

The hard function $H_\kappa(Q,\mu)$ is the same as in \eq{factorization_scet1}.
In SCET$_\II$ an additional regulator is required to handle rapidity divergences,
for which we use the $\eta$~regulator of \refscite{Chiu:2011qc,Chiu:2012ir}
as implemented at two loops in \refcite{Luebbert:2016itl},
with $\nu$ the corresponding rapidity renormalization scale.
The $B_q(\omega, \vec{k}_T, \mu, \nu)$ are the same transverse
momentum-dependent beam functions as in the single-differential $q_T$ spectrum.
The large momentum components $\omega$ in \eq{factorization_scet2} are given by
%%%
\begin{equation} \label{eq:def_omega_ab}
\omega_a = Q e^{+Y} = x_a \Ecm
\,, \qquad
\omega_b = Q e^{-Y} = x_b \Ecm
\,,\end{equation}
%%%
and we suppress the trivial dependence of the beam function on $\Ecm$.
For $k_T^2 \gg \Lambda_\QCD^2$, the beam functions satisfy a matching relation similar to \eq{matching_beam_function_scet1}~\cite{Collins:1981uw, Collins:1984kg, Becher:2010tm, GarciaEchevarria:2011rb, Chiu:2012ir},
%%%
\begin{align} \label{eq:matching_beam_function_scet2}
B_q(\omega, \vec{k}_T, \mu, \nu)
&= \sum_{j} \int_x^1 \frac{\df z}{z}\,
\cI_{qj}(\omega, \vec{k}_T, z, \mu, \nu) \, f_j\Bigl(\frac{\omega}{z \Ecm},\mu\Bigr) \bigg[1 + \ORd{\frac{\Lambda_\QCD^2}{k_T^2}}\bigg]
\,.\end{align}
%%%
The double-differential soft function $S_\kappa(k, \vec{k}_s, \mu, \nu)$ encodes
the contribution of soft radiation to both $\Tau$ and $q_T$. The RG consistency of
the cross section implies that its $\mu$ and $\nu$ RGEs do not depend on $\Tau$.
Hence, the overall RG structure of the double-differential cross section is
equivalent to the single-differential $q_T$ spectrum, with $\Tau$ acting
as an auxiliary measurement.

The factorization in \eq{factorization_scet2} separates the physics at the canonical SCET$_\II$
invariant-mass and rapidity scales
%%%
\begin{alignat}{3} \label{eq:canonical_scales_scet2}
\mu_H^\II &\sim Q
\,, \qquad
&\mu_B^\II &\sim  q_T
\,, \qquad
&\mu_S^\II &\sim q_T
\,, \nn \\
&
&\nu_B^\II &\sim Q
\,, \qquad
&\nu_S^\II &\sim q_T
\,.\end{alignat}
%%%
It has been known for a long time~\cite{Frixione:1998dw}
that directly resumming the logarithms of $q_T/Q$ in momentum space
is challenging due to the vectorial nature of $\vec q_T$, though
by now approaches for doing so exist~\cite{Monni:2016ktx, Ebert:2016gcn}.
The same complications arise here for the double-differential spectrum.
We bypass this issue, as is commonly done, by carrying out the resummation in
conjugate ($b_T$) space~\cite{Parisi:1979se, Collins:1981uk, Collins:1981va, Collins:1984kg}.
The Fourier transform from $\vec{q}_T$ to $\vec{b}_T$ turns the vectorial
convolutions in \eq{factorization_scet2} into simple products at $b_T \equiv \abs{\vec{b}_T}$.
The canonical SCET$_\II$ scales in $b_T$-space are then given by
%%%
\begin{alignat}{3} \label{eq:canonical_scales_scet2_bT_space}
\mu_H^\II &\sim Q
\,, \qquad
&\mu_B^\II &\sim  b_0/b_T
\,, \qquad
&\mu_S^\II &\sim b_0/b_T
\,, \nn \\
&
&\nu^\II_B &\sim Q
\,, \qquad
&\nu^\II_S &\sim b_0/b_T
\,,\end{alignat}
%%%
where $b_0 \equiv 2 e^{-\gamma_E} \approx 1.12291$ is conventional.
By evaluating the functions in the factorization theorem at their canonical scales
and evolving them to a common scale in both $\mu$ and $\nu$,
all logarithms of $\mu_B/\mu_H \sim \mu_S/\mu_H \sim \nu_S/\nu_B \sim (b_0/b_T) / Q$ are resummed.
In \refcite{Ebert:2016gcn} it was shown that the canonical resummation in
$b_T$ space is in fact equivalent to the exact solution of the RGE in momentum space,
except for the fact that one effectively uses
a shifted set of finite terms in the boundary conditions
(similar to the difference between renormalization schemes).
We exploit this  and require that for $q_T \ll Q$, \eq{canonical_scales_scet2_bT_space} is
exactly satisfied, such that the resummed $q_T$ spectrum in this region is obtained
from the inverse Fourier transform of the canonical $b_T$-space result.

A key feature of the resummed $q_T$ spectrum is
that the anomalous dimension $\gamma^i_\nu$, driving the $\nu$ running of the soft (or beam) function at fixed $\mu$,
is itself perturbatively renormalized at its intrinsic scale $\mu_0$
and requires resummation when $\mu_0 \neq \mu$.
Specifically, in the exponent of the $b_T$-space rapidity evolution factor we have
%%%
\begin{align} \label{eq:resummed_rapidity_anom_dim}
\tgamma^i_{\nu}(b_T, \mu)
&= -4\eta^i_\Gamma(\mu_0, \mu) + \tgamma^i_{\nu,\FO}(b_T, \mu_0) + \tgamma^i_{\nu,\np}(b_T)
\,,\end{align}
%%%
where all logarithms of $\mu/\mu_0$ are resummed inside
%%%
\begin{equation}
\eta_\Gamma^{i}(\mu_0,\mu)
= \int_{\mu_0}^\mu \! \frac{\df \mu'}{\mu'} \, \Gamma^{i}_{\rm cusp}[\as(\mu')]
\,.\end{equation}
[See eq.~(4.26) in \refcite{Ebert:2016gcn} for the analogous expressions in momentum space.]
The canonical choice of $\mu_0$ that eliminates all large logarithms in the fixed-order boundary condition $\tgamma_\nu^\FO(b_T, \mu_0)$ is
%%%
\begin{equation} \label{eq:canonical_scales_scet2_bT_space_mu0}
\mu_0 \sim b_0/b_T
\,.\end{equation}
%%%
By choosing $\mu_0$ as a function of $b_T$ such that it freezes out to a perturbative value at large $b_T$,
we avoid the Landau pole at $b_T \sim 1 / \Lambda_\QCD$.%
\footnote{In addition, this leaves fixed-order logarithms of $\mu_0 b_T$ in $\tgamma^i_{\nu, \FO}$
that lead to an exponential suppression of the $b_T$ space cross section as $b_T \to \infty$.
This increases the numerical stability of the inverse Fourier transform.}
The mismatch to the full result can in principle be captured by a nonperturbative model $\tgamma^i_{\nu,\np}(b_T)$,
which can be extracted from experimental measurements at small $q_T$.
Recently, it was shown that it could also be determined from lattice calculations~\cite{Ebert:2018gzl}.
For our purposes we set $\tgamma^i_{\nu,\np} = 0$ for simplicity.
(We similarly ignore nonperturbative effects in the SCET$_\II$ beam and soft function boundary conditions.) Our concrete choice of $\mu_0$ is given below.

\paragraph{Scale setting and fixed-order matching.}

\begin{figure*}
\centering
\includegraphics[width=\WidthTwoSubfigs]{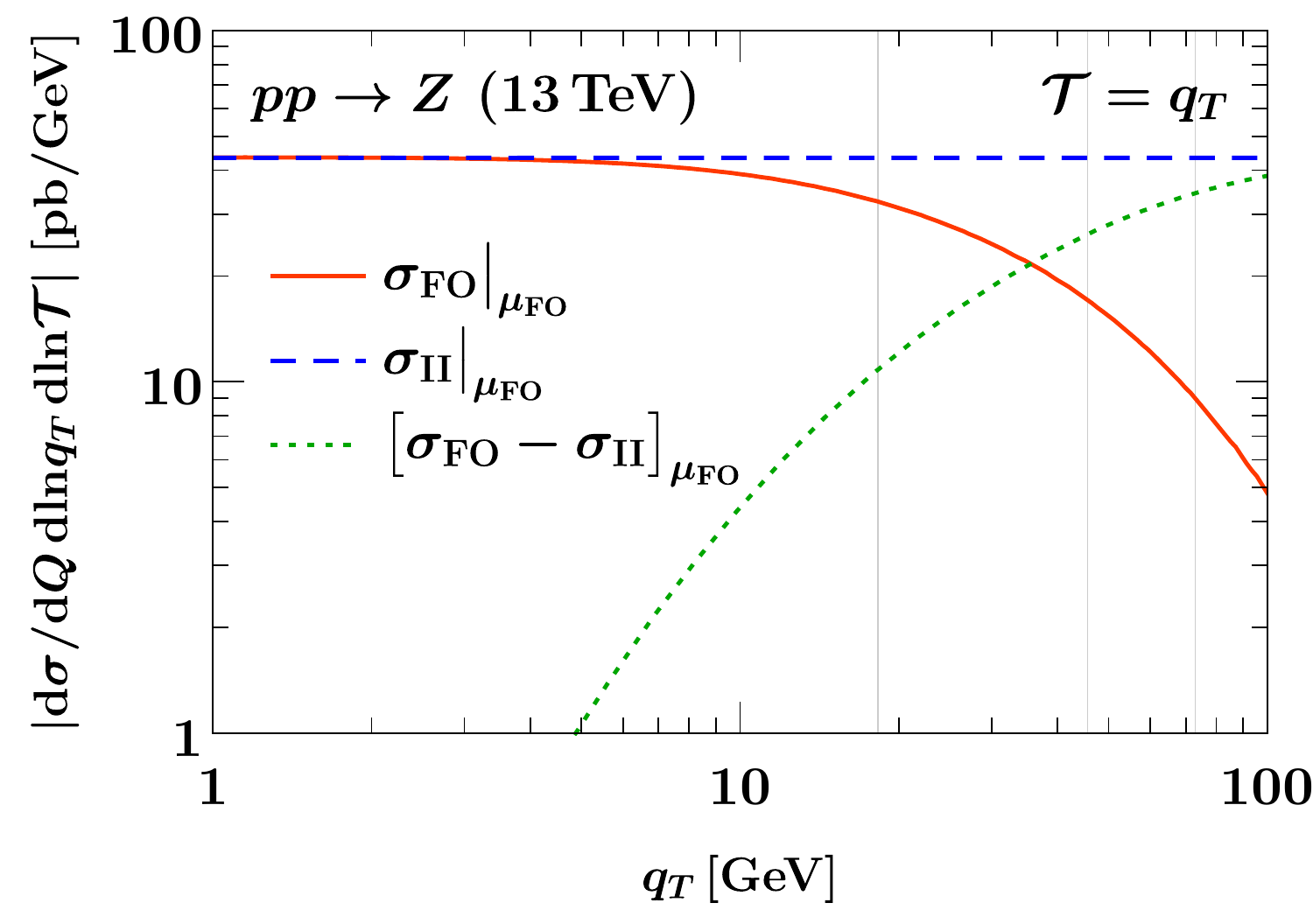}%
\hfill%
\includegraphics[width=\WidthTwoSubfigs]{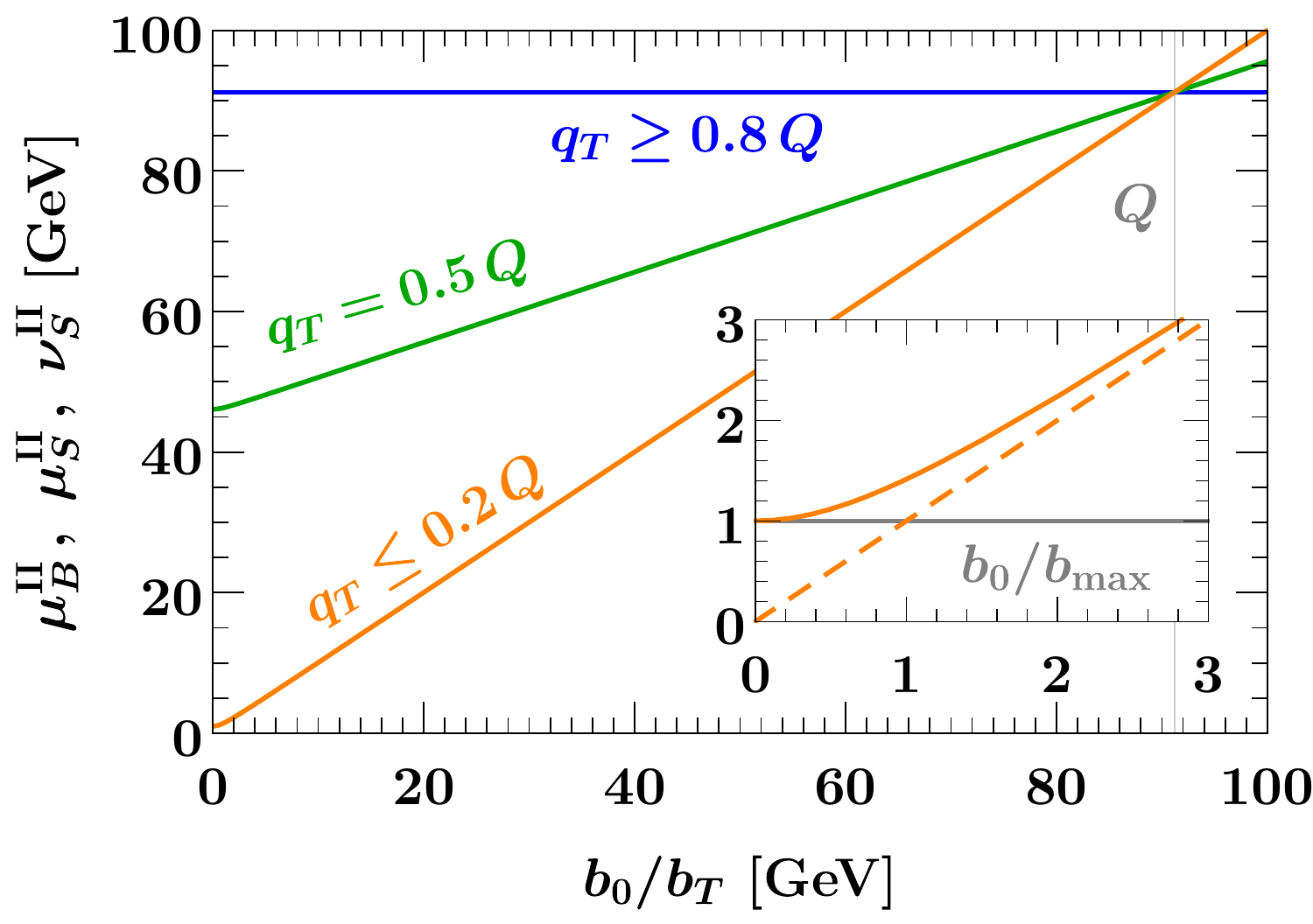}%
\caption{Left: Comparison of singular and nonsingular contributions to the fixed $\ord{\as}$ double spectrum as a function of $q_T$, with $\Tau = q_T$ kept fixed.
The orange solid line shows the full QCD result
and the dashed blue line the singular contributions contained in the SCET$_\II$ result
\eq{factorization_scet2}. The dotted green line shows their difference,
which corresponds to the power corrections indicated in \eq{power_corrections_scet2}.
The thin vertical lines correspond to the transition points $(x_1, x_2, x_3)$ given in the text.
Right: SCET$_\II$ hybrid profile scales as a function of $b_0/b_T$ for representative values of $q_T$.
The thin vertical line in the main plot corresponds to $Q$.
The inset shows the behavior of the profile in the nonperturbative region $b_0/b_T \sim \Lambda_\QCD$,
where the gray horizontal line indicates the scale $b_0/\bmax$ at which we freeze out the resummation.
The dashed orange line in the inset indicates the canonical value of $\mu_B^\II, \mu_S^\II, \nu_S^\II$.}
\label{fig:scet2_qcd_singnons}
\end{figure*}

We again extend the description of the cross section to the fixed-order region $q_T \sim \Tau \lesssim Q$ by an additive matching,
%%%
\begin{equation} \label{eq:matching_scet2}
\df \sigma^\match_\II = \df \sigma_\II \bigr|_{\mu^\II} + \bigl[ \df \sigma_\FO - \df \sigma_\II \bigr]_{\mu_\FO}
\,.\end{equation}
%%%
Here the subscript $\mu^\II$ indicates
that we evaluate $\df\sigma_\II$ at the SCET$_\II$ resummation scales
$\mu^\II$ (given below) in $b_T$ space, and take a numerical inverse Fourier transform in the end.
The subscript $\mu_\FO$ indicates that it is instead evaluated at common fixed-order scales $\mu_\FO$,
which can be done directly in momentum space.

Analogous to the discussion for SCET$_\I$, the term in square brackets in \eq{matching_scet2} is
by construction a pure nonsingular power correction at small $q_T$.
This is illustrated in the left panel of \fig{scet2_qcd_singnons}, which shows that
the difference (green dotted) between the full QCD result (solid orange) and the
SCET$_\II$ singular result (dashed blue) indeed vanishes like a power as $q_T \to 0$
along the line of fixed $\Tau = q_T$.

Approaching $q_T \sim \Tau \sim Q$, the $q_T$ resummation must again be turned off
to ensure the delicate cancellations between singular and nonsingular contributions
and to properly recover the correct fixed-order result for the spectrum.
We achieve this by constructing hybrid profile scales
that depend on both $b_T$ and $q_T$,
and undergo a continuous deformation away from the canonical $b_T$ scales in
\eq{canonical_scales_scet2} as a function of the target $q_T$ value, schematically,
%%%
\begin{equation} \label{eq:scet2_profile_requirements}
\mu_{B,S}^\II(q_T, b_T)
\,,
\nu_{B,S}^\II(q_T, b_T)
\to \mu^\II_H = \mu_\FO
\quad \text{for} \quad q_T \to Q
\,.\end{equation}
%%%
We note that $\mu_0$ does not need to asymptote to $\mu_\FO$ towards large $q_T$
because its effect on the matched result is already turned off
as $\nu_S^\II \to \nu_B^\II$. In this limit, the first and last term in
\eq{matching_scet2} exactly cancel, leaving the fixed-order result $\df\sigma_\FO$.

Since the single-differential $q_T$ resummation is not the main focus of this paper,
we strive to achieve \eq{scet2_profile_requirements} in the simplest possible way.
Specifically, we choose central scales as
%%%
\begin{align} \label{eq:central_scales_scet2}
&\mu^\II_H = \nu^\II_B = \muFO
\,, \qquad
\mu^\II_B = \mu^\II_S = \nu^\II_S =
\muFO\, f_\run^{\rm II}\Bigl( \frac{q_T}{Q}, \frac{b_0}{b^\ast(b_T)\, Q} \Bigr)
\,, \qquad
\mu_0 = \frac{b_0}{b^\ast(b_T)}
\,,\end{align}
%%%
where $f_\run^\II$ is a hybrid profile function given by
%%%
\begin{align} \label{eq:fRun2}
f_\run^\II(x,y) &= 1 + g_\run(x)(y-1)
\,.\end{align}
%%%
It controls the amount of resummation by adjusting the slope of the scales in $b_T$ space
as a function of $q_T/Q$ via the function
%%%
\begin{align}
g_\run(x) &= \begin{cases}
1 & 0 < x \leq x_1 \,, \\
1 - \frac{(x-x_1)^2}{(x_2-x_1)(x_3-x_1)} & x_1 < x \leq x_2
\,, \\
\frac{(x-x_3)^2}{(x_3-x_1)(x_3-x_2)} & x_2 < x \leq x_3
\,, \\
0 & x_3 \leq x
\,.\end{cases}
\end{align}
%%%
As a result, for $q_T \leq x_1 Q$, the slope is unity yielding the canonical resummation,
while for $q_T \geq x_3 Q$, the slope vanishes so the resummation is fully turned off.
In between, the slope smoothly transitions from one to zero, which transitions the resummation from being
canonical to being turned off. This is illustrated in the right panel of \fig{scet2_qcd_singnons}.
We use the same transition points $(x_1, x_2, x_3) = (0.2, 0.5, 0.8)$ as for SCET$_\I$,
which is supported by \fig{scet2_qcd_singnons}.

We note that our approach differs from the hybrid profile scales introduced in \refcite{Neill:2015roa}.
While the latter also satisfy the requirement in \eq{scet2_profile_requirements},
they do not reproduce the exact canonical $b_T$-space scales for $q_T \ll Q$ because they
introduce a profile shape directly in $b_T$ space.

As discussed below \eq{canonical_scales_scet2_bT_space_mu0},
we require a nonperturbative prescription
when the canonical value of $\mu_0$ (or $\mu_S^\II$, or $\mu_B^\II$)
approaches the Landau pole $b_0/b_T \sim \Lambda_\QCD$.
This is encoded in evaluating the hybrid scales at $b^\ast(b_T)$ rather than $b_T$ itself,
%%%
\begin{align} \label{eq:def_bstar}
b^\ast(b_T) = \frac{b_T}{\sqrt{1+b_T^2/\bmax^2}}
\,,\end{align}
%%%
where $b_0/\bmax \gtrsim \Lambda_\QCD$ ensures that all scales are canonical for small $b_T \approx b^\ast$,
but remain perturbative for large $b_T$ where $b^\ast \to \bmax$,
as shown in the inset in the right panel of \fig{scet2_qcd_singnons}.
In practice we pick
%%%
\begin{equation}
b_0/\bmax = 1 \GeV
\,,\end{equation}
%%%
in keeping with our choice of nonperturbative turn-off parameter in the SCET$_\I$ case.
The functional form of \eq{def_bstar} is the same as in the standard $b^\ast$ prescription~\cite{Collins:1981uk, Collins:1981va},
although any other functional form with the same asymptotic behavior is also viable.
We stress, however, that a key difference in our case is that $b^\ast$ only affects the scales,
so it essentially serves the same purpose as the $x_0$ nonperturbative cutoff in the SCET$_\I$ scales in
\eq{f_run_scet1}.
By contrast, the standard $b^\ast$ prescription corresponds to a \emph{global} replacement of $b_T$ by $b^\ast$,
including the measurement itself.
For the single-differential $q_T$ spectrum,
this global replacement induces power corrections $\ord{b_T^2 / \bmax^2}$
that scale like a generic nonperturbative contribution.
While they might complicate the extraction of nonperturbative model parameters from data~\cite{DAlesio:2014mrz}, they are not a critical issue.

For the double-differential case,
we find that a standard $b^\ast$ prescription does in fact not work.
This is because substituting $b^\ast$ for $b_T$ in the physical measurement
renders Fourier integrals of the double-differential SCET$_\II$ soft function divergent,
at least at fixed order (i.e., without Sudakov suppression).
This can be seen from \eqs{delta_soft_position_space_cumulant}{delta_soft_position_space_spectrum},
which only depend on $x = b_T \Tau$.
Substituting $b^\ast$ for $b_T$ makes them asymptote to a constant for any given $\Tau$,
which upsets their required asymptotic behavior $\sim 1/x^2$.
Physically this means that the deformation of the measurement at large $b_T$
also deforms the observable of interest, i.e., the dependence on $\Tau$.

\paragraph{Perturbative uncertainties.}

To estimate the resummation uncertainty for $\df \sigma_\II^\match$,
we adopt the set of profile scale variations introduced for the SCET$_\II$-like jet veto
in \refcite{Stewart:2013faa}. They are given by
%%%
\vspace{-1.5ex}
\begin{align}\label{eq:SCET2_vary}
\mu_S^\II
&= \muFO\, \Bigl[ f_\vary\Bigl( \frac{q_T}{Q} \Bigr) \Bigr]^{v_{\mu_S}}\,
f_\run^\II\Bigl( \frac{q_T}{Q} , \frac{b_0}{b^\ast Q} \Bigr)
\,, \nn \\
\nu_S^\II
&= \muFO\, \Bigl[ f_\vary\Bigl( \frac{q_T}{Q} \Bigr) \Bigr]^{v_{\nu_S}}\,
f_\run^\II\Bigl( \frac{q_T}{Q} , \frac{b_0}{b^\ast Q} \Bigr)
\,, \nn \\
\mu_B^\II
&= \muFO\, \Bigl[ f_\vary\Bigl( \frac{q_T}{Q} \Bigr) \Bigr]^{v_{\mu_B}}\,
f_\run^\II\Bigl( \frac{q_T}{Q} , \frac{b_0}{b^\ast Q} \Bigr)
\,, \nn \\
\nu_B^\II
&= \muFO\, \Bigl[ f_\vary\Bigl( \frac{q_T}{Q} \Bigr) \Bigr]^{v_{\nu_B}}
\,,\end{align}
%%%
where each of the four variation exponents can be $v_i = \{+1,0,-1\}$,
and $f_\vary$ was given in \eq{f_vary_def}.
The central scale choice corresponds to $(v_{\mu_S}, v_{\nu_S}, v_{\mu_B}, v_{\nu_B}) = (0,0,0,0)$,
and a priori there are 80 possible different combinations of the $v_i$.
Since the arguments of the resummed logarithms are ratios of scales,
some combinations of scale variations will lead to variations of these arguments
that are larger than a factor of two, and therefore should be excluded~\cite{Stewart:2013faa}.
After dropping these combinations we are left with 36 different scale variations for the SCET$_\II$ regime.
We add two independent variations of $b_0/\bmax = \{0.5 \GeV, 2\GeV\}$
to probe the uncertainty in our nonperturbative prescription.
The SCET$_\II$ resummation uncertainty $\Delta_\II$ is then determined
as the maximum absolute deviation from the central result among all 38 variations.
For simplicity we again refrain from variations of the transition points.
As for SCET$_\I$, $\Delta_\FO$ is estimated by overall variations of $\mu_\FO$ by a factor of two,
which is inherited by all SCET$_\II$ scales, so it probes the fixed-order uncertainties
while leaving the resummed logarithms invariant.
The total uncertainty estimate for $\df \sigma_\II^\match$ is then obtained as
%%%
\begin{equation} \label{eq:Delta_II_total}
\Delta^\II_\total = \Delta_\II \oplus \Delta_\FO
\,.\end{equation}
%%%

The matched result $\df \sigma_\II^\match$ in \eq{matching_scet2}
provides a prediction for the double-differential spectrum that covers the part of
phase space where $\Tau\sim q_T$.

\paragraph{Results for the single-differential spectrum.}

\begin{figure*}
\centering
\includegraphics[width=\WidthTwoSubfigs]{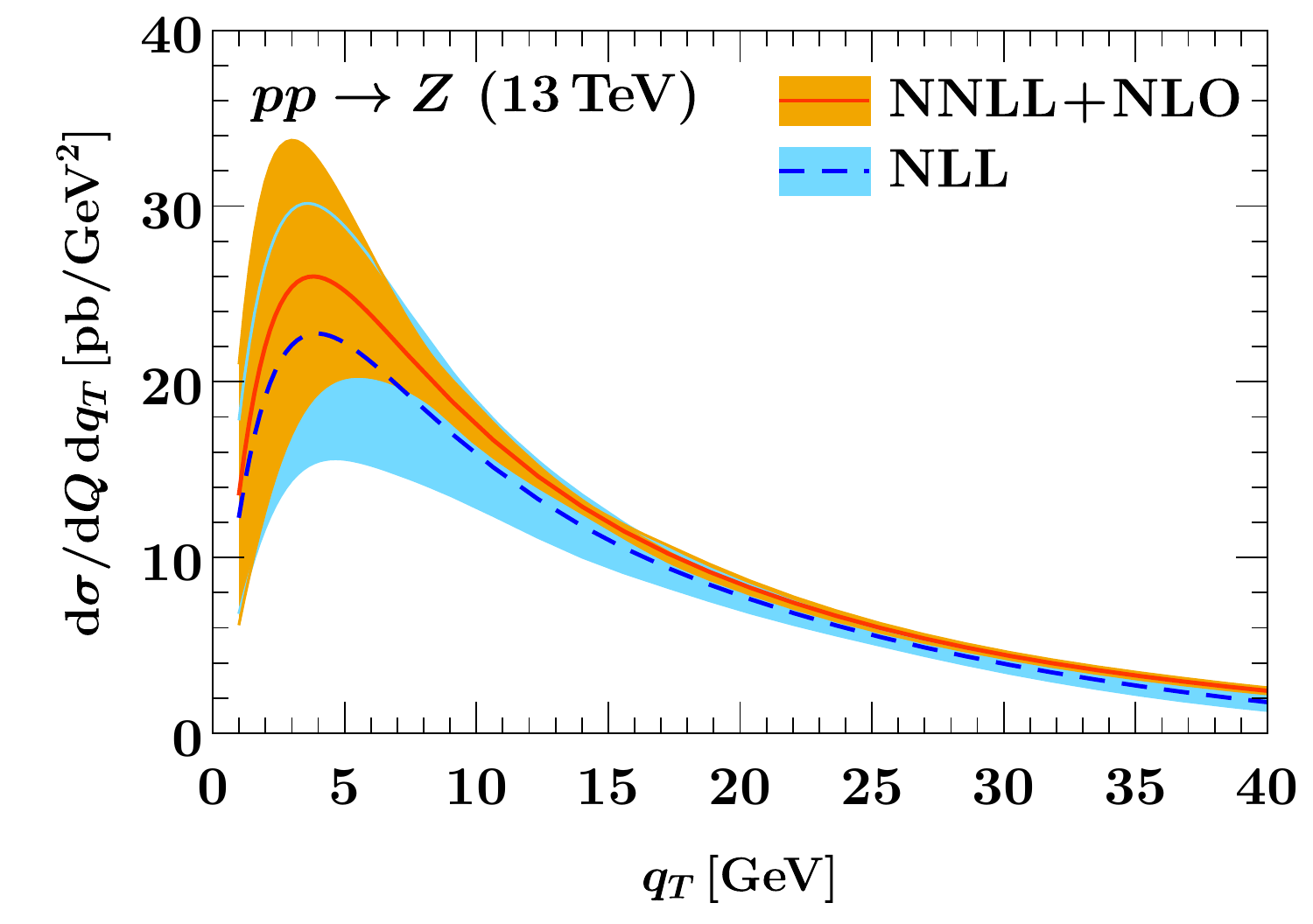}%
\hfill%
\includegraphics[width=\WidthTwoSubfigs]{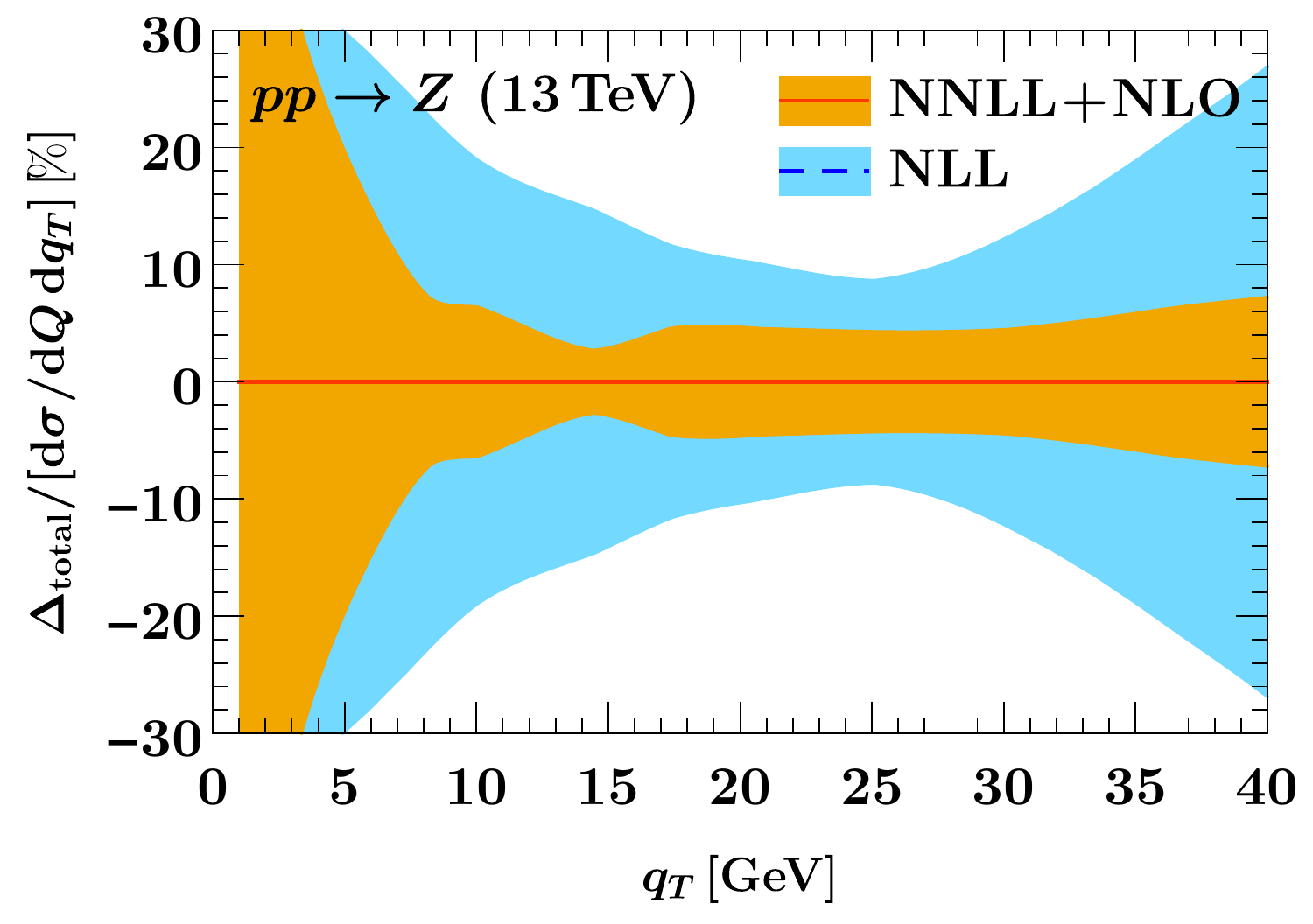}%
\caption{The single-differential $q_T$ spectrum at NLL (blue) and NNLL$+$NLO (orange),
using the $q_T$ resummation method described in the text.
The bands indicate $\Delta_\II \oplus \Delta_\FO$.
In the right panel, the uncertainties are shown as percent differences relative to the
central result at each order.}
\label{fig:single_differential_qT}
\end{figure*}

Since we are using a new method to perform the $q_T$ resummation, we also briefly
consider the single-differential $q_T$ spectrum as a sanity check of our setup.
The setup described in this section immediately carries over to the single-differential
spectrum.
In \fig{single_differential_qT} we show the $q_T$ spectrum at the
NNLL$+$NLO order we are aiming for for the double-differential spectrum, as well as
one order lower at NLL, and with the uncertainties estimated as described above.
The results look very reasonable, providing us with confidence in
our $q_T$ resummation procedure. Note that there is a slight pinch in the
uncertainty bands around $q_T = 15\GeV$, indicating that the uncertainties there
are likely a bit underestimated. This is an artifact of
scale variations that is not unusual to be seen in resummed spectrum predictions.

%===============================================================================
\subsection[\texorpdfstring{SCET$_+$: $\Tau \ll q_T \ll \sqrt{Q \Tau}$}{SCET+: Tau << qT << sqrt(Q*Tau) }]
                 {\boldmath SCET$_+$: $\Tau \ll q_T \ll \sqrt{Q \Tau}$}
\label{sec:scetp}
%===============================================================================

This regime is characterized by the presence of intermediate collinear-soft modes
that contribute both to the $q_T$ and the $\Tau$ measurement,
which uniquely fixes their scaling.
Central soft modes only contribute to $\Tau$ as in SCET$_\I$,
while the energetic collinear modes only contribute to $q_T$ as in SCET$_\II$,
%%%
\begin{align} \label{eq:modes_scetp}
n_a \text{-collinear:}
&\quad
p^\mu \sim \Bigl(\tfrac{q_T^2}{Q}, Q, q_T\Bigr)
\,, \qquad &
n_a \text{-collinear-soft:}
\quad
p^\mu \sim \Bigl(\Tau, \frac{q_T^2}{\Tau}, q_T\Bigr)
\,, \nn \\
n_b \text{-collinear:}
&\quad
p^\mu \sim \Bigl(Q, \tfrac{q_T^2}{Q}, q_T\Bigr)
\,, \qquad &
n_b \text{-collinear-soft:}
\quad
p^\mu \sim \Bigl(\frac{q_T^2}{\Tau}, \Tau, q_T\Bigr)
\,, \nn \\
\text{soft:}
&\quad
p^\mu \sim \Bigl(\Tau, \Tau, \Tau\Bigr)
\,.\end{align}
%%%
The collinear-soft modes have the same virtuality as the collinear
modes, $p^2 \sim q_T^2$, but live at more central rapidity $e^{\abs{y}} \sim q_T/\Tau$,
which is small compared to the rapidity $e^{\abs{y}} \sim Q/q_T$ of the collinear modes. Hence, the
two have a  SCET$_\II$-like relation and become a single collinear mode in the
SCET$_\I$ limit $q_T\sim \sqrt{Q\Tau}$.
At the same time, the collinear-soft and soft modes have a SCET$_\I$-like relation, being separated
in virtuality, and become a single soft mode in the SCET$_\II$ limit $\Tau\sim q_T$.
In this way, SCET$_+$ is able to connect the SCET$_\I$ and SCET$_\II$ regimes.
This is similar to the collinear-soft mode originally introduced in \refcite{Bauer:2011uc},
which instead connected two SCET$_\I$ theories.

The cross section in SCET$_+$ factorizes as~\cite{Procura:2014cba}
%%%
\begin{align} \label{eq:factorization_scetp}
\frac{\df \sigma_+}{\df Q\, \df Y\, \df q_T\, \df \Tau}
&= H_\kappa(Q, \mu)
\int\! \df^2 \vec{k}_a\, B_a(\omega_a, \vec{k}_a, \mu, \nu)
\int\! \df^2 \vec{k}_b\, B_b(\omega_b, \vec{k}_b, \mu, \nu)
\\ & \quad \times
\int\! \df \ell_a^+ \int\! \df^2 \vec{\ell}_a\, \cS_\kappa(\ell_a^+, \vec{\ell}_a, \mu, \nu)
\int\! \df \ell_b^- \int\! \df^2 \vec{\ell}_b\, \cS_\kappa(\ell_b^-, \vec{\ell}_b, \mu, \nu)
\nn \\ & \quad \times
\int\! \df k\,S_\kappa(k, \mu)\,
\delta \bigl(q_T - \abs{\vec{k}_a + \vec{k}_b + \vec{\ell}_a + \vec{\ell}_b}\bigr)\,
\delta \Bigl(\Tau - \frac{\omega_a \ell_a^+}{Q_a} - \frac{\omega_b \ell_b^-}{Q_b} - k\Bigr)
\,,\nn\end{align}
%%%
which holds up to power corrections
%%%
\begin{align} \label{eq:power_corrections_scetp}
\frac{\df \sigma}{\df Q\, \df Y\, \df q_T\, \df \Tau}
&= \frac{\df \sigma_+}{\df Q\, \df Y\, \df q_T\, \df \Tau}\,
\Big[1+ \ORd{\frac{q_T^2}{\Tau Q}, \frac{\Tau^2}{q_T^2}}\Big]
\,.\end{align}
%%%
The hard function is the same as before. The beam functions are the $q_T$-dependent
ones from SCET$_\II$, while the soft function is the $\Tau$-dependent one from SCET$_\I$.
The new ingredient is the double-differential collinear-soft function $\cS_\kappa(k,\vec{k}_T,\mu,\nu)$,
which encodes the contributions of the collinear-soft modes to both $q_T$ and $\Tau$.
Like the soft function it is defined as a matrix element of eikonal Wilson lines,
but like the beam functions it describes radiation that goes into a definite hemisphere.

Equation~\eqref{eq:factorization_scetp} can be interpreted
as a refactorization of the double-differential SCET$_\I$ and SCET$_\II$ cross sections~\cite{Procura:2014cba},
which precisely reflects the relation between the EFT modes described above.
Expanding the SCET$_\I$ double-differential beam function in the limit $q_T \ll \sqrt{Q\Tau}$,
it factorizes into the SCET$_\II$ beam function and the collinear-soft function,
%%%
\begin{align} \label{eq:refactorization_beam_function}
B_q(\omega k, \omega/\Ecm, \vec{k}_T, \mu)
&= \int\! \df^2 \vec{\ell}_T \, B_q(\omega, \vec{k}_T - \vec{\ell}_T, \mu, \nu) \,
\cS_\kappa(k, \vec{\ell}_T, \mu, \nu) \,
\Bigl[1 + \ORd{\frac{k_T^2}{\omega k}} \Bigr]
\,.\end{align}
%%%
The $\nu$ dependence of the two terms on the right-hand side must cancel, while their
$\mu$ dependence must combine into that of the left-hand side.
This allows us to derive the RGE for the collinear-soft function given
in \eq{rge_scetp_csoft}.

Similarly, expanding the SCET$_\II$ double-differential soft function in the limit $\Tau\ll q_T$,
it factorizes into the SCET$_\I$ soft function and the two $n_a$-collinear-soft and $n_b$-collinear-soft
functions,
%%%
\begin{align} \label{eq:refactorization_soft_function}
S_\kappa(k, \vec{k}_T, \mu, \nu)
&= \int\! \df^2 \vec{\ell}_T \, \int\! \df \ell_a^+ \, \cS_\kappa(\ell_a^+, \vec{\ell}_T, \mu, \nu) \,
\int\! \df \ell_b^- \, \cS_\kappa(\ell_b^-, \vec{k}_T - \vec{\ell}_T, \mu, \nu)\,
\nn \\
&\quad \times
S_\kappa\Bigl(k - \frac{\omega_a \ell_a^+}{Q_a} - \frac{\omega_b \ell_b^-}{Q_b}\Bigr) \,
\Bigl[1 + \ORd{\frac{k^2}{k_T^2}} \Bigr]
\,.\end{align}
%%%
Since the left-hand side does not depend on $\omega_{a,b}$ and $Q_{a,b}$, this dependence must
also drop out on the right-hand side, and therefore in the whole SCET$_+$ cross section in \eq{factorization_scetp}.
To see this explicitly, first recall that $\omega_a \omega_b = Q_a Q_b = Q^2$.
In addition, boost invariance at the level of the collinear-soft matrix element
implies that $\df \ell_a^+ \, \cS_\kappa(\ell_a^+, \vec{k}_T, \mu, \nu)$ can only depend on the product $\ell_a^+\nu$
(and analogously for $\ell_b^-$).%
\footnote{More explicitly, the rapidity regulator breaks the RPI-III invariance of SCET~\cite{Manohar:2002fd,Marcantonini:2008qn},
which is equivalent to boost invariance that must hold separately in each collinear sector. To restore it, $\nu$ must transform under RPI-III like $\bn \cdot p$ in each $n$-collinear-soft sector. This is most straightforward to see when strictly expanding the rapidity regulator to leading power in $\Tau^2/q_T^2$
using the soft-collinear mode scaling in \eq{modes_scetp}. The RPI-III transformation of the explicit measurement $\delta$ function in the matrix element is canceled by the corresponding integration measure in \eqs{factorization_scetp}{refactorization_soft_function}. Therefore, RPI-III invariance
implies that each collinear-soft function can only depend on the RPI-III invariant combination $\nu\, n\cdot k$.}
Hence, we can rewrite
%%%
\begin{align}
&\df \ell_a^+ \, \cS_\kappa(\ell_a^+, \vec{\ell}_a, \mu, \nu) \,
\df \ell_b^- \, \cS_\kappa(\ell_b^-, \vec{\ell}_b, \mu, \nu)\,
\delta\Bigl(\Tau - \frac{\omega_a \ell_a^+}{Q_a} - \frac{\omega_b \ell_b^-}{Q_b} - k \Bigr)
\nn \\ & \qquad
= \df k_a^+ \, \cS_\kappa\Bigl(k_a^+, \vec{\ell}_a, \mu, \frac{Q_a\nu}{\omega_a} \Bigr) \,
\df k_b^- \, \cS_\kappa\Bigl(k_b^-, \vec{\ell}_b, \mu, \frac{Q_b\nu}{\omega_b} \Bigr) \,
\delta(\Tau - k_a^+ - k_b^- - k)
\nn \\ &\qquad
= \df k_a^+ \, \cS_\kappa (k_a^+, \vec{\ell}_a, \mu, \nu) \,
\df k_b^- \, \cS_\kappa(k_b^-, \vec{\ell}_b, \mu, \nu)\,
\delta(\Tau - k_a^+ - k_b^- - k)
\,,\end{align}
%%%
where in the first step we changed variables from $\ell_{a,b}^\pm$ to $k_a^+ = \omega_a \ell_a^+ / Q_a$ and $k_b^- = \omega_b \ell_b^- / Q_b$. In the second step we performed the rapidity evolution from $\nu_{a,b} \equiv Q_{a,b}\, \nu / \omega_{a,b}$ back to a common $\nu$ at fixed $\mu$ [see \eq{rg_evolution_csoft}], for which the rapidity evolution factors exactly cancel because
%%%
\begin{equation}
\ln \frac{\nu_a}{\nu} + \ln \frac{\nu_b}{\nu} = \ln \frac{Q_a Q_b}{\omega_a \omega_b} = 0
\,.\end{equation}
%%%

The SCET$_+$ factorization in \eq{factorization_scetp} fully disentangles the physics
at the following SCET$_+$ canonical energy and rapidity scales:
%%%
\begin{alignat}{4} \label{eq:canonical_scales_scetp}
\mu^+_H &\sim Q
\,, \qquad
&\mu^+_B &\sim q_T
\,, \qquad
&\mu^+_{\cS} &\sim q_T
\,, \qquad
&\mu^+_S &\sim \Tau
\,,\nn \\
&
&\nu^+_B &\sim Q
\,, \qquad
&\nu^+_{\cS} &\sim q_T^2/\Tau
\,. \quad \end{alignat}
%%%
As for SCET$_\II$, we perform the $q_T$ resummation in $b_T$ space,
transforming the vectorial convolutions in \eq{factorization_scetp} into simple products.
In $b_T$ space, the canonical SCET$_+$ scales are
%%%
\begin{alignat}{4} \label{eq:canonical_scales_scetp_bT_space}
\mu^+_H &\sim Q
\,, \qquad
&\mu^+_B &\sim  b_0/b_T
\,, \qquad
&\mu^+_{\cS} &\sim b_0/b_T
\,, \qquad
&\mu^+_S &\sim \Tau
\,,\nn \\
&
&\nu^+_B &\sim Q
\,, \qquad
&\nu^+_{\cS} &\sim (b_0/b_T)^2/\Tau
\,. \quad \end{alignat}
%%%
By evaluating all functions at their natural scales and evolving them to common scales,
all logarithms of large scale ratios in the problem are resummed, e.g.,
%%%
\begin{equation}
\frac{(b_0/b_T)^2}{Q \Tau}
\sim \frac{\nu^+_\cS}{\nu^+_B}
\,, \qquad
\frac{\Tau}{b_0/b_T}
\sim \frac{\mu^+_S}{\mu^+_\cS}
\,, \qquad
\frac{b_0/b_T}{Q}
\sim \frac{\mu^+_B}{\mu^+_H} \sim \frac{\mu^+_{\cS}}{\mu^+_H}
\,, \qquad
\frac{\Tau}{Q}
\sim \frac{\mu^+_S}{\mu^+_H}
\,.\end{equation}
%%%
The logarithms of the first ratio appear in the double-differential SCET$_\I$ beam function
in the limit $q_T \ll \sqrt{Q\Tau}$, and are resummed in SCET$_+$ by the additional $\nu$ evolution
in the refactorization in \eq{refactorization_beam_function}.
Similarly, logarithms of the second ratio appear in the double-differential SCET$_\II$ soft function in the
limit $\Tau \ll q_T$, and are resummed in SCET$_+$ by the additional $\mu$ evolution in \eq{refactorization_soft_function}.
Our framework to match between the rich logarithmic structure predicted by \eq{factorization_scetp}
and the two boundary regimes is the subject of \sec{matching}.

%===============================================================================
\subsection{Outer space}
\label{sec:outer}
%===============================================================================

We now briefly discuss the outer phase-space regions left blank in \fig{overview_regimes}.
The region above the SCET$_\II$ regime is characterized by the hierarchy $q_T \ll \Tau \ll \sqrt{Q \Tau}$,
while the region to the right of the SCET$_\I$ regime corresponds to $\Tau \ll \sqrt{Q \Tau} \ll q_T$.
Both regions are power suppressed.

As we have discussed in \sec{scet2}, only the soft function contributes to $\Tau$ in SCET$_\II$, as the collinear contribution is power suppressed. However, for $q_T \ll \Tau$, even the soft contribution to $\Tau$ becomes power suppressed.
In particular, for a single real emission at fixed $\ord{\as}$, the region $\Tau > q_T$ is kinematically forbidden both in SCET$_\II$ as well as in full QCD. At higher orders only (soft) emissions
that are mostly back-to-back such that their transverse momenta largely cancel can fill out this region.
The cross section in this region is power suppressed by $\ord{q_T^2/\Tau^2}$.
Equivalently, expanding the SCET$_\II$ factorization of the double-differential cross section in the limit $q_T \ll \Tau$ reduces it to the single-differential $q_T$ spectrum with an overall $\delta(\Tau)$, which we exploit in our numerical implementation, cf.\ \eq{def_delta_soft}.
Physically this means that by integrating the double spectrum in SCET$_\II$ up to some $\Tau_\cut \gg q_T$,
we recover the single-differential $q_T$ spectrum,
while the effect of the cut is power suppressed in this limit.
Note that there is also a contribution from double-parton scattering~\cite{Landshoff:1978fq,Goebel:1979mi,Takagi:1979wn,Politzer:1980me} in this
region, where the two jets produced in the second interaction alongside the $Z$ boson are naturally back to back and not power suppressed.
This contribution is still not expected to much exceed the single-parton scattering contribution
because double-parton scattering itself is power suppressed by $\ord{\Lambda_{\rm QCD}^2/\Tau^2}$,
with $\Tau$ setting the scale of the second hard scatter producing the back-to-back jets.

Similarly, in the limit $\sqrt{Q \Tau} \ll q_T$, even the contribution from collinear radiation to $q_T$ becomes power suppressed in SCET$_\I$ [cf.\ \eq{def_delta_beam}], and at leading power we recover the single-differential $\Tau$ spectrum with an overall $\delta(q_T)$. This is analogous to the relation between the regimes~1 and~2 for a jet veto with a jet rapidity cut in \refcite{Michel:2018hui}, where the effect of a very forward jet rapidity cut (the auxiliary measurement) on collinear radiation becomes power suppressed.
An additional subtlety for $\sqrt{Q \Tau} \ll q_T$ is that very energetic forward radiation with energy $\sim q_T^2 / \Tau$ can theoretically contribute~\cite{Procura:2014cba}, pushing the hard scale up to $q_T^2 / \Tau \gg Q$. However, the cross section in this kinematic configuration is very strongly suppressed by the PDFs, so we choose to describe it at fixed order in this paper.

The above analysis justifies focusing on the shaded regions of phase space in \fig{overview_regimes}, corresponding to the main SCET$_\I$, SCET$_\II$, and SCET$_+$ regimes.

%%%%%%%%%%%%%%%%%%%%%%%%%%%%%%%%%%%%%%%%%%%%%%%%%%%%%%%%%%%%%%%%%%%%%%%%%%%%%%%%
\section{Matching effective theories}
\label{sec:matching}
%%%%%%%%%%%%%%%%%%%%%%%%%%%%%%%%%%%%%%%%%%%%%%%%%%%%%%%%%%%%%%%%%%%%%%%%%%%%%%%%

%===============================================================================
\subsection{Structure of power corrections}
\label{sec:power_corrections}
%===============================================================================

\begin{figure*}
\centering
\includegraphics[width=\WidthTwoSubfigs]{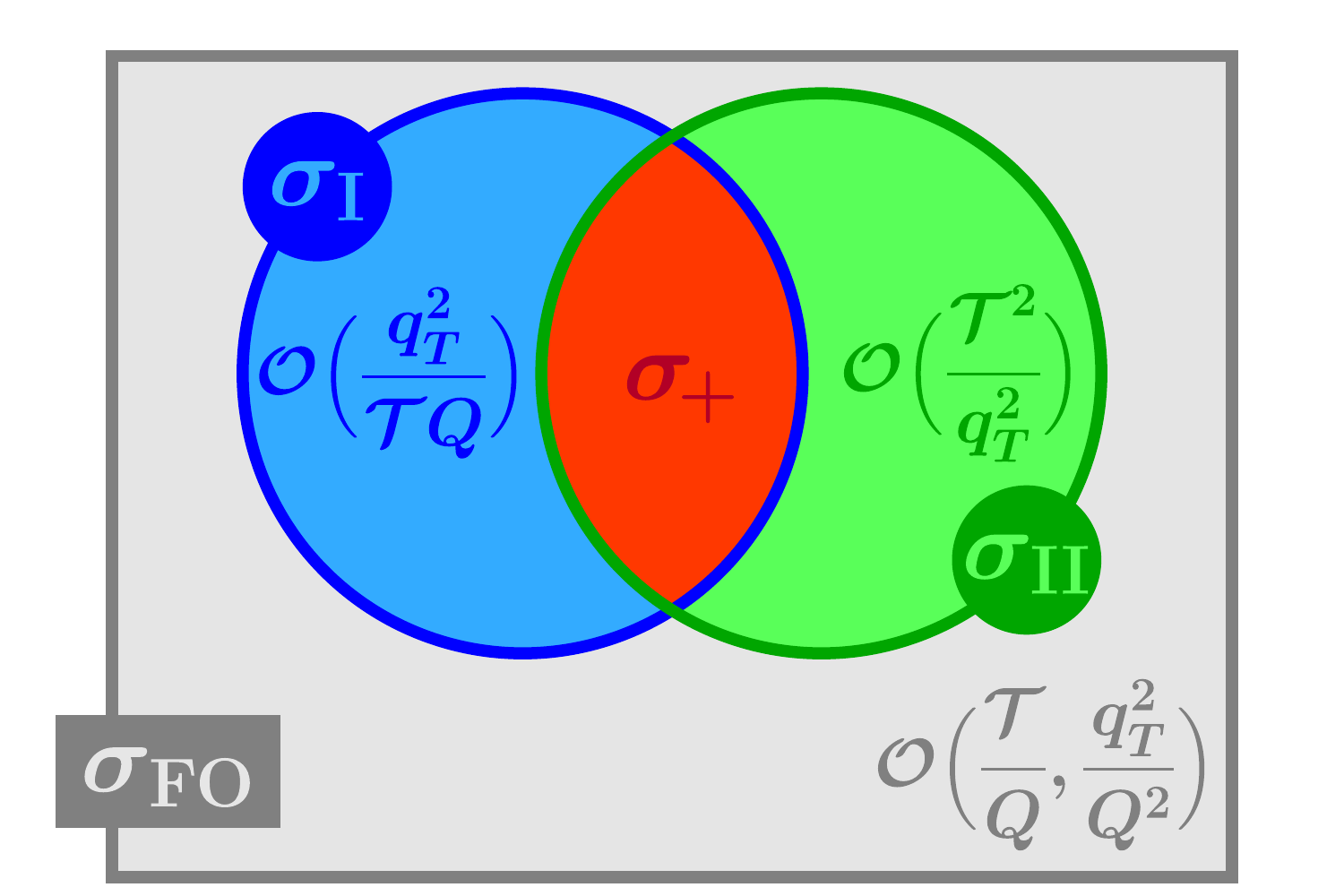}%
\caption{Venn diagram of power corrections to the factorized double-differential spectrum.
SCET$_\I$ (blue) and SCET$_\II$ (green) each capture a set of power corrections
that is expanded away in the SCET$_+$ factorization (red) and the opposite boundary regime.
A third class of power corrections to the overall soft-collinear limit
is captured by the fixed-order calculation in full QCD (gray).}
\label{fig:venn_diagram}
\end{figure*}

An important feature of our EFT setup is that the factorized cross section in SCET$_+$
differs from the ones in SCET$_\I$ and SCET$_\II$ only by a subset of the power corrections
it receives relative to the full QCD result,
%%%
\begin{align} \label{eq:power_corrections_venn}
\frac{\df \sigma_\I}{\df Q\, \df Y\, \df q_T\, \df \Tau}
&= \frac{\df \sigma_+}{\df Q\, \df Y\, \df q_T\, \df \Tau}\,
\Bigl[1+ \ORd{\frac{q_T^2}{\Tau Q}}\Bigr]
\,, \nn \\
\frac{\df \sigma_\II}{\df Q\, \df Y\, \df q_T\, \df \Tau}
&= \frac{\df \sigma_+}{\df Q\, \df Y\, \df q_T\, \df \Tau}\,
\Bigl[1+ \ORd{\frac{\Tau^2}{q_T^2}}\Bigr]
\,.\end{align}
%%%
This is illustrated in \fig{venn_diagram},
and follows from comparing \eq{power_corrections_scetp} to \eq{power_corrections_scet1} and \eq{power_corrections_scet2}, respectively.
Crucially, \eq{power_corrections_venn} also holds when
the cross sections are evaluated at common (but not necessarily fixed-order) scales.

\begin{figure*}
\centering
\includegraphics[width=\WidthTwoSubfigs]{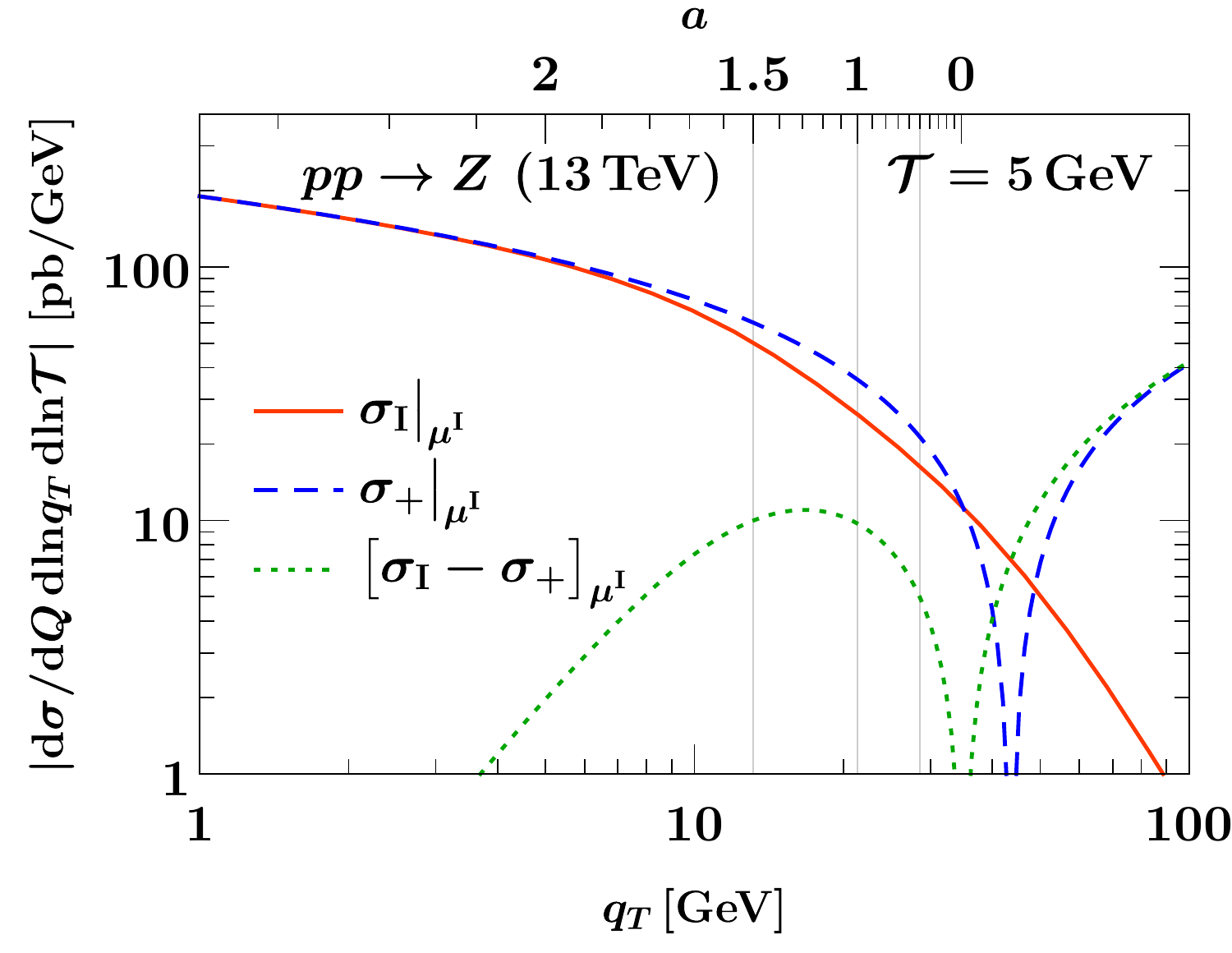}%
\hfill%
\includegraphics[width=\WidthTwoSubfigs]{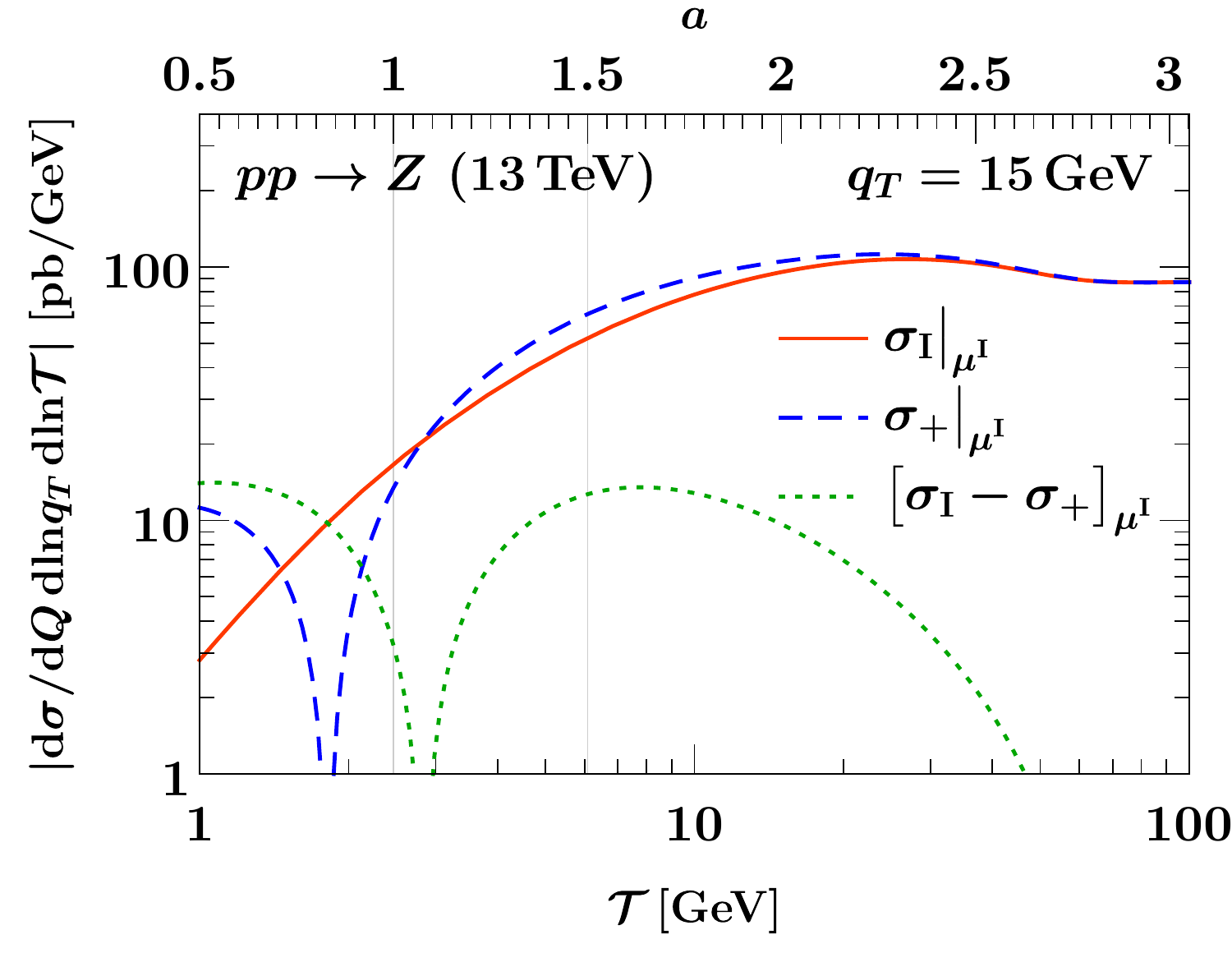}%
\caption{Singular/nonsingular comparison between SCET$_\I$ and SCET$_+$ at NNLL as a function of $q_T$ for fixed $\Tau = 5 \GeV$ (left) and as a function of $\Tau$ for fixed $q_T = 15 \GeV$ (right).
The orange solid lines show the full SCET$_\I$ result including resummation.
The dashed blue lines show the corresponding SCET$_+$ singular limit with \emph{only} SCET$_\I$ resummation. The dotted green lines show their difference,
corresponding to the power corrections indicated in \eq{power_corrections_venn}.
The thin vertical lines indicate our choice of transition points $(a_1, a_2, a_3)$ with respect to the regime parameter $a$ (upper horizontal axis), as discussed in \sec{profiles}.}
\label{fig:scetp_scet1_singnons}
\end{figure*}

For example, both $\sigma_\I$ and $\sigma_+$ share a logarithmic singularity with respect to $\Tau/Q$,
which can be resummed by running between the scales
of the hard, soft, and (refactorized) beam functions.
In SCET$_+$ this amounts to  setting the $\mu^+$ scales to be equal to their $\mu^\I$
counterparts,
%%%
\begin{equation} \label{eq:mu_1}
\mu_B^+ = \mu_\cS^+ = \mu_B^\I
\,, \qquad
\nu_B^+ = \nu_\cS^+ = \muFO
\,, \qquad
\mu_S^+ = \mu_S^\I
\,,\end{equation}
%%%
such that any large logarithms inside the refactorized beam function in \eq{refactorization_beam_function}
are treated at fixed order.
We write $\df \sigma_+ \big|_{\mu^\I}$ to indicate that $\df \sigma_+$ is evaluated at scales
that satisfy \eq{mu_1}.
A natural way to judge the size of the power corrections in \eq{power_corrections_venn} then is
to compare $\df \sigma_+ \big|_{\mu^\I}$ to $\df \sigma_\I \big|_{\mu^\I}$,
with our choices for $\mu^\I$ as given in \sec{scet1},
i.e., including the whole set of all-order terms from the $\Tau$ resummation in both of them.
This comparison is shown in \fig{scetp_scet1_singnons} for representative choices of fixed $\Tau$ and $q_T$ at NNLL.
We can clearly read off a power-like behavior of the difference
$\bigl[ \df \sigma_\I - \df \sigma_+ \bigr]_{\mu^\I}$ (dotted green)
as either $q_T \to 0$ for fixed $\Tau$ (left panel) or $\Tau \to \infty$ for fixed $q_T$ (right panel).
This also provides a nontrivial check on our implementation of $\sigma_\I$ and $\sigma_+$.
This comparison in \fig{scetp_scet1_singnons}
is analogous to the usual procedure of comparing the full-theory result for a cross section
with its singular EFT limit at a common scale $\mu_\FO$.
Here, SCET$_\I$ takes on the role of the full theory, while SCET$_+$ provides the singular limit, and the comparison is performed at common scales $\mu^\I$.

\begin{figure*}
\centering
\includegraphics[width=\WidthTwoSubfigs]{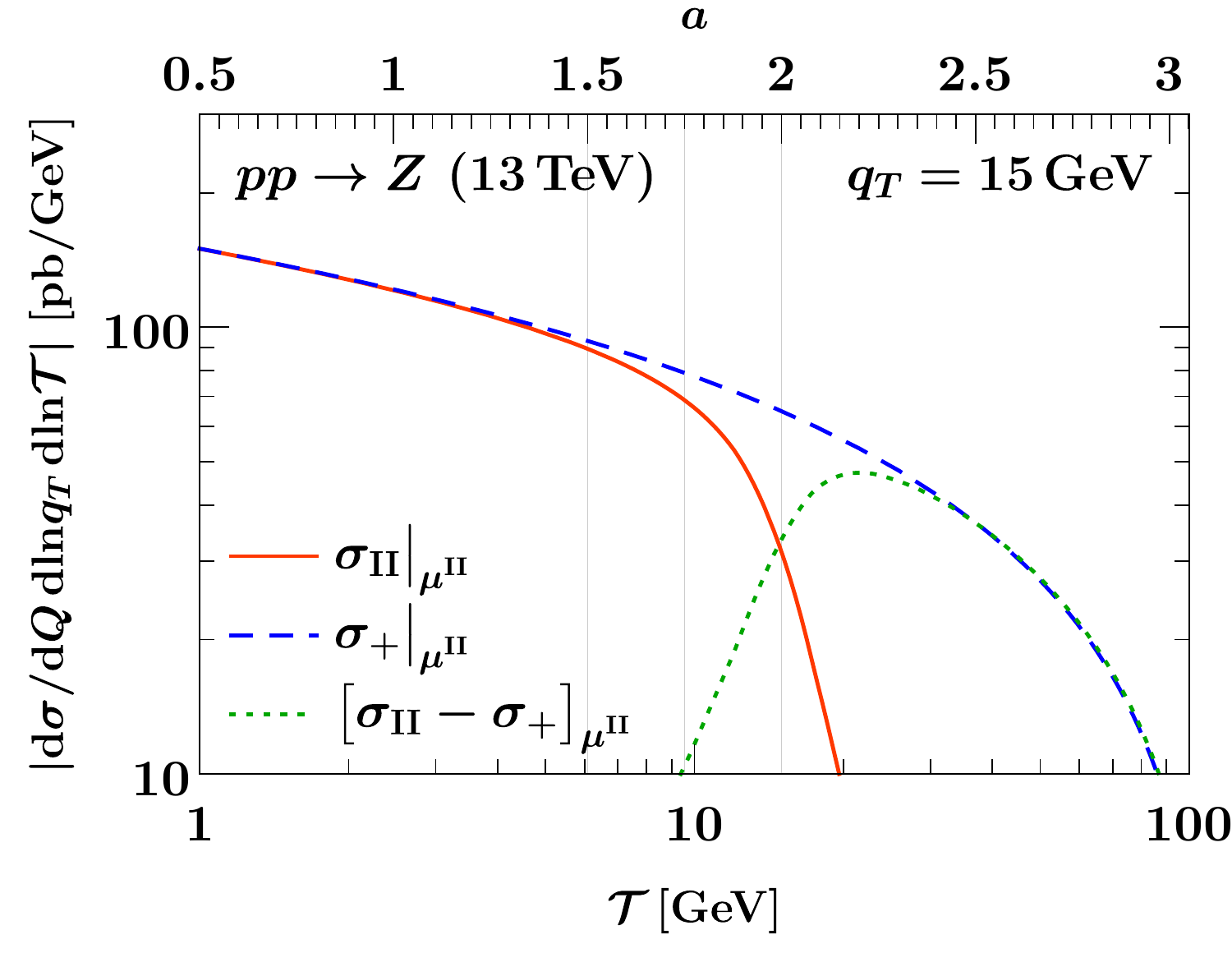}%
\hfill%
\includegraphics[width=\WidthTwoSubfigs]{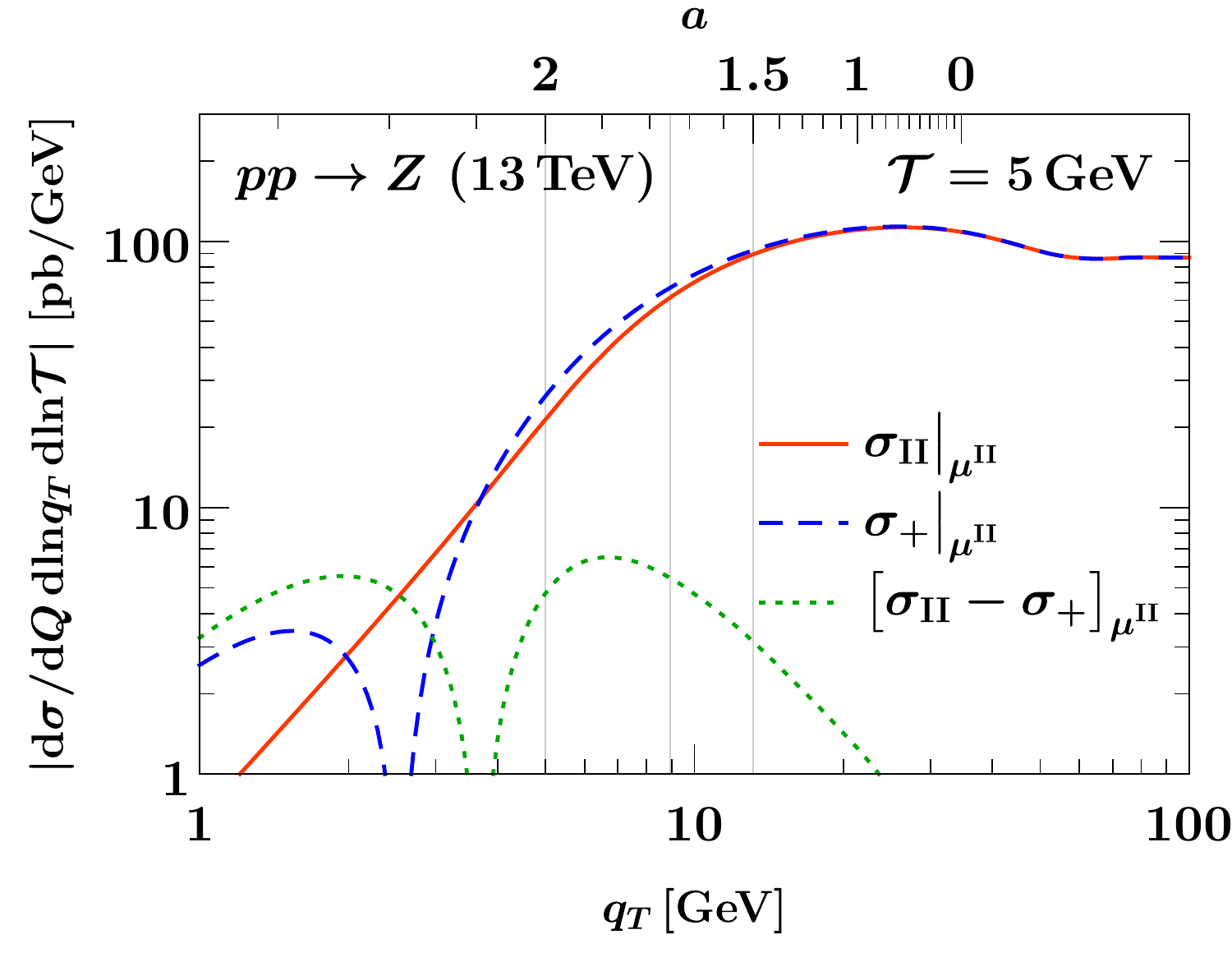}%
\caption{Singular/nonsingular comparison between SCET$_\II$ and SCET$_+$ at NNLL as a function of $\Tau$ for fixed $q_T = 15 \GeV$ (left) and as a function of $q_T$ for fixed $\Tau = 5 \GeV$ (right).
The orange solid lines show the full SCET$_\II$ result including resummation.
The dashed blue lines show the corresponding SCET$_+$ singular limit with \emph{only} SCET$_\II$ resummation. The dotted green lines show their difference,
corresponding to the power corrections indicated in \eq{power_corrections_venn}.
The thin vertical lines indicate our choice of transition points $(a_4, a_5, a_6)$ with respect to the regime parameter $a$ (upper horizontal axis), as discussed in \sec{profiles}.}
\label{fig:scetp_scet2_singnons}
\end{figure*}

Similarly, both $\sigma_\II$ and $\sigma_+$
have a common singular structure as $q_T/Q \to 0$.
In this case, resumming the shared logarithmic terms requires running between the hard, beam, and (refactorized) soft function. In SCET$_+$ this amounts to  setting the $\mu^+$ scales to be equal to their $\mu^\II$ counterparts,
%%%
\begin{equation}
\mu_\cS^+ = \mu_S^+ = \mu_S^\II
\,, \qquad
\nu_\cS^+ = \nu_S^\II
\,,\end{equation}
%%%
which treats the large logarithms in the refactorized double-differential soft function
in \eq{refactorization_soft_function} at fixed order.
We denote this choice of scales by $\df \sigma_+ \bigr|_{\mu^\II}$,
with scale setting in $b_T$ space and the inverse Fourier transform understood as in \sec{scet2}.
In \fig{scetp_scet2_singnons} we compare $\df \sigma_+ \bigr|_{\mu^\II}$ to $\df \sigma^\II \bigr|_{\mu^\II}$ at NNLL as a function of $\Tau$ at fixed $q_T$ (left) and vice versa (right).
It is clear that even when evaluated at its intrinsic scales,
$\df \sigma_\II \bigr|_{\mu^\II}$ (solid orange) exhibits an unresummed singularity as $\Tau/q_T \ll 1$,
which, as expected, is captured by $\df \sigma_+ \bigr|_{\mu^\II}$ (dashed blue) up to power corrections (dotted green).
This check is highly nontrivial as it involves an additional Fourier transform on both sides of the comparison.
We note that the strong kinematic suppression of the double spectrum for $\Tau \gtrsim q_T$
is correctly captured by SCET$_\II$, where central soft modes resolve the phase-space boundary.
In SCET$_+$, soft modes have too little energy and collinear-soft modes are too forward to resolve it,
leading to large power corrections in this region.

\begin{figure*}
\centering
\includegraphics[width=\WidthTwoSubfigs]{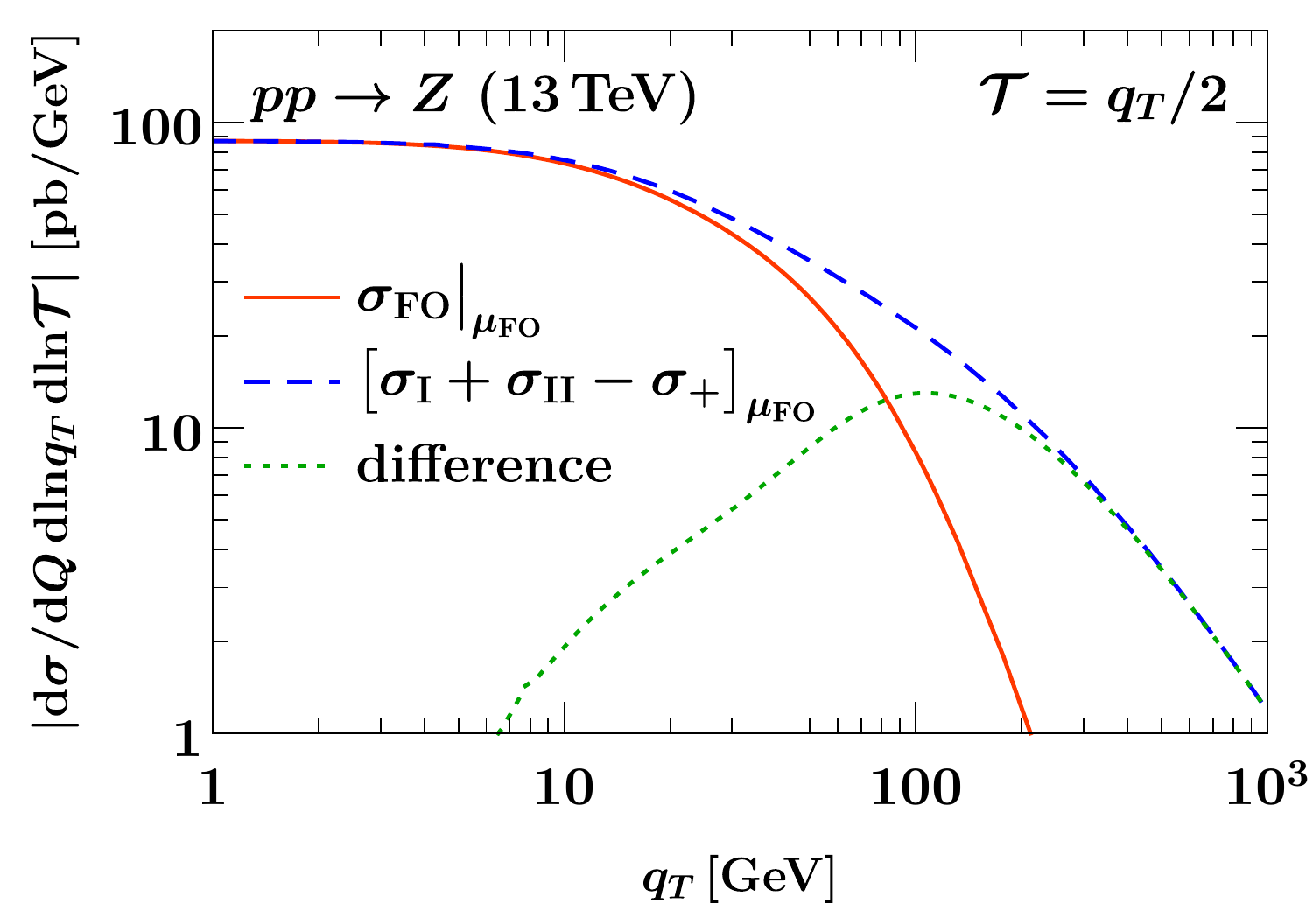}%
\hfill%
\includegraphics[width=\WidthTwoSubfigs]{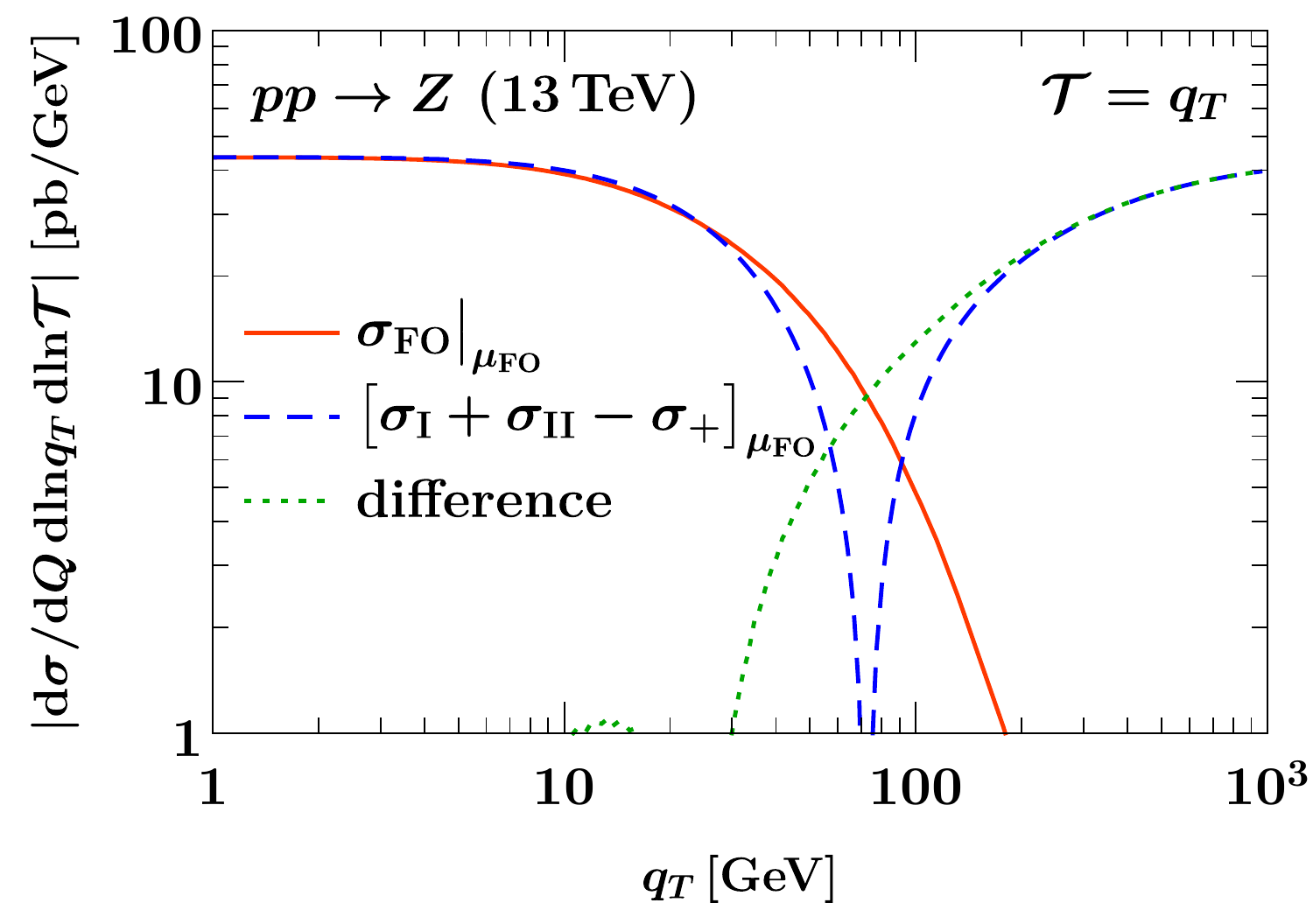}%
\caption{Singular/nonsingular comparison between the matched SCET descriptions and QCD at fixed $\ord{\as}$ as a function of $q_T$ for $\Tau = q_T/2$ (left) and $\Tau = q_T$ (right).
The orange solid line shows the fixed-order QCD  double spectrum,
the dashed blue the matched SCET result in \eq{power_corrections_matched}, and the dotted green the difference.}
\label{fig:matched_qcd_singnons}
\end{figure*}

As a final important consequence of \fig{venn_diagram},
the complete infrared structure of the double-differential spectrum for $q_T \ll Q$ and $\Tau \ll Q$, i.e., for any hierarchy between $q_T$ and $\Tau$,
is described by adding the SCET$_\I$ and SCET$_\II$ cross sections
and removing the overlap between the two by subtracting the SCET$_+$ cross section,
%%%
\begin{align} \label{eq:power_corrections_matched}
\frac{\df \sigma}{\df Q\, \df Y\, \df q_T\, \df \Tau}
&= \Bigl[ \frac{\df \sigma_\I}{\df Q\, \df Y\, \df q_T\, \df \Tau} + \frac{\df \sigma_\II}{\df Q\, \df Y\, \df q_T\, \df \Tau} - \frac{\df \sigma_+}{\df Q\, \df Y\, \df q_T\, \df \Tau} \Bigr]
\nn \\
&\quad \times
\Bigl[1+ \ORd{\frac{q_T^2}{Q^2}, \frac{\Tau}{Q}}\Bigr]
\,.\end{align}
%%%
In \fig{matched_qcd_singnons} we numerically check this relation at fixed $\ord{\as}$,
which requires setting all scales equal to a common $\mu_\FO$.
We plot the comparison as a function of $q_T$ along lines of fixed $\Tau/q_T = 1/2$ (left) and $\Tau/q_T = 1$ (right),
finding excellent agreement between the full result (solid orange) and the first line on the right-hand side of \eq{power_corrections_matched} (dashed blue),
as evident from the power-like behavior of their difference (dotted green) as $q_T, \Tau \to 0$.

This singular/nonsingular comparison is qualitatively different from the structure of power corrections in either SCET$_\I$ or SCET$_\II$ alone, which we already verified in \fig{scet1_qcd_singnons} and \fig{scet2_qcd_singnons}.
Because SCET$_\I$ and SCET$_\II$ both involve an additional expansion about a specific hierarchy between
$q_T$ and $\Tau$, they incur power corrections $\ord{\Tau^2/q_T^2}$ or $\ord{q_T^2/(Q\Tau)}$, respectively.
Accordingly, they only recover the singular limit of full QCD
when approaching it along specific lines in the $(q_T,\Tau)$ plane.
This is different from \fig{matched_qcd_singnons}, where
the combined expression in \eq{power_corrections_matched} (dashed blue) describes the singular limit $q_T,\Tau \to 0$ along an arbitrary line of approach,
with the ratio $q_T/\Tau$ effectively controlling the ``admixture'' of power corrections $\ord{q_T^2/Q^2}$ and $\ord{\Tau/Q}$, respectively.
We have verified that also for other fixed ratios of $q_T$ and $\Tau$,
the singular behavior of full QCD is correctly described.

As a final remark, as noted in \refcite{Gaunt:2015pea},
this property actually qualifies the expression $\df \sigma_\I + \df \sigma_\II - \df \sigma_+$
for use as a double-differential subtraction term to treat infrared divergences in fixed-order
calculations.

%===============================================================================
\subsection{Matching formula}
\label{sec:matching_formula}
%===============================================================================

The structure of power corrections discussed in the previous section,
together with the all-order resummation shared between SCET$_+$ and SCET$_\I$ or SCET$_\II$,
suggests the following matching formula to describe all regions of the double-differential spectrum:
%%%
\begin{align} \label{eq:matching}
\df \sigma^\match
= \df \sigma_+ \bigr|_{\mu^+}
&+ \bigl[ \df \sigma_\I - \df \sigma_+ \bigr]_{\mu^\I}
+ \bigl[ \df \sigma_\II - \df \sigma_+ \bigr]_{\mu^\II}
\nn \\
&+ \bigl[ \df \sigma_\FO - \df \sigma_\I - \df \sigma_\II + \df \sigma_+ \bigr]_{\mu_\FO}
\,.\end{align}
%%%
The only ingredient in this matching formula we have not yet discussed is $\df \sigma_+\bigr|_{\mu^+}$,
for which all ingredients in the SCET$_+$ factorization are evaluated at the SCET$_+$ scales $\mu^+$,
such that the full RGE of SCET$_+$ is in effect.
In the following we describe the requirements on $\mu^+$
to ensure the best possible prediction across phase space.
Our precise construction of $\mu^+$ to satisfy all requirements is the subject of \sec{profiles}.

In the simplest case, i.e., when the power corrections in \eq{power_corrections_scetp} are small, and thus the SCET$_+$ parametric assumptions are satisfied,
$\mu^+$ is given by the canonical SCET$_+$ scales in \eq{canonical_scales_scetp_bT_space}.
As for $\mu^\II$, these scales are set in $b_T$ space, followed by an inverse Fourier transform.

As we approach the SCET$_\I$ region,
the resummation inside the refactorization of the beam function in \eq{refactorization_beam_function} must be turned off,
%%%
\begin{equation} \label{eq:scetp_to_scet1_profile_requirements_beam}
\left.\begin{aligned}
\mu_B^+(q_T, \Tau, b_T) &\to \mu_B^\I(\Tau)
\\
\mu_\cS^+(q_T, \Tau, b_T) &\to \mu_B^\I(\Tau)
\\
\nu_\cS^+(q_T, \Tau, b_T) &\to \nu_B^+(q_T, \Tau, b_T)
\end{aligned}\quad\right\rbrace
\quad \text{for} \quad q_T \to \sqrt{Q \Tau}
\,.\end{equation}
%%%
In addition we can identify the soft scales in SCET$_\I$ and SCET$_+$
because the soft functions are identical,
%%%
\begin{equation} \label{eq:scetp_to_scet1_profile_requirements_soft}
\mu_S^+(q_T, \Tau, b_T) \to \mu_S^\I(\Tau)
\quad \text{for} \quad q_T \to \sqrt{Q \Tau}
\,.\end{equation}
%%%
These relations must hold for every value of the $b_T$ argument of the scale.

Similarly, as we approach the SCET$_\II$ region,
the scales inside the refactorized soft function \eq{refactorization_soft_function}
must become equal
%%%
\begin{equation} \label{eq:scetp_to_scet2_profile_requirements_soft}
\left.\begin{aligned}
\mu_S^+(q_T, \Tau, b_T) &\to \mu_S^\II(q_T, b_T)
\\
\mu_\cS^+(q_T, \Tau, b_T) &\to \mu_S^\II(q_T, b_T)
\\
\nu_\cS^+(q_T, \Tau, b_T) &\to \nu_S^\II(q_T, b_T)
\end{aligned}\quad\right\rbrace
\quad \text{for} \quad q_T \to \Tau
\,,\end{equation}
%%%
and we can identify the scales
of the common beam function in SCET$_\II$ and SCET$_+$,
%%%
\begin{equation} \label{eq:scetp_to_scet2_profile_requirements_beam}
\left.\begin{aligned}
\mu_B^+(q_T, \Tau, b_T) &\to \mu_B^\II(q_T, b_T)
\\
\nu_B^+(q_T, \Tau, b_T) &\to \nu_B^\II(q_T, b_T)
\end{aligned}\quad\right\rbrace
\quad \text{for} \quad q_T \to \Tau
\,.\end{equation}
%%%

Some of the above requirements for the behavior at the boundary
are already satisfied by the canonical SCET$_+$ scales,
e.g., the canonical soft scales in SCET$_+$ and SCET$_\I$ are simply equal.
The challenge in these cases is to extend the scale choice onto the opposite boundary,
where they are constrained in a nontrivial way.
The nontrivial all-order information in SCET$_+$ is mostly encoded in the canonical choice of
%%%
\begin{equation} \label{eq:scetp_canonical_profile_requirements}
\nu_\cS^+(q_T, \Tau, b_T) = \frac{(b_0/b_T)^2}{\Tau}
\quad \text{for} \quad \Tau \ll q_T \ll \sqrt{Q\Tau}
\,,\end{equation}
%%%
which does not coincide with any scale on either boundary.

It is instructive to explicitly consider the behavior of \eq{matching}
on the SCET$_\I$ and SCET$_\II$ phase-space boundaries, as well as in the fixed-order region.
By construction, for any choice of $\mu^+$ scales
satisfying \eqs{scetp_to_scet1_profile_requirements_beam}{scetp_to_scet1_profile_requirements_soft}
we have
%%%
\begin{equation} \label{eq:matching_asymptotic_scet1}
\df \sigma_+ \bigr|_{\mu^+} \to \df \sigma_+ \bigr|_{\mu^\I}
\quad \text{for} \quad q_T \to \sqrt{Q \Tau}
\,.\end{equation}
%%%
It follows that
%%%
\begin{align} \label{eq:asymptote_scet_1}
\df \sigma^\match \to \df \sigma_\I \bigr|_{\mu^\I}
&+ \bigl[ \df \sigma_\FO - \df \sigma_\I\bigr]_{\mu_\FO}
\nn \\
&+ \bigl[ \df \sigma_\II - \df \sigma_+ \bigr]_{\mu^\II}
- \bigl[ \df \sigma_\II - \df \sigma_+ \bigr]_{\mu_\FO}
\quad \text{for} \quad q_T \to \sqrt{Q \Tau}
\,.\end{align}
%%%
This mostly coincides with the result in \eq{matching_scet1} of matching $\df\sigma_\I$ to the
fixed-order result $\df\sigma_\FO$,
and is guaranteed to capture all large logarithms of $\Tau/Q$ captured by the SCET$_\I$ RGE.
It improves over \eq{matching_scet1} by also resumming logarithms of $q_T/Q$ in the power corrections $\ord{\Tau^2/q_T^2}$, encoded in $\bigl[ \sigma_\II - \sigma_+ \bigr]_{\mu^\II}$.
This is not a numerically large effect and cannot be exploited to achieve the resummation of $\Tau$ at next-to-leading power, as it is only a subset of all power corrections.

Similarly, \eqs{scetp_to_scet2_profile_requirements_soft}{scetp_to_scet2_profile_requirements_beam} imply that
%%%
\begin{equation} \label{eq:matching_asymptotic_scet2}
\df \sigma_+ \bigr|_{\mu^+} \to \df \sigma_+ \bigr|_{\mu^\II}
\quad \text{for} \quad q_T \to \Tau
\,,\end{equation}
%%%
and consequently
%%%
\begin{align} \label{eq:asymptote_scet_2}
\df \sigma^\match \to \df \sigma_\II \bigr|_{\mu^\II}
&+ \bigl[ \df \sigma_\FO - \df \sigma_\II \bigr]_{\mu_\FO}
\nn \\
&+ \bigl[ \df \sigma_\I - \df \sigma_+ \bigr]_{\mu^\I}
- \bigl[ \df \sigma_\I - \df \sigma_+ \bigr]_{\mu_\FO}
\quad \text{for} \quad q_T \to \Tau
\,.\end{align}
%%%
This mostly coincides with the result in \eq{matching_scet2} of matching $\df\sigma_\II$ to the fixed-order result $\df\sigma_\FO$, and thus captures all large logarithms of $q_T/Q$ captured by the SCET$_\II$ RGE.
In addition, it resums logarithms of $\Tau/Q$ in the $\mathcal{O}(q_T^2/(\Tau Q))$ power corrections encoded in $[\df \sigma_\I - \df \sigma_+]_{\mu^\I}$.

Finally, in the fixed-order region, all $\mu^+$, $\mu^\I$, and $\mu^\II$ scales become equal to $\mu_\FO$.
Thus as desired, the matched prediction reduces to the fixed-order result,
%%%
\begin{equation}
\df \sigma^\match \to \df \sigma_\FO \bigr|_{\mu_\FO}
\quad \text{for} \quad q_T, \Tau \to Q
\,.\end{equation}
%%%

%===============================================================================
\subsection{Profile scales}
\label{sec:profiles}
%===============================================================================

In this section, we describe our choice of the central $\mu^+$ scales for the various ingredients in the SCET$_+$ factorized cross section,
taking into account the transition to the SCET$_\I$ and SCET$_\II$ boundary theories
as well as the transition to the fixed-order region.
The SCET$_+$ scales are obtained using a regime parameter
that selects the appropriate combination of scales from the boundary theories in each region of phase space,
and selects a third, independent choice in the SCET$_+$ ``bulk'' when necessary.
The profile functions that handle the transition to fixed order can conveniently be reused from SCET$_\I$ and SCET$_\II$.

\begin{table}
\centering
\begin{tabular}{l|c|c|c}
   \hline \hline
   Scale &  SCET$_\I$  & SCET$_+$ & SCET$_\II$ \\ \hline
   $\mu_H$ & $Q$ & $Q$ & $Q$ \\
   $\mu_B$ & $\sqrt{\Tau Q} $ & $b_0/b_T$ & $b_0/b_T$ \\
   $\nu_B$ & & $Q$ & $Q$ \\
   $\mu_\cS$ & & $b_0/b_T$ & \\
   $\nu_\cS$ & & $(b_0/b_T)^2/\Tau$ &\\
   $\mu_S $ & $\Tau$ & $\Tau$ & $b_0/b_T$ \\
   $\nu_S$ & & & $b_0/b_T$ \\
   \hline \hline
\end{tabular}
\caption{Summary of canonical scales in SCET$_\I$, SCET$_+$, and SCET$_\II$
[see \eq{canonical_scales_scet1}, \eqref{eq:canonical_scales_scet2_bT_space} and \eqref{eq:canonical_scales_scetp_bT_space}].
For SCET$_+$ and SCET$_\II$ we give the canonical scales in $b_T$ space.
}
\label{tab:scales}
\end{table}

\begin{figure*}
\centering
\hfill%
\includegraphics[height=\HeightTwoSubfigs]{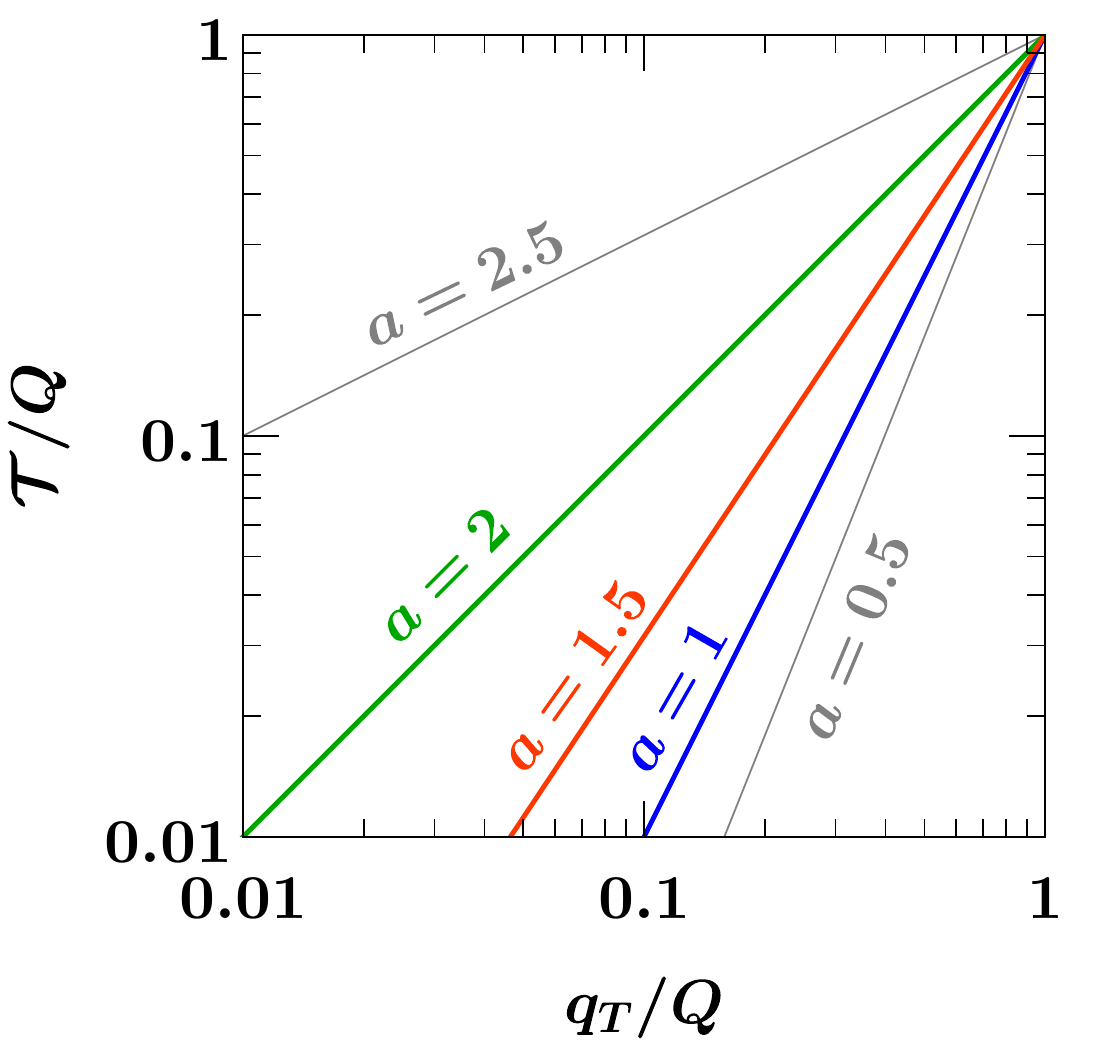}%
\hfill%
\includegraphics[height=\HeightTwoSubfigs]{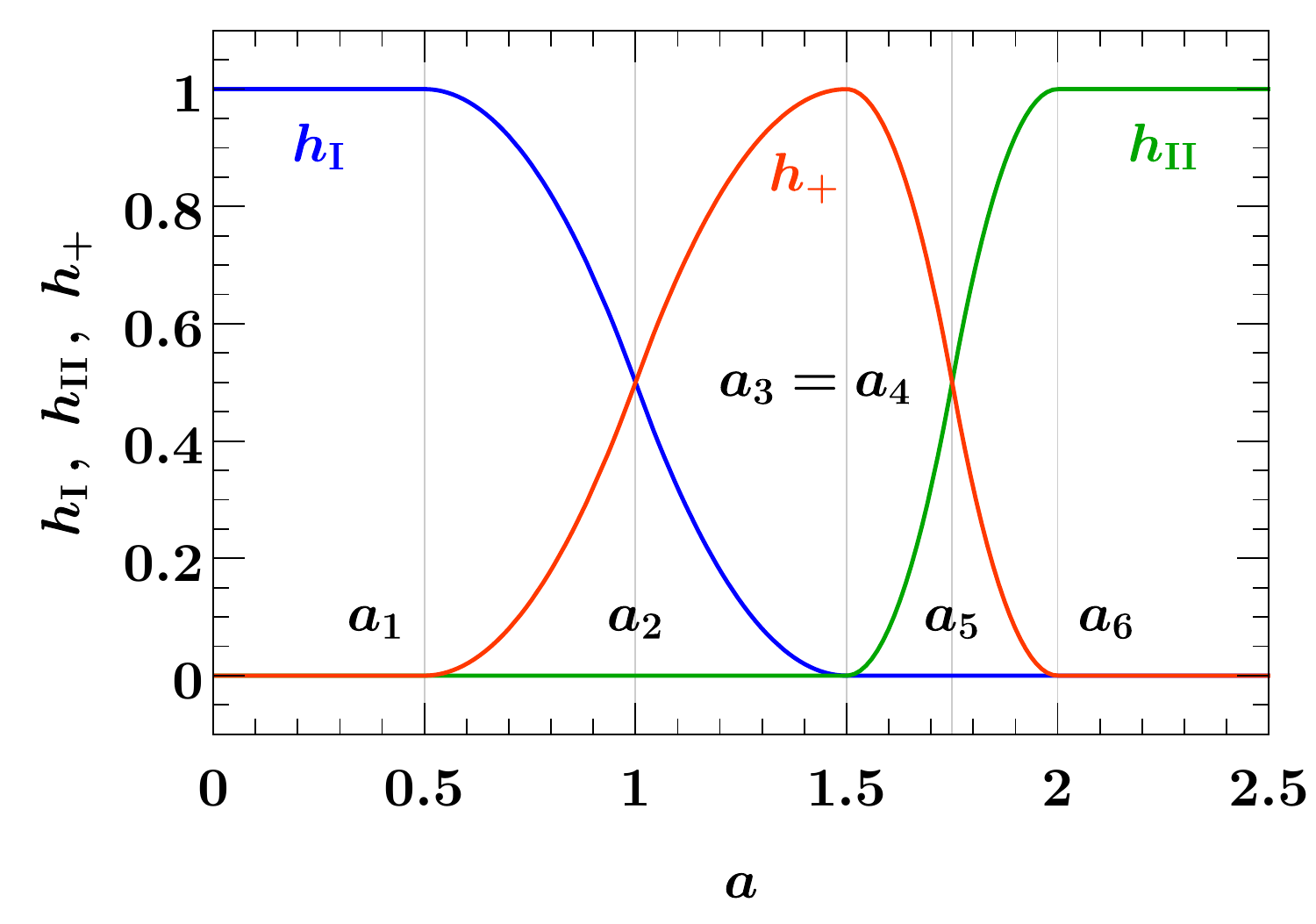}%
\caption{Left: Illustration of the regime parameter $a$ that governs the matching between EFTs.
We show lines of constant $a$ in the $(q_T, \Tau)$ plane.
For $a = 1$ the SCET$_\I$ parametric relation is exactly satisfied, $q_T = \sqrt{Q\Tau}$,
whereas for $a = 2$, the SCET$_\II$ parametric relation is exactly satisfied, $q_T = \Tau$.
Right: Helper functions used to interpolate between scales on the boundaries (SCET$_\I$, SCET$_\II$) and in the bulk (SCET$_+$).
The helper functions have continuous derivatives and always sum to one.
The individual helper functions are exactly one in their respective canonical regions.
See the text for a detailed discussion.}
\label{fig:helper_functions}
\end{figure*}

We start by summarizing the canonical scales for SCET$_\I$, SCET$_\II$, SCET$_+$ in \tab{scales}.
At these scales, the arguments of logarithms in the ingredients of the factorized cross section are order one, i.e., all large logarithms are resummed by RG evolution.
To interpolate between the canonical scales in different regimes,
we find it convenient to introduce the regime parameter
%%%
\begin{align} \label{eq:def_regime_parameter}
a = 3 - \frac{\abs{\ln(\Tau/Q)}}{\abs{\ln(q_T/Q)}}
\,.\end{align}
%%%
Its definition is carefully chosen
such that $a=1$ when the SCET$_\I$ parametric relation is exactly satisfied, $q_T = \sqrt{\Tau Q}$,
and $a=2$ on the SCET$_\II$ boundary of phase space, $q_T = \Tau$.
As illustrated in the left panel of \fig{helper_functions},
the canonical SCET$_+$ region lies at intermediate $a \sim 1.5$.
The requirements on the SCET$_+$ scales were given in \eqs{scetp_to_scet1_profile_requirements_beam}{scetp_to_scet1_profile_requirements_soft} for the transition to SCET$_\I$,
and in \eqs{scetp_to_scet2_profile_requirements_soft}{scetp_to_scet2_profile_requirements_beam} for SCET$_\II$.
To satisfy these requirements, we take weighted products of scales on the boundary and in the bulk,
schematically,
%%%
\begin{align} \label{eq:scetp_weighted_product_schematic}
\mu^+ = \bigl[\mu^\I\bigr]^{h_\I(a)} \, \bigl[\mu^+_{\bulk} \bigr]^{h_{+}(a)}
\bigl[\mu^\II \bigr]^{h_\II(a)}
\,.\end{align}
%%%
The weights in the exponent are given by helper functions that depend on $a$,
as illustrated in the right panel of \fig{helper_functions}.
They satisfy
%%%
\begin{equation}
h_\I(a) + h_+(a) + h_\II(a) = 1
\,,\end{equation}
%%%
for any $a$ and are given explicitly in \eq{helper_functions} below.
The helper functions ensure that the appropriate scales are used in each region,
e.g., $h_\II(a)$ is one in the vicinity of $a=2$
and vanishes for $a < 1.5$, with a smooth transition between regions.

For the soft and collinear-soft scales, \eq{scetp_weighted_product_schematic} takes the following concrete form:
%%%
\begin{alignat}{5} \label{eq:scetp_weighted_product_csoft_soft}
\mu_\cS^+
&= \bigl[\mu_{B}^\I\bigr]^{h_\I(a)} \, &&\bigl[\mu_{\cS,\bulk}^+\bigr]^{h_+(a)} \, &&\bigl[\mu_{S}^\II\bigr]^{h_\II(a)}
\,, \nn \\
\nu_{\cS}^+
&=\bigl[ \nu \bigr]^{h_\I(a)} \, &&\bigl[\nu_{\cS,\bulk}^+\bigr]^{h_{+}(a)} \, &&\bigl[\nu_{S}^\II\bigr]^{h_\II(a)} \,
\,, \nn \\
\mu_{S}^+
&=\bigl[ \mu_S^\I \bigr]^{h_\I(a)} \, &&\bigl[\mu_{S,\bulk}^+\bigr]^{h_{+}(a)} \, &&\bigl[\mu_{S}^\II\bigr]^{h_\II(a)} \,
\,.\end{alignat}
%%%
The most nontrivial of these cases is $\nu_{\cS}$,
which must be equal to the overall $\nu$ in the SCET$_\I$ region to turn off the rapidity resummation there,
has a distinct canonical value in the SCET$_+$ bulk,
and must asymptote to yet another value on the SCET$_\II$ boundary.
We note that $\mu_S^+$ also requires a distinct treatment in the bulk to ensure
that the hierarchy $\mu_S^+ < \mu_\cS^+$ inside the refactorized soft function,
as implied by the SCET$_+$ power counting, is not upset by variations (see next subsection).
Our central choices for the above scales in the bulk are
%%%
\begin{align} \label{eq:scetp_independent_bulk_scales}
\mu_{\cS,\bulk}^+ &= \mu_\FO \, f_\run^\II\Bigl( \frac{q_T}{Q}, \frac{b_0}{b^\ast(b_T) \, Q} \Bigr)
\,, \quad
\nu_{\cS,\bulk}^+ = \mu_\FO \frac{ \Bigl[ f_\run^\II\Bigl( \frac{q_T}{Q}, \frac{b_0}{b^\ast(b_T) Q} \Bigr) \Bigr]^2}{ f_\run^\I \Bigl( \frac{\Tau}{Q} \Bigr)}
\,, \nn \\
\mu_{S,\bulk}^+ &= \mu_\FO \, f_\run^\I \Bigl( \frac{\Tau}{Q} \Bigr)
\,.\end{align}
%%%
The profile function $f_\run^\I$ was introduced for the transition between SCET$_\I$ and fixed-order QCD in \eq{f_run_scet1},
and similarly for the hybrid profile $f_\run^\II$ in \eq{fRun2}
and the nonperturbative $b^\ast(b_T)$ prescription in \eq{def_bstar}.
These functions turn off the resummation of logarithms involving $q_T$ ($b_T$) and $\Tau$, respectively,
as the fixed-order regime is approached,
and also ensure that scales are frozen in the nonperturbative regime to avoid the Landau pole.
It is straightforward to check that away from the nonperturbative region, the above bulk scales
all assume their canonical values for $q_T, \Tau \ll Q$ as given in \tab{scales},
and asymptote to $\mu_\FO$ when simultaneously taking $q_T, \Tau \to Q$.
The beam function scales in the bulk can simply be associated with their SCET$_\II$ counterparts
and only require a transition towards the SCET$_\I$ boundary,
%%%
\begin{align} \label{eq:scetp_weighted_product_beam}
\mu_B^{+} &= \bigl[\mu_{B}^\I\bigr]^{h_\I(a)} \, \bigl[\mu_{B}^\II\bigr]^{h_{+}(a) + h_\II(a)}
\,, \nn \\
\nu_B^{+} &= \bigl[ \nu \bigr]^{h_\I(a)} \, \bigl[\nu_{B}^\II\bigr]^{h_{+}(a) + h_\II(a)}
\,.\end{align}
%%%

In our numerical implementation, we choose the helper functions $h_{\I, \II, +}$ as
%%%
\begin{align} \label{eq:helper_functions}
h_\I(a) &\equiv \begin{cases}
1 & a < a_1\,, \\
1-c_{123}(a) & a_1 \leq a < a_2\,, \\
c_{312}(a) & a_2 \leq a < a_3\,, \\
0 & a_3 \leq a\,,
\end{cases} \qquad
h_\II(a) \equiv \begin{cases}
0 & a < a_4\,, \\
c_{456}(a) & a_4 \leq a < a_5\,, \\
1-c_{645}(a) & a_5 \leq a < a_6\,, \\
1 & a_6 \leq a \,,
\end{cases} \nn\\[0.5em]
h_{+}(a) &\equiv 1 - h_\I(a) - h_\II(a)
\,,\end{align}
%%%
where the polynomials governing the interpolation between zero and one are
%%%
\begin{align}
c_{ijk}(a) = \frac{(a-a_i)^2}{(a_i-a_j)(a_i-a_k)}
\, .\end{align}
%%%
The transition points $a_{1,\ldots,6}$ determine the transition
between the different regions, as can be seen from the helper functions in \fig{helper_functions}:
For values $a_3 \leq a < a_4$, the exact canonical SCET$_+$ scales are selected,
implying that the resummation of logarithms of both $q_T$ and $\Tau$ is fully turned on.
For lower values $a_1 \leq a < a_3$, the additional $q_T$ resummation is smoothly turned off
and for $a< a_1$, SCET$_\I$ scales are used so that only logarithms of $\Tau$ are resummed. Conversely, for higher values of the regime parameter $a_4 \leq a < a_6$,
the resummation of $\Tau$ logarithms is smoothly turned off.
At values $a_6 \leq a$, SCET$_\II$ scales are selected by the helper functions,
and the additional resummation of logarithms of $\Tau$ is completely turned off.

In practice we use $(a_1,a_2,a_3,a_4,a_5,a_6) = (0.5,\, 1.0,\, 1.5,\, 1.5,\, 1.75,\, 2.0)$.
This choice ensures that for $a \geq a_6 = 2$, we fully recover SCET$_\II$ resummation
and faithfully describe the kinematic edge at $q_T \sim \Tau$
by preserving the $\ord{1}$ cancellation between $\sigma_+\bigr|_{\mu^\II}$ and the SCET$_\II$ nonsingular contribution
visible at $a \sim 2$ in the left panel of \fig{scetp_scet2_singnons}.
(In both \figs{scetp_scet1_singnons}{scetp_scet2_singnons},
corresponding values of $a$ are indicated on the horizontal axis at the top of the panels.)
On the other hand, from \fig{scetp_scet1_singnons} we observe
that power corrections from SCET$_\I$ are smaller
and tend to set in at values of $a$ lower than the naively expected $a = 1$.
E.g., an $\ord{1}$ cancellation between $\sigma_+\bigr|_{\mu^\I}$ and the SCET$_\I$ nonsingular only is in effect
around $a \sim 0.5$ in the right panel of \fig{scetp_scet1_singnons},
leaving more room for slowly turning off the SCET$_+$ resummation down towards $a_1 = 0.5$.
This is expected because the SCET$_\I$ nonsingular encodes the suppression of collinear recoil
beyond the naive phase-space boundary at $a \sim 1$ ($q_T \sim \sqrt{Q\Tau}$)
that is washed out by the PDFs,
unlike the sharp kinematic edge at $q_T \sim \Tau$ encoded in the SCET$_\II$ nonsingular.
For simplicity we set $a_3 = a_4$ for our central prediction,
i.e., we shrink the canonical SCET$_+$ region to a point at $a = 1.5$,
and fix $a_2$ ($a_5$) to be the midpoint between $a_1$ and $a_3$ ($a_4$ and $a_6$).
Variations of the transition points,
including independent variations of $a_3$ and $a_4$,
are considered as part of our uncertainty estimate described in the next section.

%===============================================================================
\subsection{Perturbative uncertainties}
\label{sec:variations}
%===============================================================================

In this section we describe how we assess perturbative uncertainties
by varying the scales entering the matched prediction in \eq{matching}.
Following the same approach as for SCET$_\I$ and SCET$_\II$ on their own (see \secs{scet1}{scet2}),
we distinguish between resummation uncertainties and a fixed-order uncertainty.
The fixed-order uncertainty  $\Delta_\FO$ is estimated by varying $\mu_\FO$ up and down by a factor of two,
i.e., by setting $\muFO = \{Q/2, 2Q\}$.
Since all scales (in any piece of the matching formula) include an overall factor of $\mu_\FO$,
the ratios between the various scales remain unchanged
and the same logarithms are resummed.
The fixed-order uncertainty $\Delta_\FO$ is then taken to be the maximum deviation from the central cross section.

We consider several sources of resummation uncertainty entering the matched prediction in \eq{matching}.
To probe the tower of logarithms of $\Tau/Q$ predicted by the SCET$_\I$ RGE,
we perform variations of $\mu_B^\I$ and $\mu_S^\I$
parametrized by $\alpha$ and $\beta$ as in \eq{SCET1_vary}.
This directly affects the resummed power corrections $\bigl[ \df \sigma_\I - \df \sigma_+ \bigr]_{\mu_\I}$ captured by SCET$_\I$.
In addition, however, $\df \sigma_+ \bigr|_{\mu_+}$ near the SCET$_\I$ boundary also undergoes variations
because for large $h_\I$, the SCET$_+$ scales in \eqs{scetp_weighted_product_csoft_soft}{scetp_weighted_product_beam} strongly depend on their SCET$_\I$ counterparts
and inherit their variations.
Our setup thus ensures that in (or near) the SCET$_\I$ region,
variations probing resummed logarithms of $\Tau/Q$ are properly treated as correlated
between the SCET$_+$ cross section and the SCET$_\I$ matching correction.
When referring to the matched prediction in \eq{matching},
we take $\Delta_\I$ to be the maximum deviation of $\df \sigma_\match$
from its central value under these correlated variations of $\alpha$, $\beta$.

In complete analogy, we define $\Delta_\II$ as the maximum deviation
under correlated variations of $\mu_\II$ as described in \sec{scet2}.
These variations act on both $\bigl[ \df \sigma_\II - \df \sigma_+ \bigr]_{\mu_\II}$
and $\df \sigma_+ \bigr|_{\mu_+}$,
where now the SCET$_+$ scales inherit variations from $\mu_\II$ near the SCET$_\II$ boundary (where $h_\II$ is large).
As a result, $\Delta_\II$ probes an all-order set of logarithms of $(b_0/b_T)/Q$ predicted and resummed by the SCET$_\II$ RGE, and properly captures the correlated tower of logarithms in SCET$_+$.
We like to stress that our setup is fully general
with respect to the method chosen to perform scale variations for the boundary theories,
as any variation will automatically be inherited by SCET$_+$.

As a final source of uncertainty,
we consider the uncertainty inherent in our matching procedure
and in our choice of SCET$_+$ scales in the bulk.
To estimate this we perform the following 8 variations of the (in principle arbitrary) transition points $(a_1, a_3, a_4, a_6)$,
%%%
\begin{alignat}{7}
(\uparrow, -, -, -)
\,, \qquad &&
(-, \downarrow, -, -)
\,, \qquad &&
(-, -, -, \downarrow)
\,, \qquad &&
(-, \uparrow, \uparrow, -)
\,, \nn \\
(\downarrow, -, -, -)
\,, \qquad &&
(-, -, \uparrow, -)
\,, \qquad &&
(-, -, -, \uparrow)
\,, \qquad &&
(-, \downarrow, \downarrow, -)
\,,\end{alignat}
%%%
where $\uparrow \, (\downarrow)$ indicates a variation by $+0.2 \, (-0.2)$,
a dash indicates keeping the transition point fixed,
and we always maintain $a_2 = (a_1 + a_3)/2$ and $a_5 = (a_4 + a_6)/2$.
In addition, we perform the following two variations of the SCET$_+$ bulk scales,
%%%
\begin{align} \label{eq:}
\mu_{\cS,\bulk}^+ &= \mu_\FO \, \Bigl( \frac{q_T}{\Tau} \Bigr)^{+\gamma/2} \, f_\run^\II\Bigl( \frac{q_T}{Q}, \frac{b_0}{b^\ast Q} \Bigr)
\,, \nn \\
\mu_{S,\bulk}^+ &= \mu_\FO \, \Bigl( \frac{q_T}{\Tau} \Bigr)^{-\gamma/2} \, f_\run^\I \Bigl( \frac{\Tau}{Q} \Bigr)
\,, \qquad \qquad
\gamma = \{ + 1/6, - 1/6 \}
\,,\end{align}
%%%
where $\gamma = 0$ corresponds to the central scales in \eq{scetp_independent_bulk_scales}.
Similarly to the role of $\beta$ in the SCET$_\I$ variations [see \eq{SCET1_vary}],
making the strength of the $\gamma$~variations depend on the ratio $q_T/\Tau$
ensures that the hierarchy $\mu_S < \mu_\cS$ implied by the SCET$_+$ power counting
is not upset by variations, counting $b_0/b_T \sim q_T$.
We note that the third independent bulk scale $\nu_{\cS,\bulk}^+$ does not require
independent variation because it only enters through rapidity logarithms of $\nu_B^+/\nu_\cS^+$,
which are already being probed by variations of $\nu_B^+$ inherited from SCET$_\II$.
Taking the envelope of the eight transition point variations and the two bulk scale variations,
we obtain a third contribution to the resummation uncertainty denoted by $\Delta_+$.
The total uncertainty assigned to the matched prediction is then given by adding all contributions in quadrature,
%%%
\begin{align} \label{eq:Delta_total}
\Delta_\total = \Delta_+ \oplus \Delta_\I \oplus \Delta_\II \oplus \Delta_\FO
\,.\end{align}
%%%

%===============================================================================
\subsection{Differential and cumulant scale setting}
\label{sec:diff_cumul_scale_setting}
%===============================================================================

We will now discuss the issue of differential versus cumulant scale setting,
starting with the simpler case of a cross section differential in a single observable and using 0-jettiness $\Tau$ as an example.
There are two equivalent quantities of interest in this case,
namely the spectrum $\df \sigma / \df \Tau$ with respect to $\Tau$
and the cumulant cross section $\sigma(\TauCut)$ with a cut on $\Tau$.
The two quantities are related by
%%%
\begin{equation} \label{eq:general_relation_spectrum_cumulant}
\sigma(\TauCut) = \int_0^{\TauCut} \! \df \Tau \, \frac{\df \sigma}{\df \Tau}
\,,\end{equation}
%%%
where we suppress the dependence on $Q^2$ and $Y$ for the purposes of this subsection.
Accordingly, in a resummation analysis
one can implement the resummation scales
either in terms of the differential variable $\Tau$ to directly predict the spectrum,
or in terms of the cumulant variable $\Tau_\cut$ to predict the cross section integrated up to $\TauCut$.
The other observable then follows from \eq{general_relation_spectrum_cumulant}.

Explicitly, with differential scale setting (indicated by the subscript), the differential and cumulant cross section are given by
%%%
\begin{align} \label{eq:diff_scales}
\frac{\df \sigma_\diff}{\df \Tau} &=\frac{\df \sigma}{\df \Tau} \Big|_{\mu(\Tau)}
\,, \nn \\
\sigma_\diff(\Tau_\cut) &= \int^{\Tau_\cut}\! \df \Tau\,
\biggl[
   \theta(\Tau > \Tau_\np) \frac{\df \sigma}{\df \Tau} \Big|_{\mu(\Tau)}
   + \theta(\Tau \leq \Tau_\np)  \frac{\df \sigma}{\df \Tau} \Big|_{\mu(\Tau_\np)}
\biggr]
\,.\end{align}
 %%%
In the first term under the integral in the cumulant cross section, all scales $\mu$ entering the resummed and matched prediction depend on the integration variable $\Tau$.
Because our setup only reliably predicts the spectrum away from the nonperturbative region,
we choose to integrate the resummed spectrum with differential scale setting up from some small cutoff $\Tau_\np$,
and include an ``underflow'' contribution given by the second term under the integral.
For the underflow contribution for $\Tau \leq \Tau_\np$, the spectrum is evaluated at fixed scales corresponding
to $\Tau_\np$, such that the integral can be done analytically.
The underflow contribution is Sudakov suppressed and thus typically small.

Using cumulant scale setting, we instead use
%%%
\begin{align} \label{eq:cumul_scales}
\sigma_\cumul(\Tau_\cut)  &= \int^{\Tau_\cut}\! \df \Tau\, \frac{\df \sigma}{\df \Tau} \Big|_{\mu(\Tau_\cut)}
\,, \nn \\
\frac{\df \sigma_\cumul}{\df \Tau} &= \frac{\df \sigma}{\df \Tau} \Big|_{\mu(\Tau)} + \sum_i
\bigg[ \frac{\df}{\df \mu_i} \int^{\Tau}\! \df \Tau'\, \frac{\df \sigma}{\df \Tau'}\bigg]_{\mu(\Tau)} \, \frac{\df \mu_i(\Tau)}{\df \Tau}
\,.\end{align}
%%%
In this case, the scales in the cumulant cross section depend on $\TauCut$, and not the integration variable $\Tau$,
so the integral up to $\TauCut$ can easily be performed analytically.
The expression for the differential cross section arises from taking the derivative of the cumulant cross section,
where the chain rule leads to the sum of derivatives of each of the scales $\mu_i$ in $\mu$ with respect to $\Tau$.

Cumulant scale setting ensures that for $\Tau_{\rm cut} \to Q$,
the resummed and matched cumulant cross section exactly reproduces the inclusive fixed-order cross section.
This follows from the generic requirement on profile scales in the fixed-order region,
%%%
\begin{align} \label{eq:FO_limit}
  \mu_i(\TauCut) \to  \mu_\FO
\quad \text{for} \quad \TauCut \to Q
\,.\end{align}
%%%
Thus for cumulant scale setting, the spectrum has the correct (fixed-order) normalization.
However, the additional derivatives of the scales in  \eq{cumul_scales} tend to produce artifacts in the spectrum
if the profile functions $\mu_i(\Tau)$ used to interpolate between the resummation region $\Tau \ll Q$ to the fixed-order region $\Tau \sim Q$ undergo a rapid transition. In particular, a smooth
matching to the fixed-order prediction at the level of the differential spectrum typically
requires differential scale setting. Moreover, the scale variations using cumulant scale
setting tend to produce unreliable uncertainties for the spectrum.

If instead differential scale setting is used, the spectrum is free from such artifacts.
However, the integral of the spectrum will not exactly recover the inclusive fixed-order cross section,
and the uncertainties obtained for the cumulant by integrating the spectrum scale variations
tend to accumulate and end up being much larger than they should be for the total cross section.
As in the case of the spectrum with cumulant scale setting,
this mismatch purely arises from residual scale dependence,
and therefore is formally beyond the working order. It can however still be numerically significant.

Therefore, in general one should use the scale setting that is appropriate for the quantity of interest,
i.e., one should use cumulant scale setting when making predictions for the cumulant, and differential scale setting when one is interested in the spectrum.
This issue of differential versus cumulant scale setting
is well appreciated in the literature for the single-differential case, see e.g.\ \refscite{Abbate:2010xh, Almeida:2014uva, Alioli:2015toa, Bertolini:2017eui}.
It fundamentally results from the fact that long-range correlations across the spectrum
are not accounted for by the profile scales used for the differential predictions.
Conversely, profile scales for the cumulant do not correctly capture the slope of the
cumulant and its uncertainty.
An elaborate procedure for obtaining a spectrum with differential scales that still produce the
exact cross section and uncertainties was developed in \refcite{Bertolini:2017eui}.
In the \textsc{Geneva} Monte Carlo generator,
the mismatch between differential and cumulant scales is accounted for by adding explicit higher-order terms \cite{Alioli:2015toa}.

For a simultaneous measurement of $q_T$ and $\Tau$, there are in principle four quantities of interest,
namely the double-differential spectrum $\df \sigma / \df q_T \, \df \Tau$,
the single-differential spectra $\df \sigma(\qTcut) / \df \Tau$ and $\df \sigma(\TauCut) / \df q_T$ with a cut on the other variable,
and the double cumulant $\sigma(\qTcut, \TauCut)$.
They are all related by integration or differentiation,
allowing for four different ways of setting scales in each case.
For our explicit numerical results in \sec{results}, we take a pragmatic approach
and use the appropriate combination of differential or cumulant scale setting with respect to either $q_T$ or $\Tau$ for each of these quantities.
This is achieved by evaluating the resummed prediction at profile scales
given by the setup described in \secs{scet1}{scet2} as well as \sec{profiles},
but with $q_T$ ($\Tau$) replaced by $\qTcut$ ($\TauCut$) as appropriate.
In this way we are guaranteed to avoid artifacts from profile functions in spectrum observables,
and on the other hand ensure that cumulant observables have the correct limiting behavior;
e.g., $\sigma(\qTcut, \TauCut)$ will by construction recover the inclusive fixed-order cross section when lifting both cuts,
while $\df \sigma(\qTcut) / \df \Tau$ and $\df \sigma(\TauCut) / \df q_T$ exactly recover
the resummed and matched prediction for the respective inclusive spectrum at large values of the cut.

Nevertheless, it is interesting to ask how well the different combinations of differential and cumulant scale setting fare for observables other than the one they are designed to describe.
In particular we should ask how well the $(q_T, \Tau)$ scale setting we described in earlier sections
performs at the level of cumulant observables and their inclusive limit.
To do so, we can always promote a spectrum using differential scale setting in $q_T$ ($\Tau$)
to a prediction for the cumulant up to $\qTcut$ ($\TauCut$) using the analogue of \eq{diff_scales}.
The only nontrivial new procedure is computing the double cumulant directly from $(q_T, \Tau)$ scales,
where we need to account for an overlap in underflow contributions as
%%%
\begin{align} \label{eq:diff_scales_2D}
\sigma_{\diff,\diff}(\qTcut, \TauCut)
&= \int^{\qTcut}\! \df q_T\, \int^{\Tau_\cut}\! \df \Tau\,
\Bigl[
\theta(q_T > q_T^\np) \, \theta(\Tau > \Tau_\np) \frac{\df \sigma}{\df q_T \, \df \Tau} \Big|_{\mu(q_T, \Tau)}
\\ & \qquad\qquad\qquad\qquad\quad
+ \theta(q_T \leq q_T^\np) \, \theta(\Tau > \Tau_\np) \frac{\df \sigma}{\df q_T \, \df \Tau} \Big|_{\mu(q_T^\np, \Tau)}
\nn \\ & \qquad\qquad\qquad\qquad\quad
+ \theta(q_T > q_T^\np) \, \theta(\Tau \leq \Tau_\np) \frac{\df \sigma}{\df q_T \, \df \Tau} \Big|_{\mu(q_T, \Tau_\np)}
\nn \\ & \qquad\qquad\qquad\qquad\quad
- \theta(q_T \leq q_T^\np) \, \theta(\Tau \leq \Tau_\np) \frac{\df \sigma}{\df q_T \, \df \Tau} \Big|_{\mu(q_T^\np, \Tau_\np)}
\Bigr]
\,. \nn\end{align}
%%%
The distinction between differential or cumulant scale setting is only relevant for $q_T$ versus $\qTcut$
but not for the underlying resummation in $b_T$ space,
so we suppress the dependence of the hybrid scales on $b_T$.
In practice we use $q_T^\np = \Tau_\np = 1 \GeV$,
and implement the integrals in \eqs{diff_scales}{diff_scales_2D}
as sums over logarithmically spaced bins with bin size $\Delta(\log_{10} q_T) = \Delta(\log_{10} \Tau) = 0.08$,
where the spectrum is evaluated at the logarithmic midpoint of the bin.
Scale variations in the integrated results are performed
by integrating each instance of the spectrum separately
and computing maximum deviations from central in the end.
The final results are interpolated for clarity.

\begin{figure*}
\centering
\includegraphics[width=\WidthTwoSubfigs]{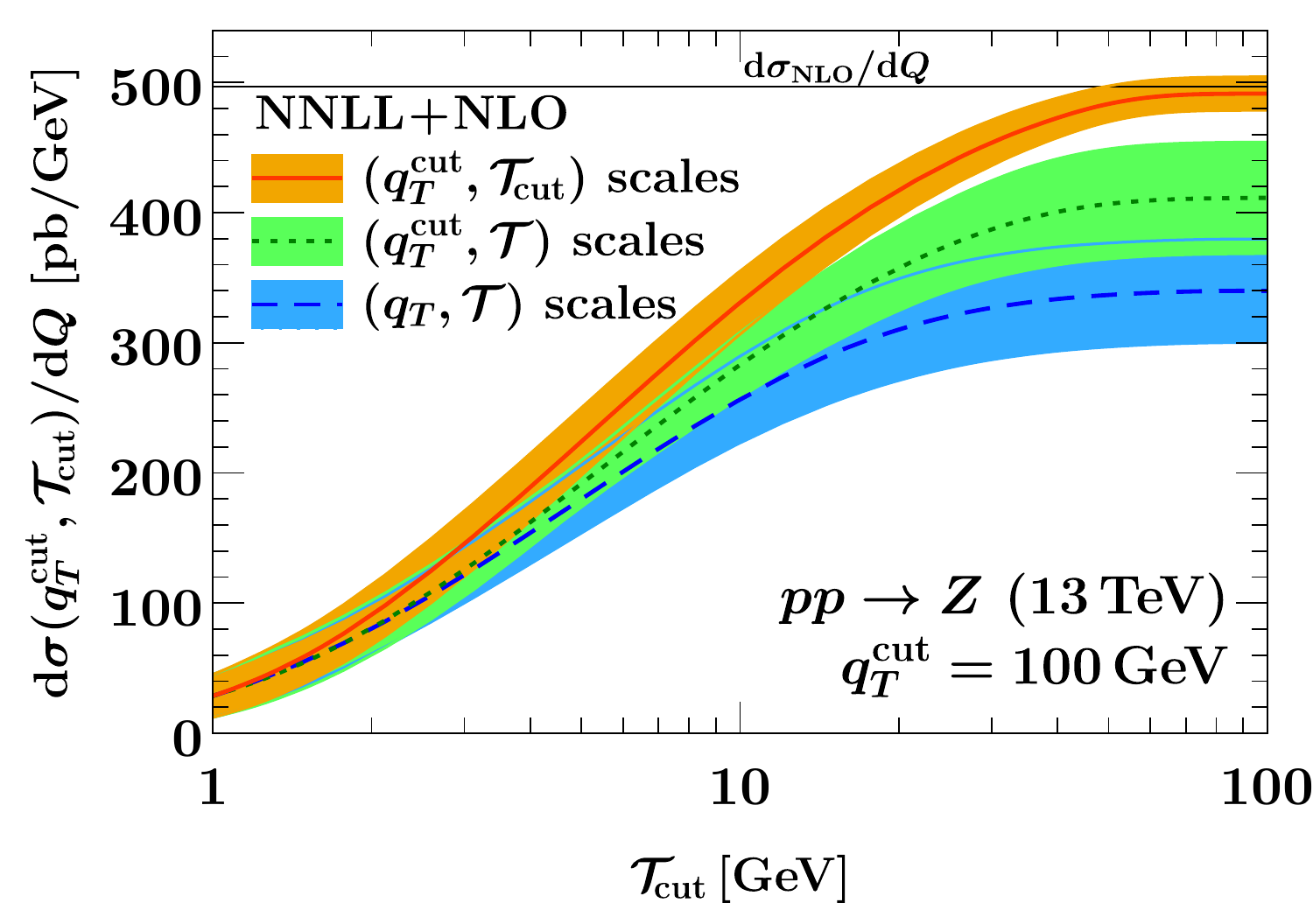}%
\hfill%
\includegraphics[width=\WidthTwoSubfigs]{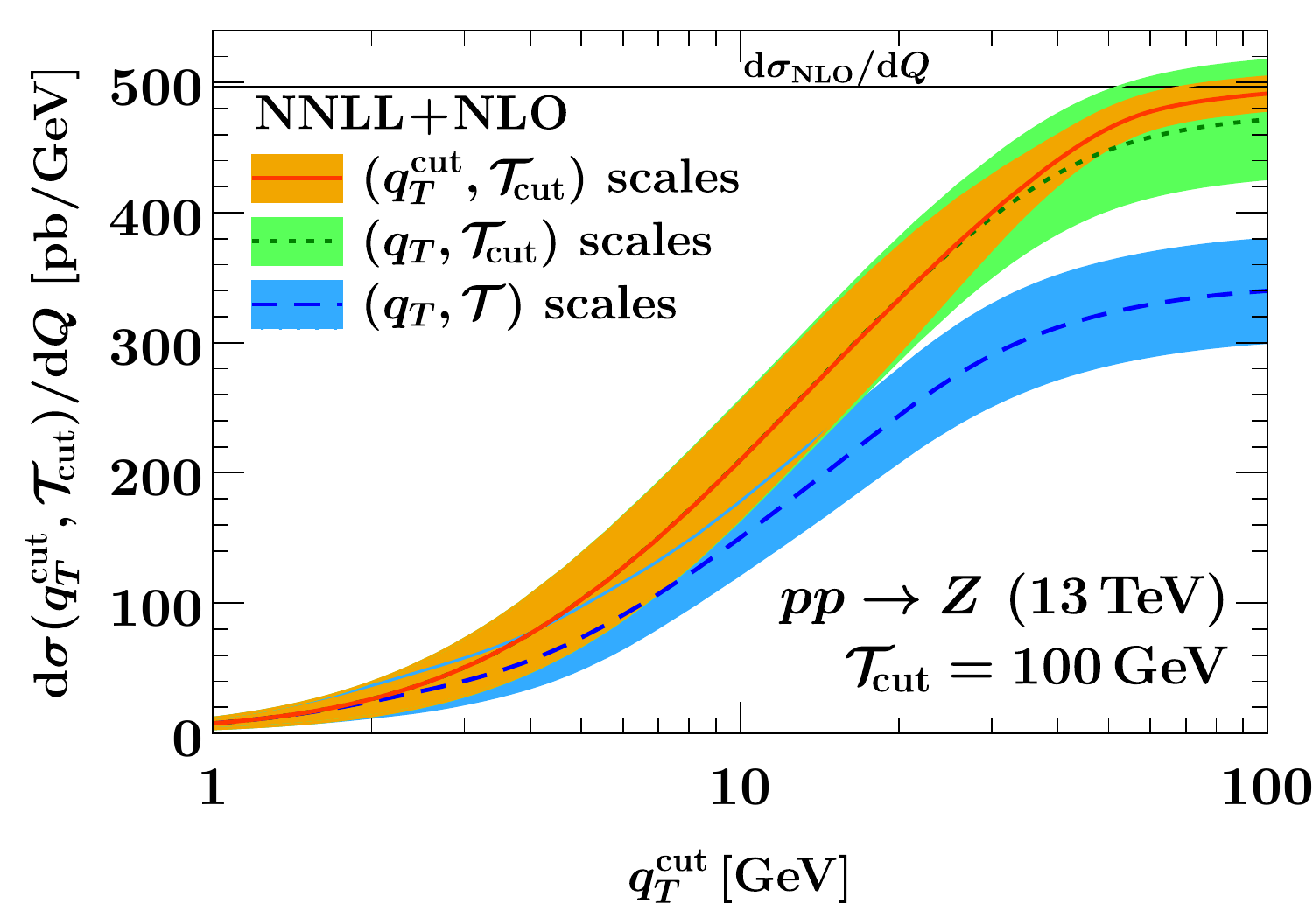}%
\caption{The double cumulant cross section
as a function of $\TauCut$ for $\qTcut = 100 \GeV$ (left)
and as a function of $\qTcut$ for $\TauCut = 100 \GeV$ (right).
The bands indicate the total perturbative uncertainty $\Delta_\total$, see \sec{variations}.
The colors correspond to different scale setting prescriptions (default: solid orange); see the text for details.}
\label{fig:compare_scales_sigma_qTcut_TauCut}
\end{figure*}

\begin{figure*}
\centering
\includegraphics[width=\WidthTwoSubfigs]{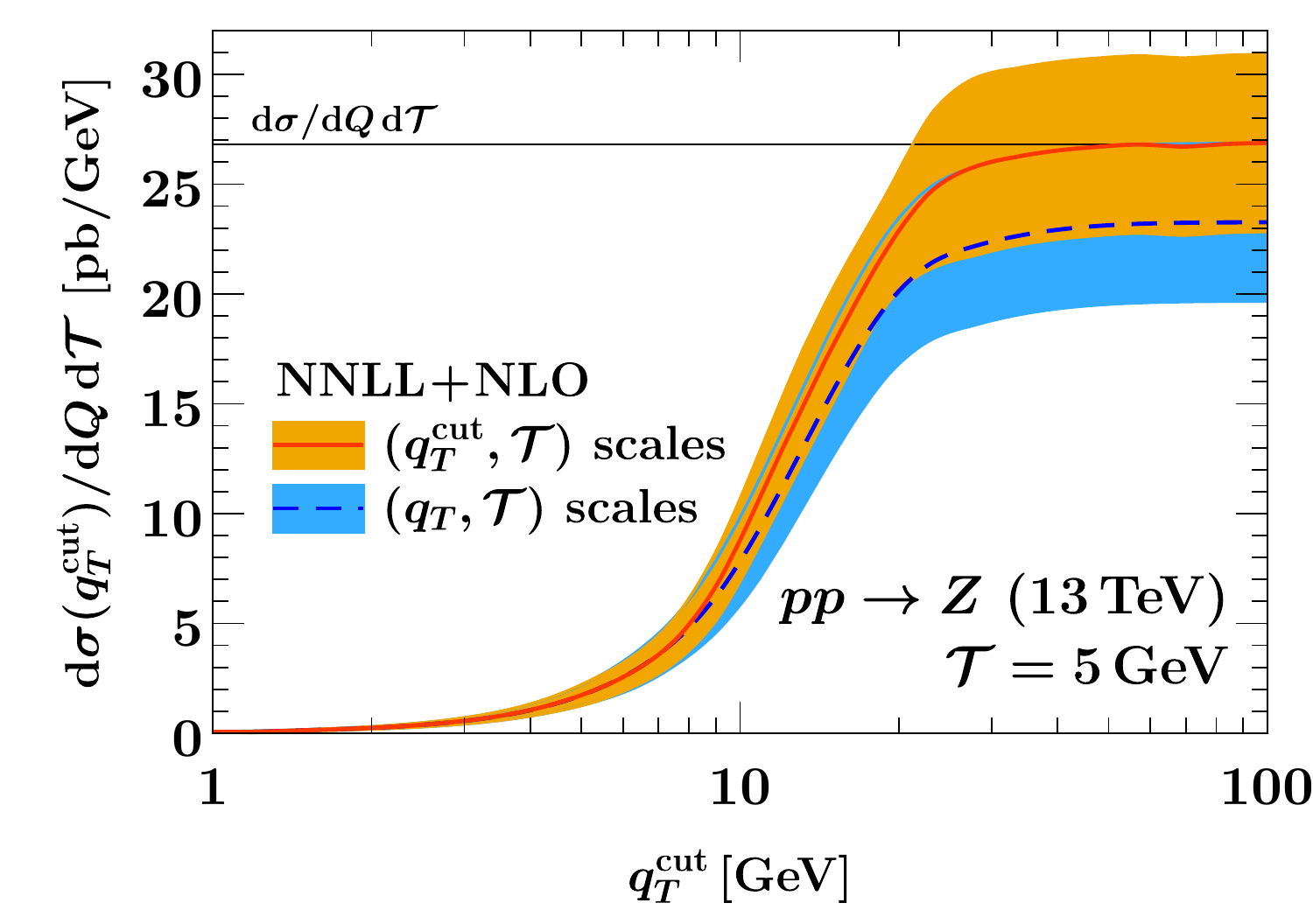}%
\hfill%
\includegraphics[width=\WidthTwoSubfigs]{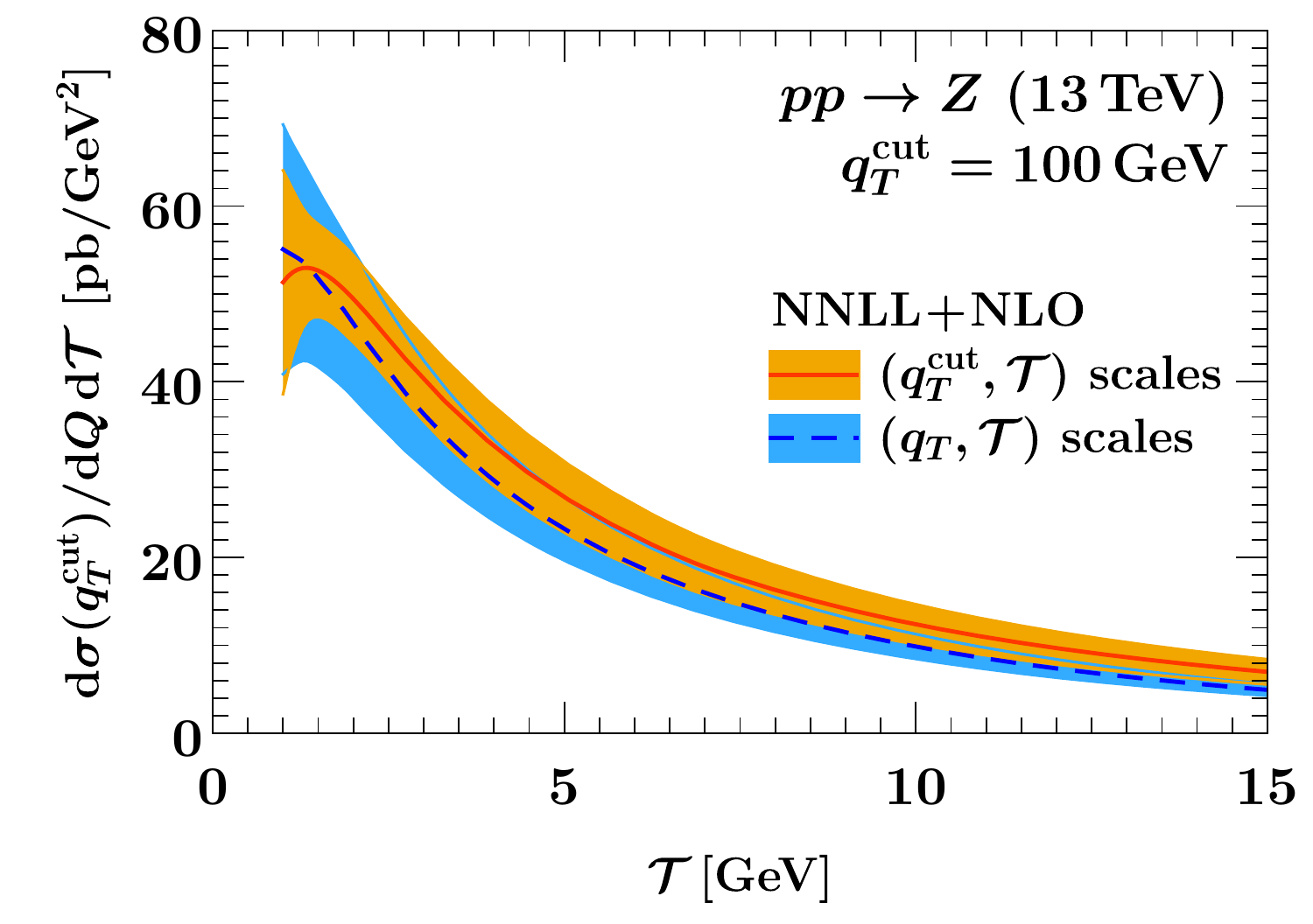}%
\caption{The $\Tau$ spectrum with a cut on $q_T$
as a function of $\qTcut$ at fixed $\Tau = 5 \GeV$ (left)
and as a function of $\Tau$ at fixed $\qTcut = 100 \GeV$ (right).
The bands indicate the total perturbative uncertainty $\Delta_\total$, see \sec{variations}.
The colors correspond to different scale setting prescriptions (default: solid orange); see the text for details.}
\label{fig:compare_scales_sigma_qTcut_Tau}
\end{figure*}

\begin{figure*}
\centering
\includegraphics[width=\WidthTwoSubfigs]{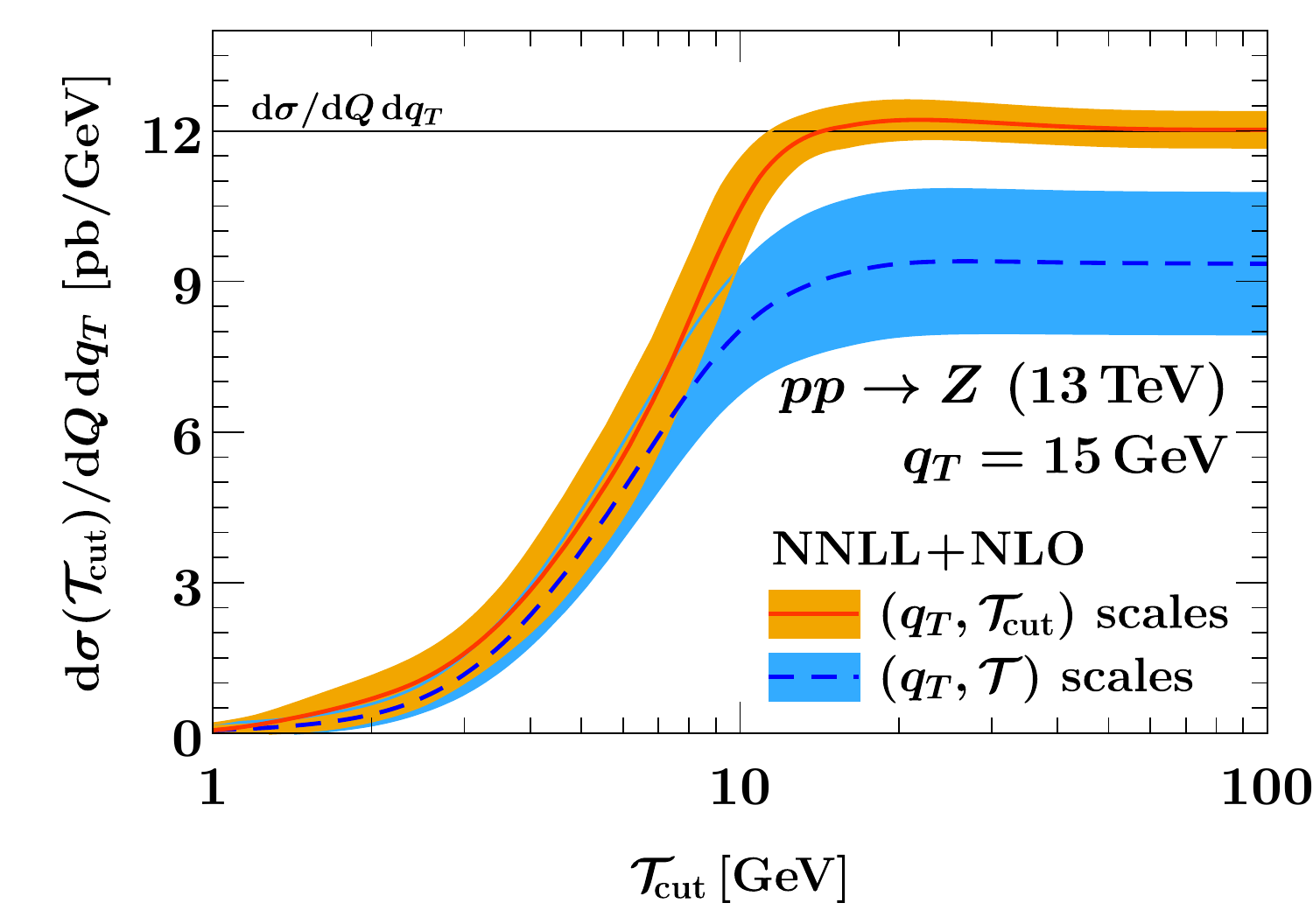}%
\hfill%
\includegraphics[width=\WidthTwoSubfigs]{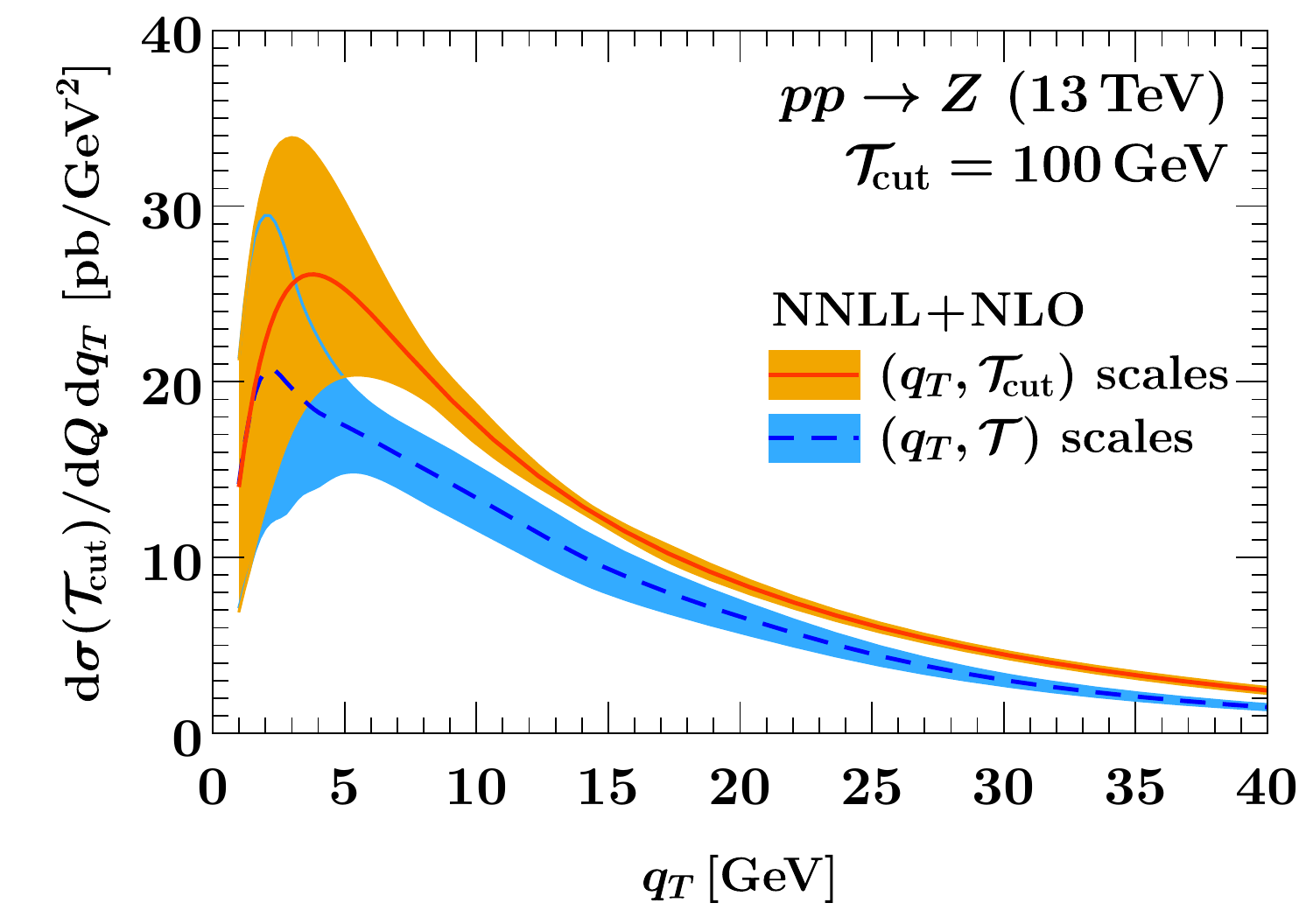}%
\caption{The $q_T$ spectrum with a cut on $\Tau$
as a function of $\TauCut$ for $q_T = 15 \GeV$ (left)
and as a function of $q_T$ for $\TauCut = 100 \GeV$ (right).
The bands indicate the total perturbative uncertainty $\Delta_\total$, see \sec{variations}.
The colors correspond to different scale setting prescriptions (default: solid orange); see the text for detail.}
\label{fig:compare_scales_sigma_qT_TauCut}
\end{figure*}

In figures~\ref{fig:compare_scales_sigma_qTcut_TauCut} to \ref{fig:compare_scales_sigma_qT_TauCut},
we compare our default scale setting for various cumulant observables (solid orange)
to more differential scale setting (dashed blue and dotted green),
i.e., choosing $\mu$ in terms of $q_T$ rather than $\qTcut$ and/or $\Tau$ rather than $\TauCut$.
In \fig{compare_scales_sigma_qTcut_TauCut}, we show the double cumulant cross section,
for which our default is to use scales in terms of $q_T^\cut$ and $\Tau_\cut$.
The horizontal reference line indicates the inclusive fixed-order cross section.
In \fig{compare_scales_sigma_qTcut_Tau} we show the $\Tau$ spectrum with a cut on $q_T$,
for which our default scales are in terms of $q_T^{\cut}$ and $\Tau$,
and the converse for \fig{compare_scales_sigma_qT_TauCut}.
In \figs{compare_scales_sigma_qTcut_Tau}{compare_scales_sigma_qT_TauCut} the left panel shows the dependence on the cut at a representative point along the spectrum,
with the reference line indicating the resummed prediction for the inclusive (strictly single-differential) spectrum.
The right panel shows the spectrum at a representative choice of the cut.

We start by observing that in all cases,
the predictions obtained using the default scale setting (solid orange)
cleanly asymptote to the respective target observable (the reference line) for large values of the cut.
The central double-differential prediction in the left panel of \fig{compare_scales_sigma_qT_TauCut} slightly overshoots the inclusive result
beyond the phase-space boundary $\TauCut \gtrsim q_T$ (where our calculation is effectively a leading-order calculation),
but is monotonic within uncertainties.
Furthermore, the uncertainty obtained using our default is smaller than any of the ones obtained from more differential scale setting.
This is expected because differential scale setting cannot account for correlations between different bins of the spectrum,
giving rise to a larger band in the cumulant cross sections.

We further note that predictions obtained using $q_T$ or $\qTcut$ scale setting are mutually compatible,
i.e., their uncertainty bands (very nearly) overlap, as long as the scale setting with respect to $\Tau$ is done the same way in both cases.
This can be seen from the right panel of \fig{compare_scales_sigma_qTcut_TauCut}
by contrasting the default $(\qTcut, \TauCut)$ scales (solid orange) and $(q_T, \TauCut)$ scales (dotted green).
Similarly, in \fig{compare_scales_sigma_qTcut_Tau} we find that the default $(\qTcut, \Tau)$ scales (solid orange)
and $(q_T, \Tau)$ scale setting (dashed blue) roughly differ by their respective uncertainties.
In principle these relations are expected
since the unphysical scale dependence is canceled by higher-order corrections,
which our scale variations are designed to probe.
For the case of $q_T$ versus $\qTcut$ scales in particular,
we note that due to our specific choice of hybrid profile scales in \eq{fRun2},
differences between the two prescriptions only start to appear when turning off the resummation, such that $g_\run$ is nonzero.
E.g.\ for a high $\TauCut = 100 \GeV$, which is also a good proxy for the inclusive $q_T$ spectrum,
the two prescriptions fully agree in the canonical region $\qTcut \leq 20 \GeV$ (see the right panel of \fig{compare_scales_sigma_qTcut_TauCut}).
This is responsible for the good overall agreement because most of the cross section is concentrated in the canonical region.

The comparison of $\Tau$ versus $\TauCut$ scales is much less favorable,
with the former failing to reproduce the latter's inclusive limit within uncertainties in all cases.
This is in line with the discrepancy reported in \refcite{Bertolini:2017eui}
for a single-differential measurement of thrust in $e^+ e^-$ collisions and at a comparable working order (NLL$'+$NLO).
The mismatch is most striking between the default scales (solid orange)
and $(q_T, \Tau)$ scales (dashed blue) in \figs{compare_scales_sigma_qTcut_TauCut}{compare_scales_sigma_qT_TauCut},
implying that more effort is required to ensure both a correct integral
and the best possible prediction for the shape of the double-differential spectrum.

From our previous discussion we conclude that the mismatch mostly reduces to the question of differential versus cumulant scale setting in $\Tau$ alone,
so that the methods developed for the single-differential case in \refscite{Bertolini:2017eui, Alioli:2015toa} can be brought to bear here as well if desired. However,
since this is a well-known issue that is merely inherited from the single-differential
case, we do not pursue this further here.

Instead, we consider a modification of our profile scales to illustrate that the
issue is indeed a correlated higher-order effect related to scale choices.
Specifically, we can consider lowering the canonical low scale $\mu_S^\I \sim (\mu_B^\I)^2/\mu_H^\I \sim \Tau$ in SCET$_\I$ by a factor of $c = 0.5$,
which does not parametrically violate the canonical scaling.
Including a smooth interpolation to the fixed-order and nonperturbative region,
this can be achieved by replacing \eq{f_run_scet1} with
%%%
\begin{align} \label{eq:f_run_scet1_modified}
f_\run^\I(c; x)
&= \begin{cases}
x_0 \Bigl(  1 + \frac{c^2 x^2}{4 x_0^2} \Bigr) &x \leq 2x_0 / c\,, \\
cx & 2x_0/c < x \leq x_1\,, \\
cx + \frac{(2-cx_2-cx_3)(x-x_1)^2}{2(x_2-x_1)(x_3-x_1)} & x_1 < x \leq x_2\,, \\
1 - \frac{(2-cx_1-cx_2)(x-x_3)^2}{2(x_3-x_1)(x_3-x_2)} & x_2 < x \leq x_3\,, \\
1 & x_3 < x\,,
\end{cases}
\end{align}
%%%
and keeping the entire remaining profile setup unchanged;
setting $c = 1$ recovers \eq{f_run_scet1}.

\begin{figure*}
\centering
\includegraphics[width=\WidthTwoSubfigs]{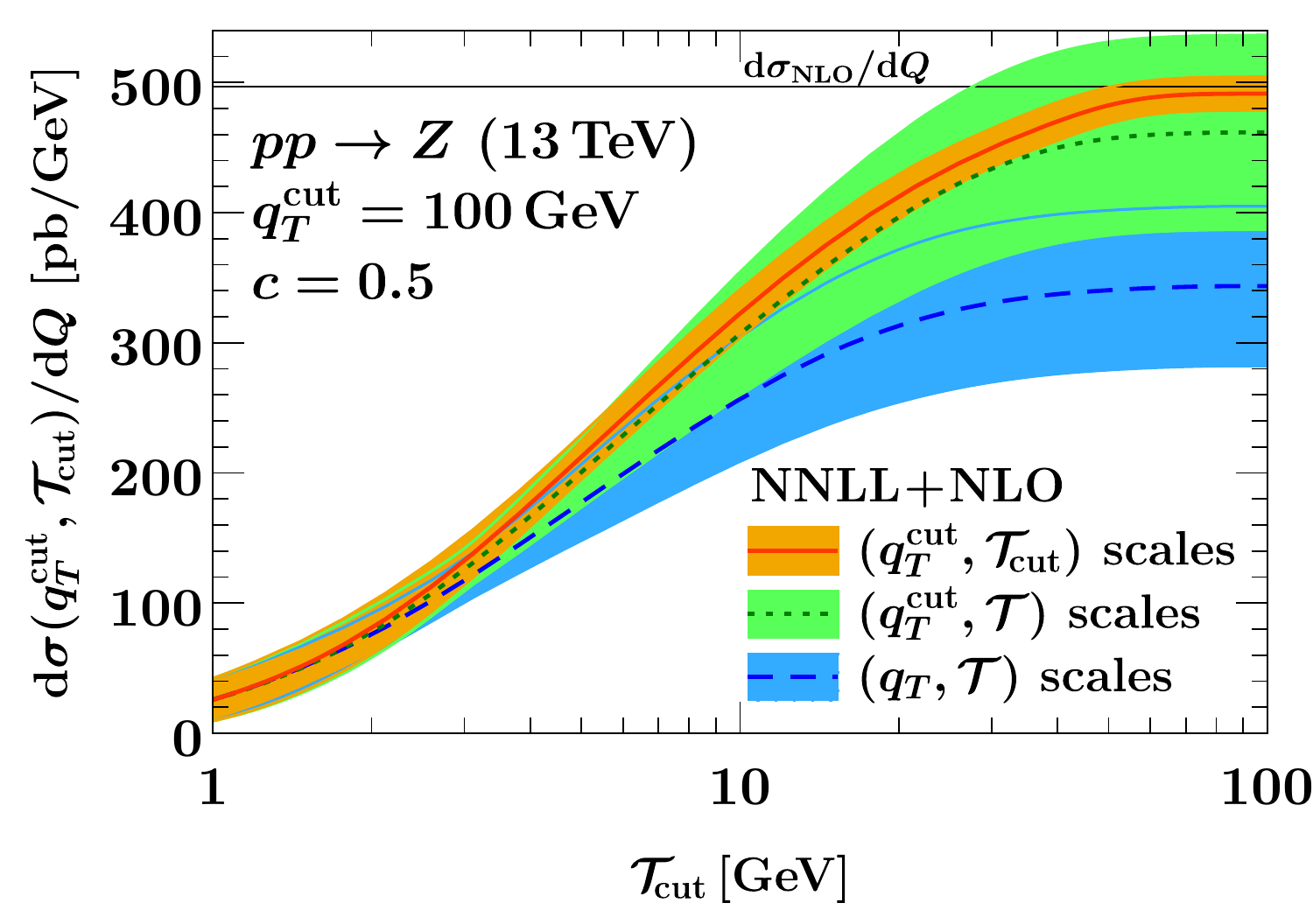}%
\hfill%
\includegraphics[width=\WidthTwoSubfigs]{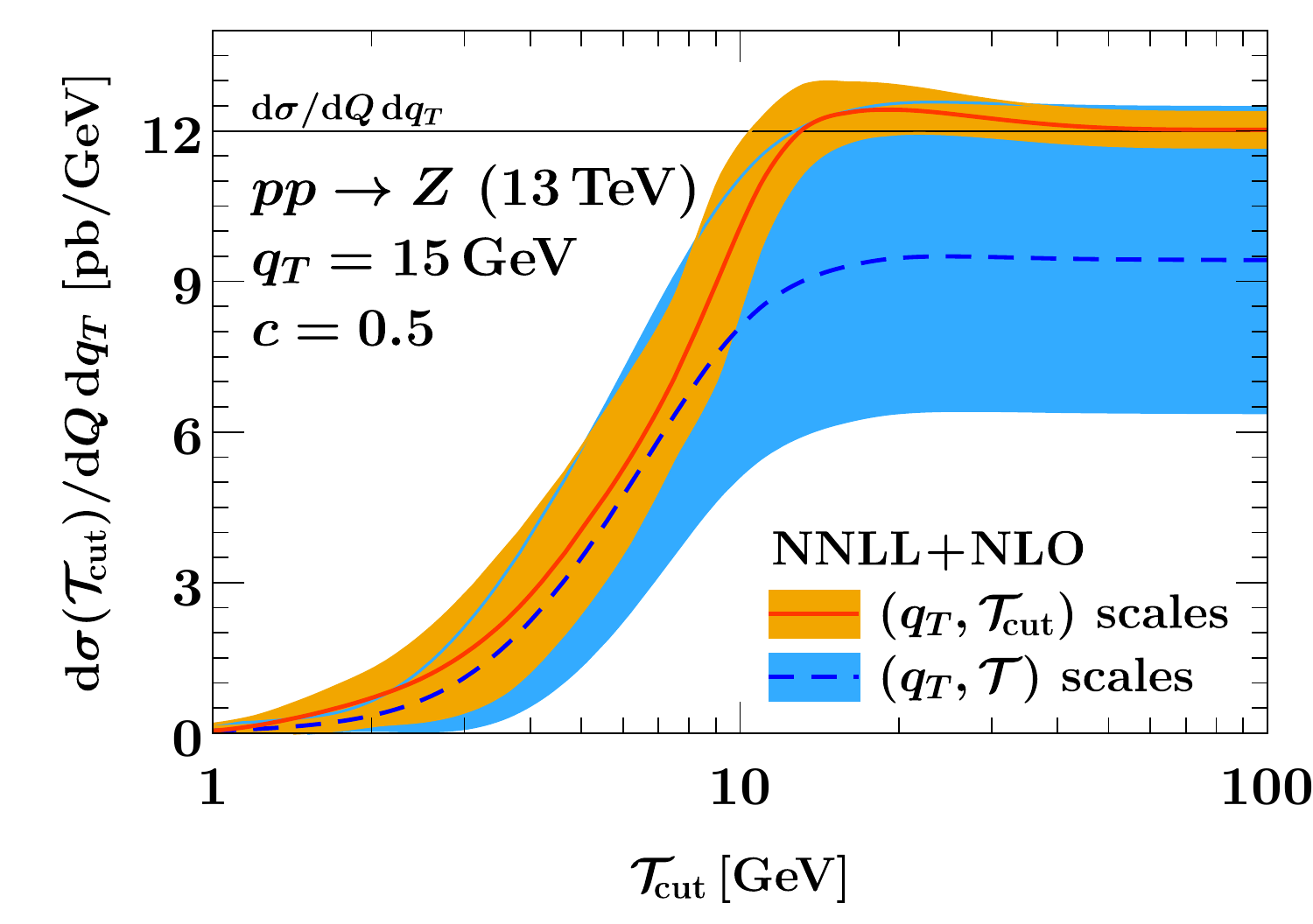}%
\caption{Left: The double cumulant cross section
as a function of $\TauCut$ for $\qTcut = 100 \GeV$
for different scale setting prescriptions,
with a modified slope $c = 0.5$ of the SCET$_\I$ profile scales, see \eq{f_run_scet1_modified}.
Right: The $q_T$ spectrum with a cut on $\Tau$ as a function of $\TauCut$
for different scale setting prescriptions,
also using modified SCET$_\I$ profile scales with $c = 0.5$.
The bands indicate the total perturbative uncertainty $\Delta_\total$, see \sec{variations}.}
\label{fig:compare_scales_c_05}
\end{figure*}

Our results using \eq{f_run_scet1_modified} are shown in \fig{compare_scales_c_05},
where we repeat the left panels of \figs{compare_scales_sigma_qTcut_TauCut}{compare_scales_sigma_qT_TauCut}
using the modified setup.
Note that for simplicity, we use \eq{f_run_scet1_modified} for all results in this figure,
i.e., for both differential and cumulant scale setting.
We find that the simple modification \eq{f_run_scet1_modified}
already substantially improves the agreement between differential and cumulant scale setting,
with the result from $(\qTcut, \Tau)$ scales (dotted green, left panel) covering the inclusive fixed-order cross section
and the result from $(q_T, \Tau)$ scales (dashed blue, right panel) covering the result from single-differential $q_T$ resummation, at the price of much larger uncertainties.

We conclude that with additional effort, e.g.\ applying the methods used in \refscite{Alioli:2015toa, Bertolini:2017eui}, it would be possible to fully reconcile the best possible predictions for both the differential shape and the cumulant of the double-differential spectrum.
However, for our purposes we can simply use the appropriate scale setting
for the observable of interest. In particular, if the experimental observable of interest has cumulant-like character in either $q_T$ or $\Tau$,
e.g.\ if large bins in either observable are considered,
the double-differential profile setup given in this paper,
using $(\qTcut, \Tau)$ or $(q_T, \TauCut)$ scales as appropriate,
will be completely sufficient.

%%%%%%%%%%%%%%%%%%%%%%%%%%%%%%%%%%%%%%%%%%%%%%%%%%%%%%%%%%%%%%%%%%%%%%%%%%%%%%%%
\section{Results}
\label{sec:results}
%%%%%%%%%%%%%%%%%%%%%%%%%%%%%%%%%%%%%%%%%%%%%%%%%%%%%%%%%%%%%%%%%%%%%%%%%%%%%%%%

In this section we present our results for Drell-Yan production $pp \to Z/\gamma^\ast \to \ell^+ \ell^-$ at the LHC,
with a simultaneous measurement of the transverse momentum $q_T$ of the lepton pair and the 0-jettiness event shape $\Tau$.
The center-of-mass energy is taken to be $\Ecm = 13 \TeV$.
We assume that in addition, the invariant mass $Q$ of the lepton pair is measured,
and write $pp \to Z$ for short if $Q = m_Z$, and $pp \to Z^\ast$ otherwise.
The subsequent decay and the contribution from the virtual photon are included in either case.

To obtain numerical results for the SCET$_\I$, SCET$_\II$, and SCET$_+$ contributions,
we have implemented all pieces of the relevant double-differential factorized cross sections to
$\ord{\as}$ and their RGEs to NNLL in \texttt{SCETlib}~\cite{scetlib}.
The fixed NLO contributions in full QCD are obtained from \texttt{MCFM~8.0}~\cite{Campbell:1999ah,Campbell:2011bn,Campbell:2015qma}.
We make use of the \texttt{MMHT2014nnlo68cl}~\cite{Harland-Lang:2014zoa} NNLO PDFs
with five-flavor running and $\as(m_Z)= 0.118$.
Since we focus on the perturbative calculation and do not include any nonperturbative effects,
we provide the results down to $1 \GeV$ in $q_T$ and $\Tau$.

The outline of this section is as follows:
In \sec{results_double_spectrum} we present our fully resummed prediction for the double-differential spectrum,
both as surface plots over the $(q_T, \Tau)$ plane
and for selected slices along lines of constant $q_T$ or $\Tau$.
We demonstrate that our prediction smoothly interpolates between the SCET$_\I$ and SCET$_\II$ boundary theories,
i.e., we show that our matching formula in \eq{matching} recovers the matched predictions on either boundary
and improves over them by an additional resummation of power-suppressed terms.
Finally, in \sec{results_spectrum_with_cut} we present our predictions
for the single-differential spectra $\df\sigma(\qTcut)/\df\Tau$ and $\df\sigma(\TauCut)/\df q_T$ with a cut on the other variable,
and show how they recover the inclusive single-differential $\Tau$ and $q_T$ spectra for large values of $\qTcut$ and $\TauCut$, respectively.

%===============================================================================
\subsection{Double spectrum and comparison with boundary theories}
\label{sec:results_double_spectrum}
%===============================================================================

\begin{figure*}
\centering
\includegraphics[width=\textwidth]{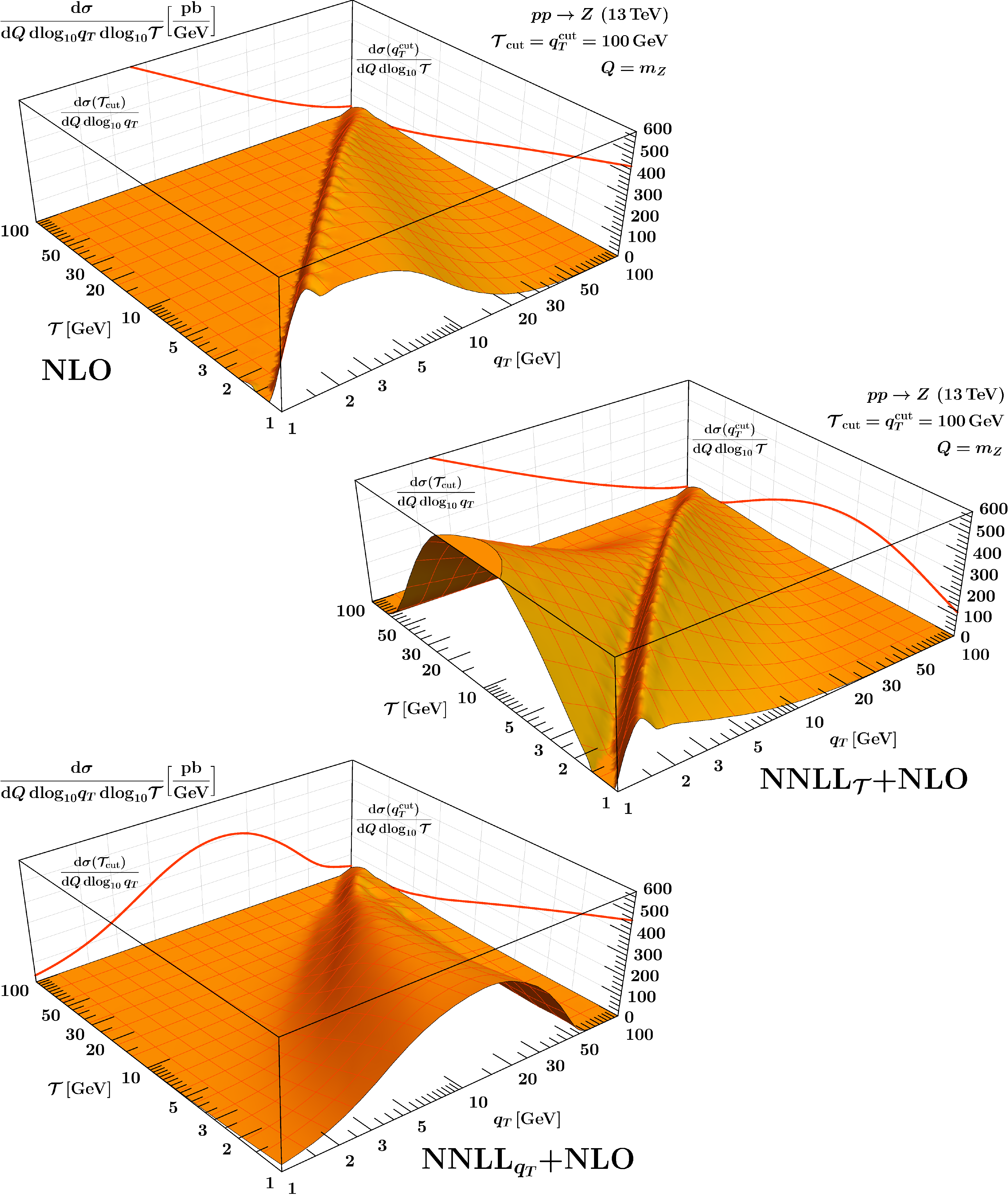}%
\caption{The double-differential Drell-Yan cross section at fixed NLO (top), resummed NNLL$_\Tau+$NLO (center), and NNLL$_{q_T}+$NLO (bottom).
The resummed predictions are obtained by using only SCET$_\I$ (SCET$_\II$) renormalization group evolution
to resum logarithms of $\Tau$ ($q_T$), as outlined in \sec{scet1} (\sec{scet2}),
and matching the result to the fixed-order cross section.
For better visibility we show the spectrum with respect to $\log_{10} q_T$ and $\log_{10} \Tau$.
On the rear walls we show the result of integrating the double spectrum over either variable up to a cut at $100 \GeV$.}
\label{fig:3d_plot_fo_scet1_scet2_only}
\end{figure*}

To highlight the necessity of the simultaneous resummation of large logarithms of both $q_T$ and $\Tau$,
we start by showing results for the double spectrum (the cross section double-differential in $q_T$ and $\Tau$) where only some of the logarithms are resummed.
These results are shown as surface plots in \fig{3d_plot_fo_scet1_scet2_only},
where we plot the double-differential spectrum with respect to $\log_{10} q_T$ and $\log_{10} \Tau$ for better visibility.
In each case the left rear wall of the surface plot shows the result of integrating the double-differential spectrum up to $\TauCut = 100 \GeV$,
but staying differential in $\log_{10} q_T$.
Similarly, the right rear wall shows the projection onto the single-differential spectrum in $\log_{10} \Tau$, with a cut at $\qTcut = 100 \GeV$.%
\footnote{We refer the reader to \sec{diff_cumul_scale_setting} for the precise way we perform these integrals.}

The top left panel of \fig{3d_plot_fo_scet1_scet2_only} shows the spectrum evaluated at fixed $\ord{\as}$, without any resummation.
The double-differential fixed-order spectrum diverges logarithmically for small $\Tau$ at any value of $q_T$,
while its projections onto the single-differential spectra in $q_T$ and $\Tau$
feature double-logarithmic singularities.
Notably, the double-differential spectrum has a sharp kinematic edge along $q_T = \Tau$.
This sharp edge is unphysical because it only reflects
the kinematics of a single on-shell emission with transverse momentum $k_T$ at rapidity $\eta$,
which contributes at most $\Tau = k_T \, e^{-\abs{\eta}} \leq k_T = q_T$.
Due to the vectorial nature of $q_T$, however,
back-to-back emissions can populate the region $\Tau > q_T$ at higher orders,
and the kinematic edge must be smeared out.

Next, we consider the cases in which only logarithms of one variable are resummed,
while logarithms involving the auxiliary variable are treated at fixed order.
In the middle panel of \fig{3d_plot_fo_scet1_scet2_only}, we show the result of
resumming logarithms of $\Tau$ using the SCET$_\I$ matched result in \eq{matching_scet1}.
The resummation is performed at NNLL and is matched to full NLO, which we refer to as NNLL$_\Tau+$NLO.
As discussed in \sec{scet1}, this prediction is valid
as long as the parametric relation $\Tau \ll q_T \sim \sqrt{Q\Tau}$ is satisfied.
This corresponds to the SCET$_\I$ phase-space boundary (blue) in \fig{overview_regimes},
running from the region of small $\Tau$ and intermediate $q_T$ towards the fixed-order region where $q_T \sim \Tau \sim Q$.
It is clear that away from its region of validity, the NNLL$_\Tau+$NLO result contains unresummed logarithms of $q_T$
because at any point in $\Tau$, the prediction diverges for very small $q_T$.
In particular, power corrections of $\mathcal{O}(\Tau^2/q_T^2)$ are only captured by the fixed-order matching.
They become $\mathcal{O}(1)$ as one approaches the diagonal $\Tau = q_T$ (the green line in \fig{overview_regimes}),
and encode the phase-space boundary at $q_T \sim \Tau$.
As in the NLO case, treating this phase-space boundary at fixed order leads to the sharp kinematic edge along the diagonal;
physically, the all-order tower of collinear emissions that contribute to $q_T$ in SCET$_\I$
cannot resolve the boundary because it arises from the dynamics at central rapidities.
The projections onto the rear walls highlight that only $\Tau$ is resummed.
The single-differential $q_T$ spectrum still diverges as $q_T \to 0$,
while the $\Tau$ spectrum features a physical Sudakov peak.

\begin{figure*}
\centering
\begin{minipage}[t]{\WidthDiagonalSubfig}%
   \vspace{0pt}%
   \includegraphics[width=\textwidth]{plots/3d_plots/resummed_matched_Q_mZ}%
\end{minipage}%
\hfill%
\begin{minipage}[t]{\WidthThreeSubfigs}%
   \vspace{0pt}%
   \includegraphics[width=\textwidth]{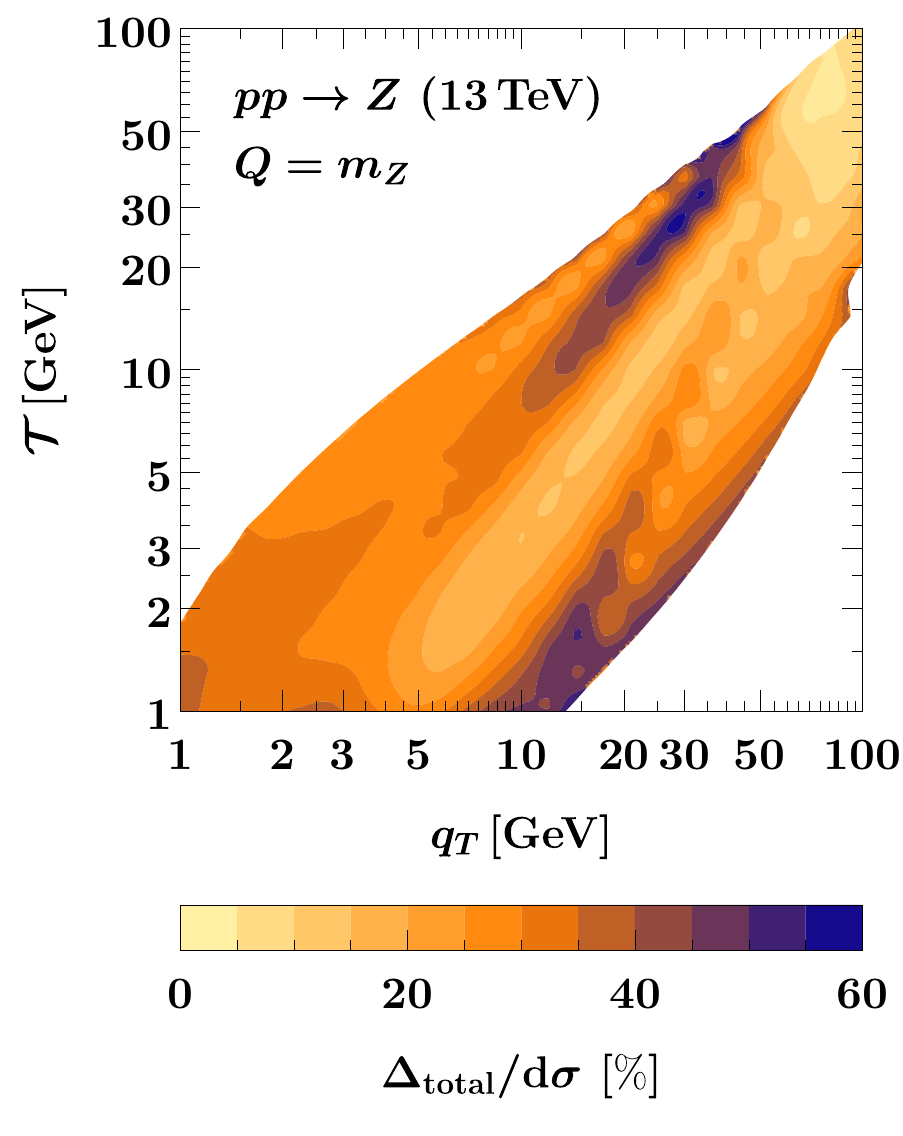}%
\end{minipage}%
\\
\begin{minipage}[t]{\WidthDiagonalSubfig}%
   \vspace{0pt}%
   \includegraphics[width=\textwidth]{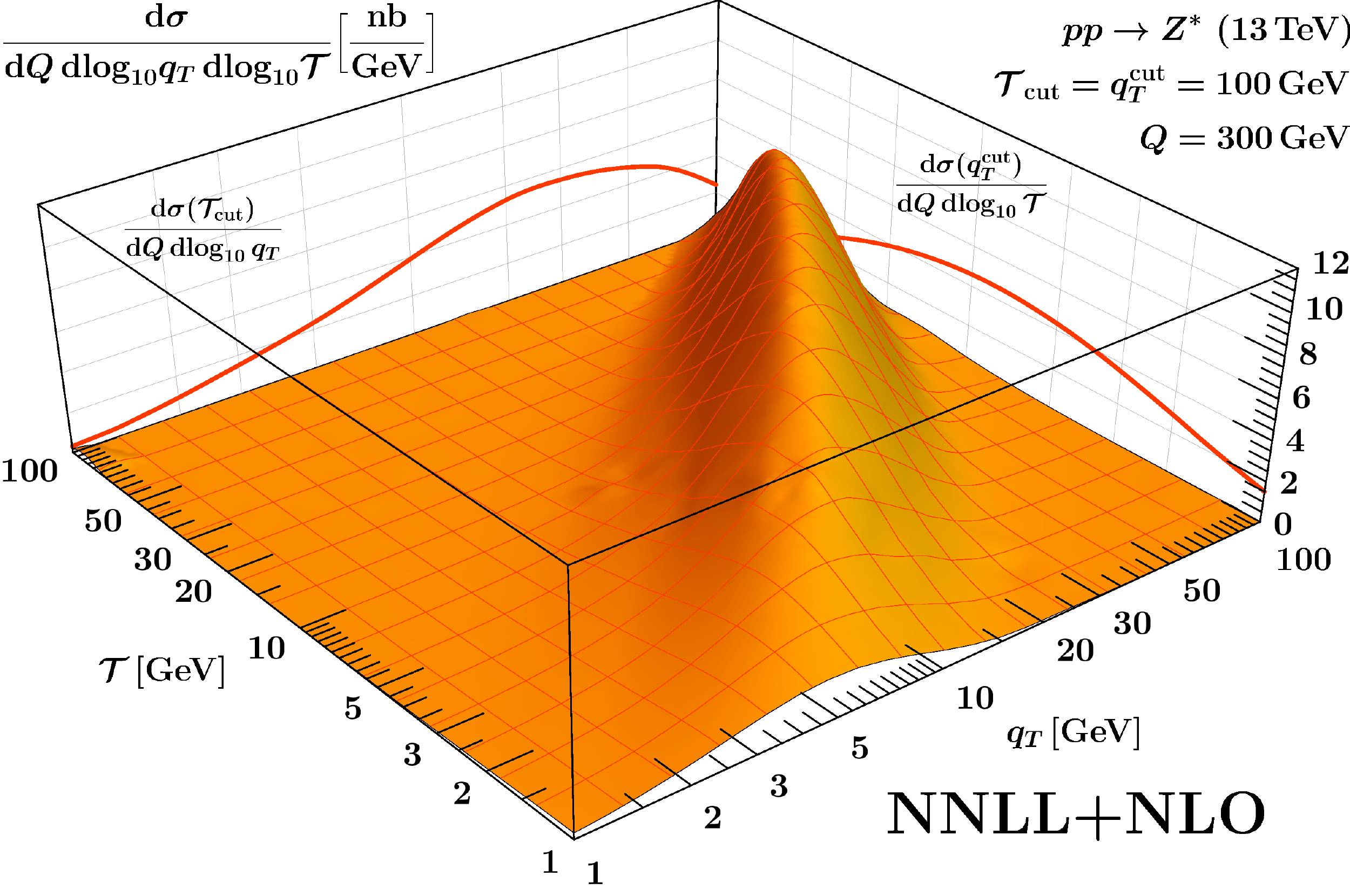}%
\end{minipage}%
\hfill%
\begin{minipage}[t]{\WidthThreeSubfigs}%
   \vspace{0pt}%
   \includegraphics[width=\textwidth]{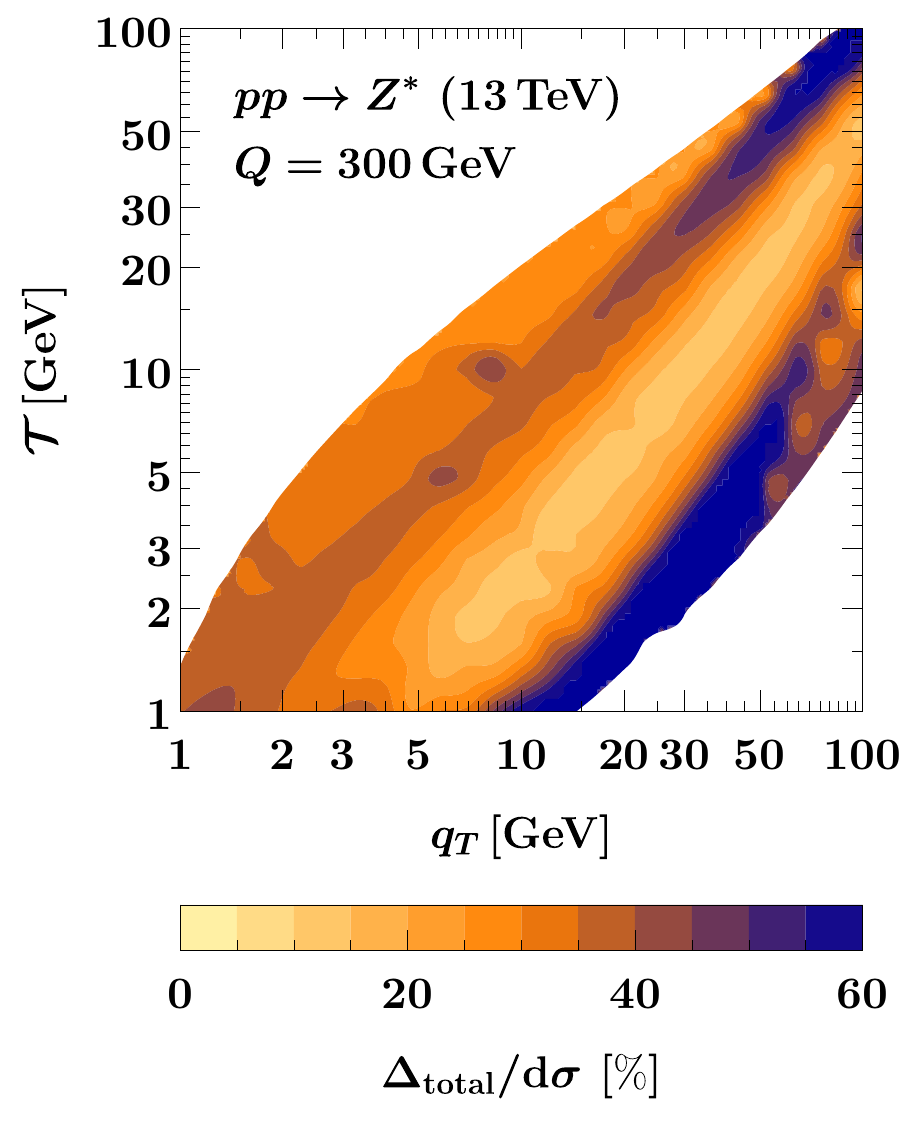}%
\end{minipage}%
\caption{The double-differential Drell-Yan cross section at NNLL$+$NLO, at $Q = m_Z$ (top) and $Q = 300 \GeV$ (bottom), with respect to $\log_{10} q_T$ and $\log_{10} \Tau$.
On the rear walls we show the result of integrating the double spectrum over either variable up to a cut at $100 \GeV$.
The contour plots indicate total perturbative uncertainties relative to the cross section, $\Delta_\total = \Delta_+ \oplus \Delta_\I \oplus \Delta_\II \oplus \Delta_\FO$.
The contour plots are left blank in the region where $\df \sigma / (\df Q \, \df \log_{10} q_T \, \df \log_{10} \Tau)$ is less than $3\%$ of its peak height.}
\label{fig:3d_plot}
\end{figure*}

In the bottom panel of \fig{3d_plot_fo_scet1_scet2_only},
we show the result of resumming logarithms of (the $b_T$ variable conjugate to) $q_T$ to NNLL and matching to fixed NLO, using the SCET$_\II$ matched result in \eq{matching_scet2}.
We denote this order by NNLL$_{q_T}$+NLO.
This result is valid for $\Tau \sim q_T \ll \sqrt{Q\Tau}$,
i.e., around the SCET$_\II$ phase-space boundary (green) in \fig{overview_regimes},
where we find the onset of a Sudakov peak from the $q_T$ resummation and a smooth kinematic suppression towards $\Tau \gg q_T$.
However, the NNLL$_{q_T}$+NLO result diverges for smaller values of $\Tau$.
This is due to unresummed logarithms of $\Tau$ in both the factorized cross section in SCET$_\II$
and terms of $\mathcal{O}(q_T^2/(Q\Tau))$ that are treated at fixed order as part of the matching correction.
In this case the single-differential projections show a Sudakov peak in $q_T$, but a logarithmic divergence at small $\Tau$.

\begin{figure*}
\centering
\includegraphics[width=\WidthThreeSubfigs]{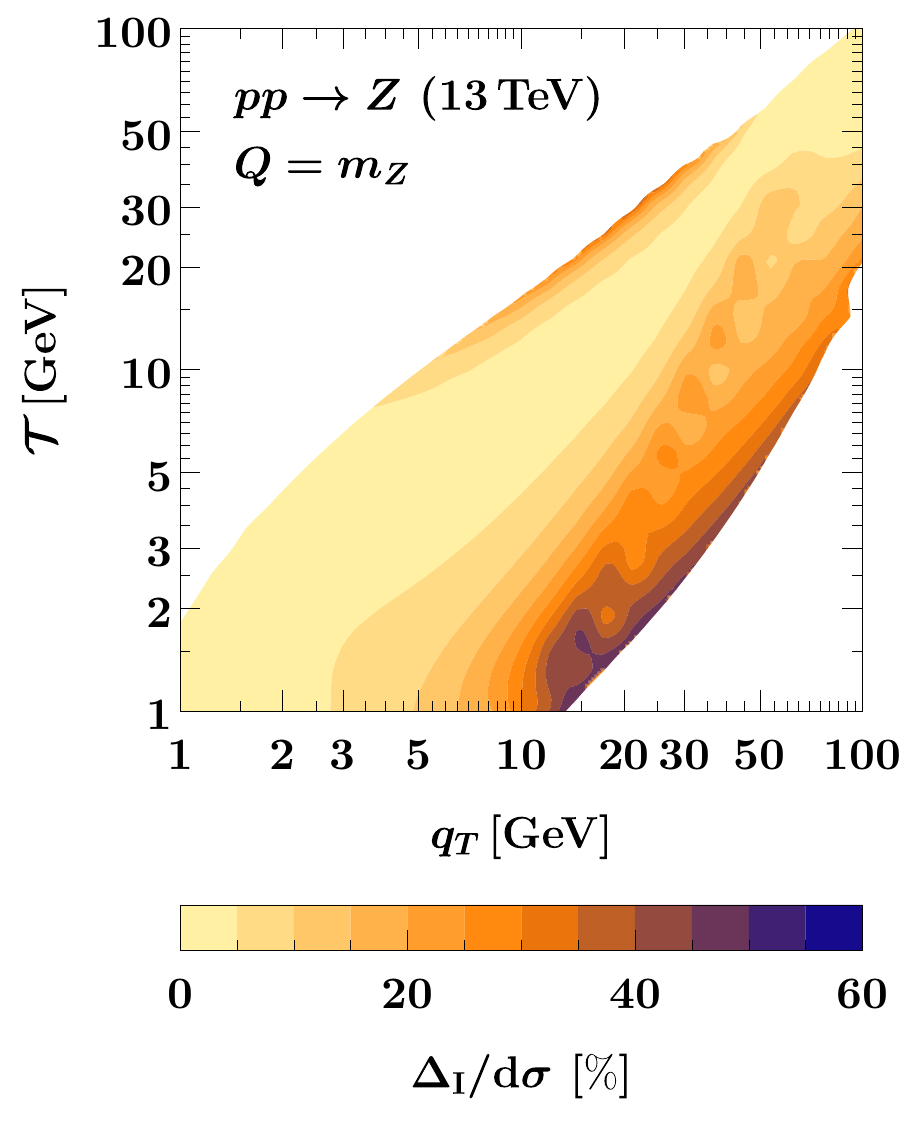}%
\hfill%
\includegraphics[width=\WidthThreeSubfigs]{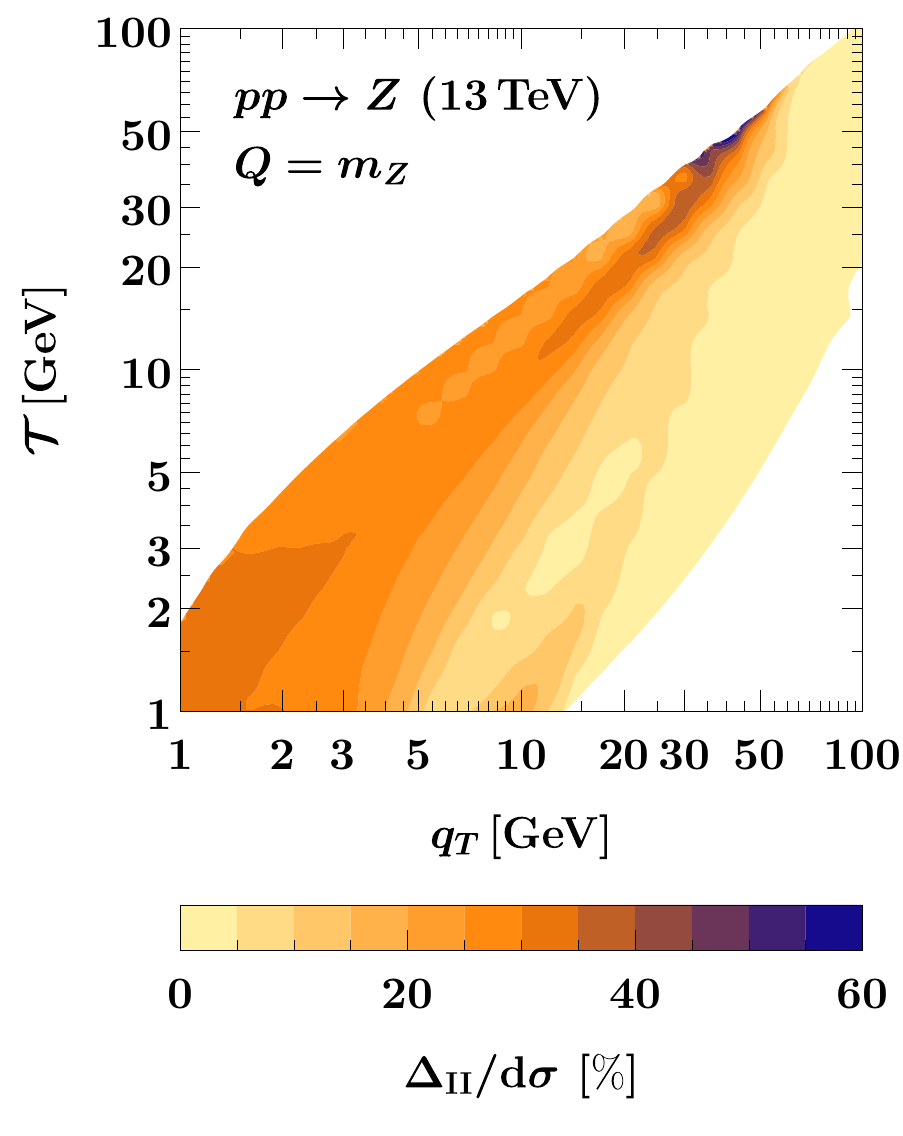}%
\hfill%
\includegraphics[width=\WidthThreeSubfigs]{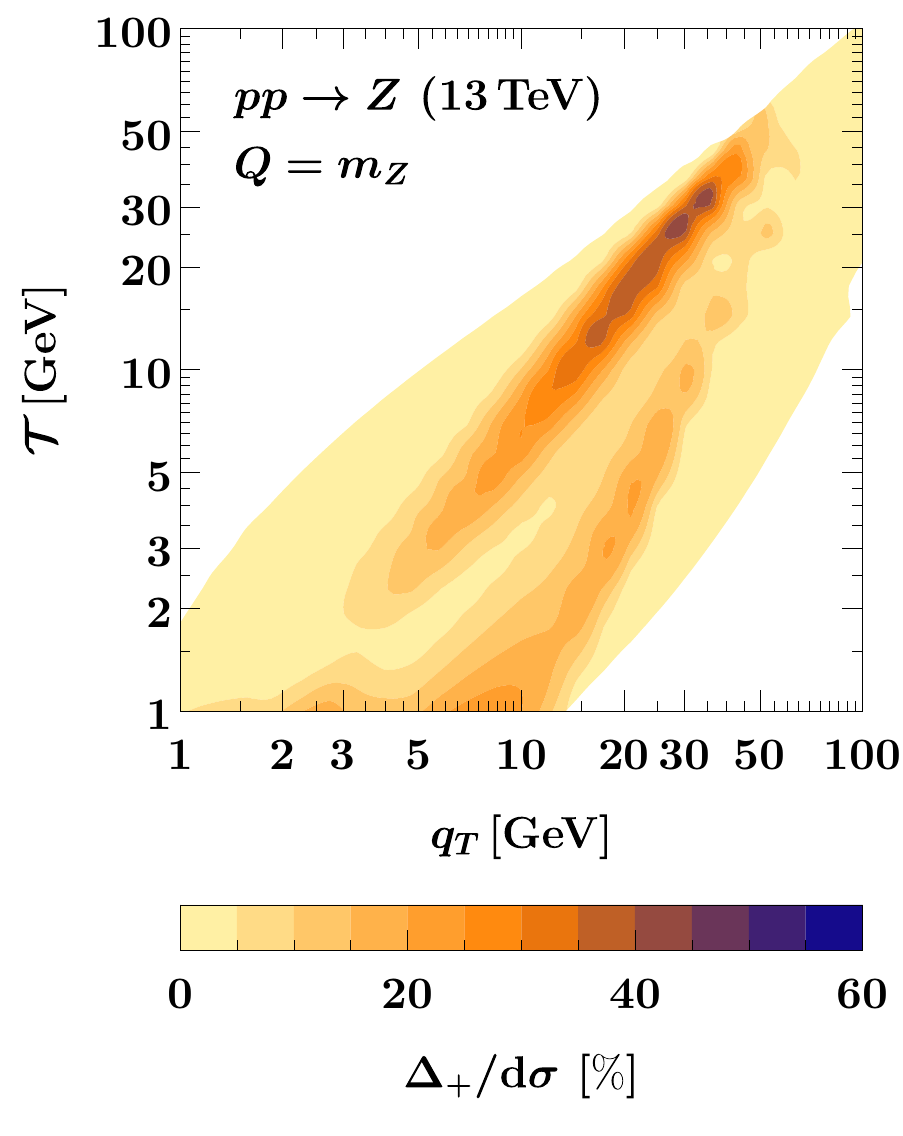}%
\caption{Breakdown of resummation uncertainties contributing to the relative uncertainty in the top right panel of \fig{3d_plot},
showing (from left to right) $\Delta_\I$, $\Delta_\II$, and $\Delta_+$.
As in \fig{3d_plot} we leave regions blank where the cross section is small.}
\label{fig:3d_plot_breakdown_uncertainties}
\end{figure*}

Our final results for the Drell-Yan double spectrum are shown in \fig{3d_plot},
as given by the fully matched prediction in \eq{matching}.
Here all resummed contributions are evaluated at NNLL, and we match to fixed NLO.
This achieves, for the first time, the complete resummation of all large logarithms in the double spectrum,
so we simply refer to this order as NNLL$+$NLO.
The top row of plots is for $Q = m_Z$, i.e., for Drell-Yan production at the $Z$ pole.
In the bottom row we consider $Q = 300 \GeV$ as a representative phase-space point at higher production energies.
Our matched and fully resummed double spectrum features a two-dimensional Sudakov peak
that is situated between the two parametric phase-space boundaries (cf.\ \fig{overview_regimes}),
is smoothly suppressed beyond,
and shifts towards higher values of $q_T$ and $\Tau$ for $Q = 300 \GeV$, as expected.
Integrating the double spectrum over either variable also results in a physical Sudakov peak,
as can be seen from the projections onto the rear walls.
Up to small differences in scale setting discussed in \sec{diff_cumul_scale_setting},
the left and right rear walls agree with the result of integrating
the NNLL$_{q_T}+$NLO and NNLL$_\Tau+$NLO results in \fig{3d_plot_fo_scet1_scet2_only} over $\Tau$ and $q_T$, respectively.
The contour plots in \fig{3d_plot} show the total perturbative uncertainties $\Delta_\total$ as percent deviations from the central result for the double spectrum.
As described in \sec{variations}, $\Delta_\total$ combines estimates of all sources of resummation uncertainty in the prediction.

In \fig{3d_plot_breakdown_uncertainties}, we break down the uncertainty for the Drell-Yan double-differential spectrum at $Q=m_Z$
into its contributions from SCET$_\I$, SCET$_\II$ and SCET$_+$ resummation uncertainties, respectively.
As expected, the SCET$_\I$ resummation uncertainty dominates in the SCET$_\I$ region of phase space, and similarly for SCET$_\II$.
The SCET$_+$ resummation uncertainty is largest along the phase-space boundaries,
indicating that it is mostly sensitive to variations of the transition points,
i.e., the points where the intrinsic SCET$_+$ resummation is turned off in our matched prediction.

\begin{figure*}
\centering
\includegraphics[width=\WidthTwoSubfigs]{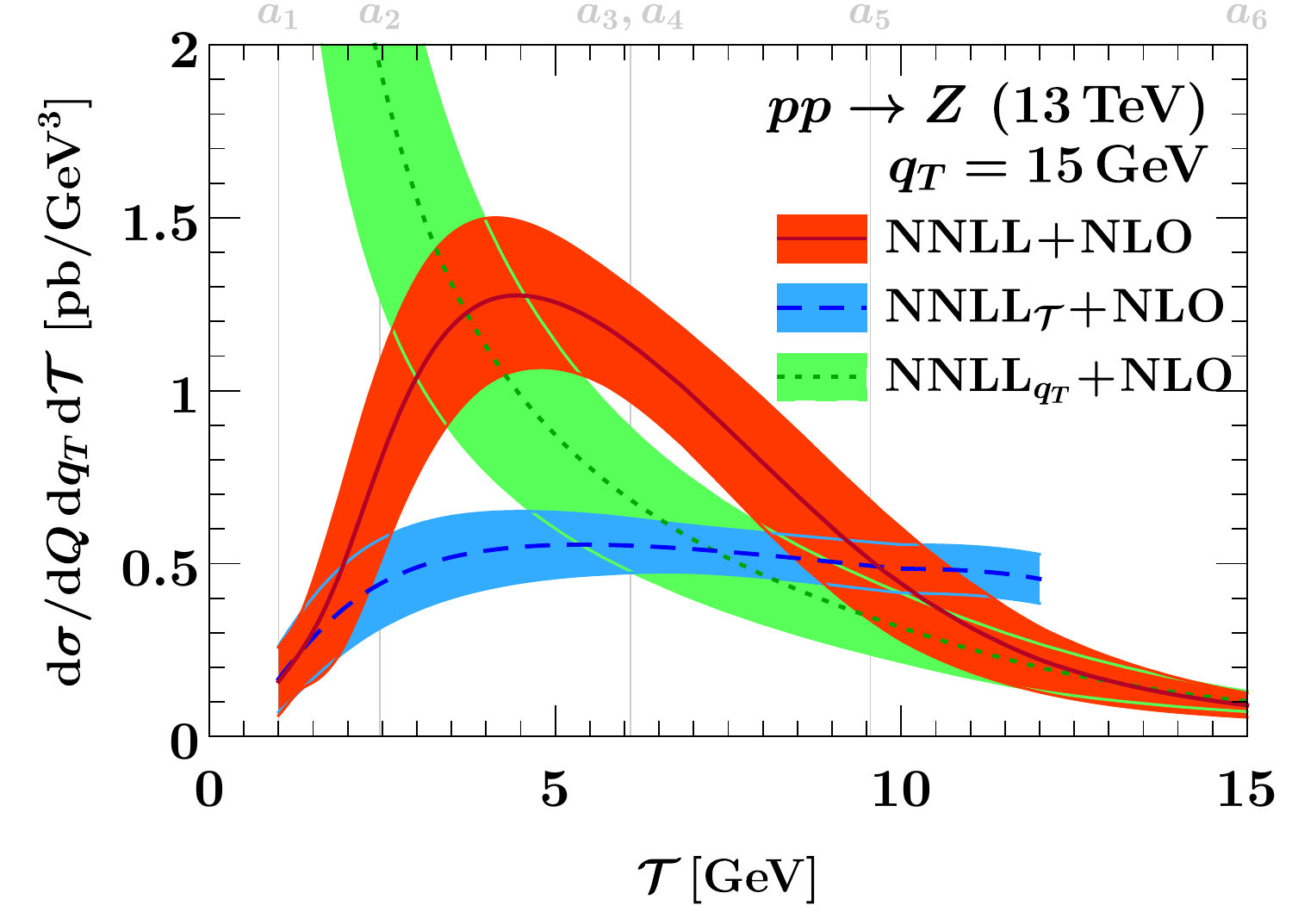}%
\hfill%
\includegraphics[width=\WidthTwoSubfigs]{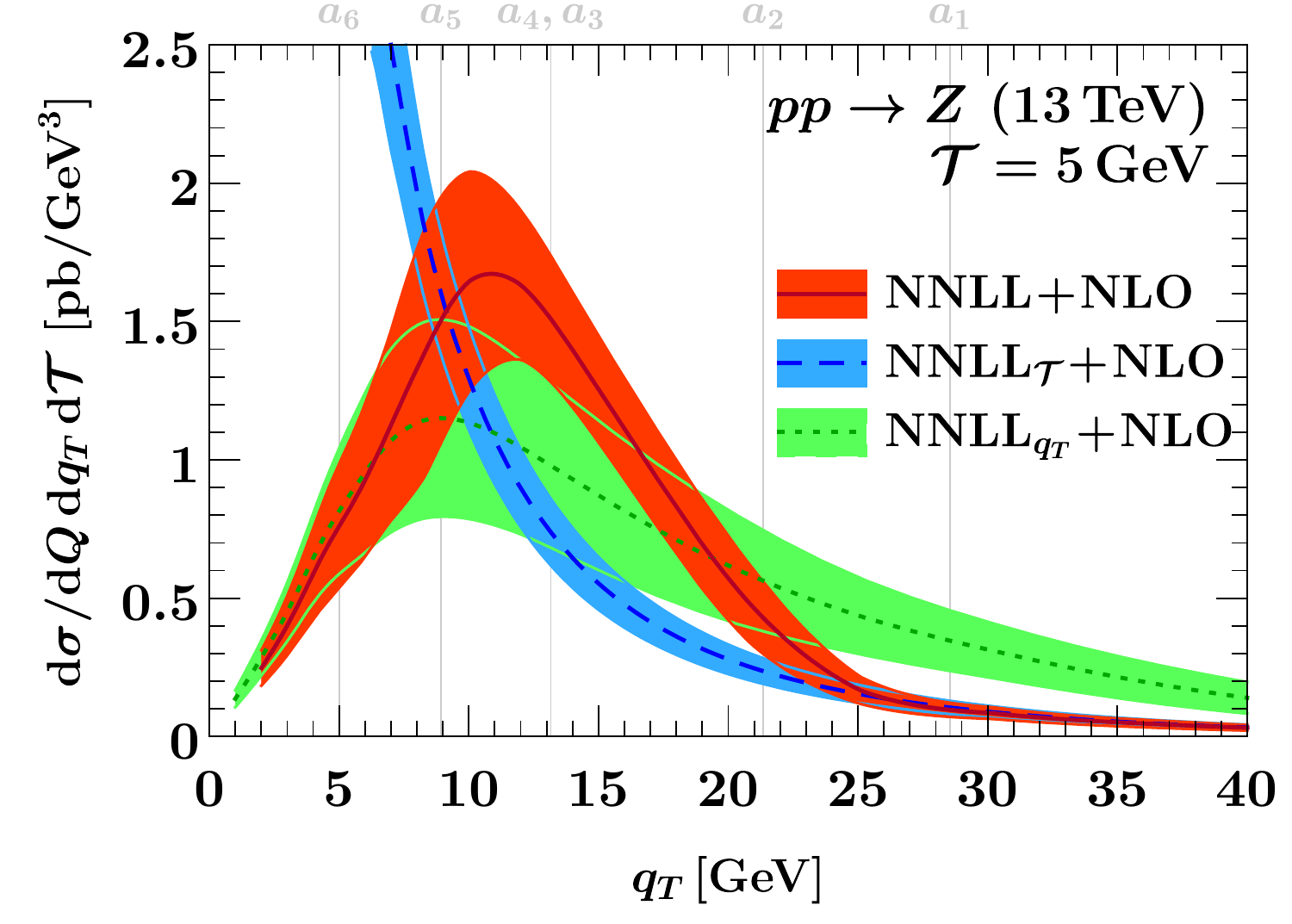}%
\\
\includegraphics[width=\WidthTwoSubfigs]{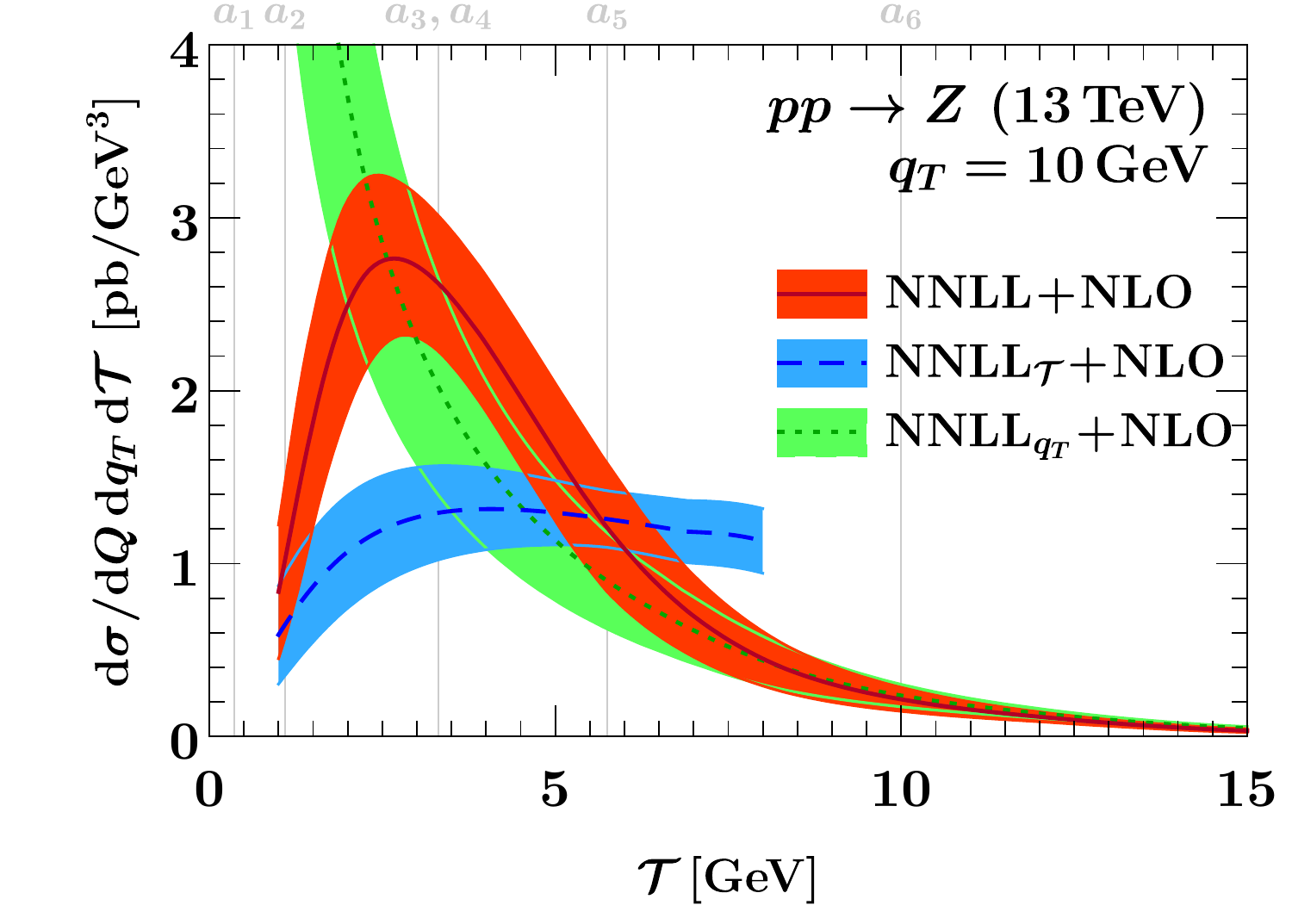}%
\hfill%
\includegraphics[width=\WidthTwoSubfigs]{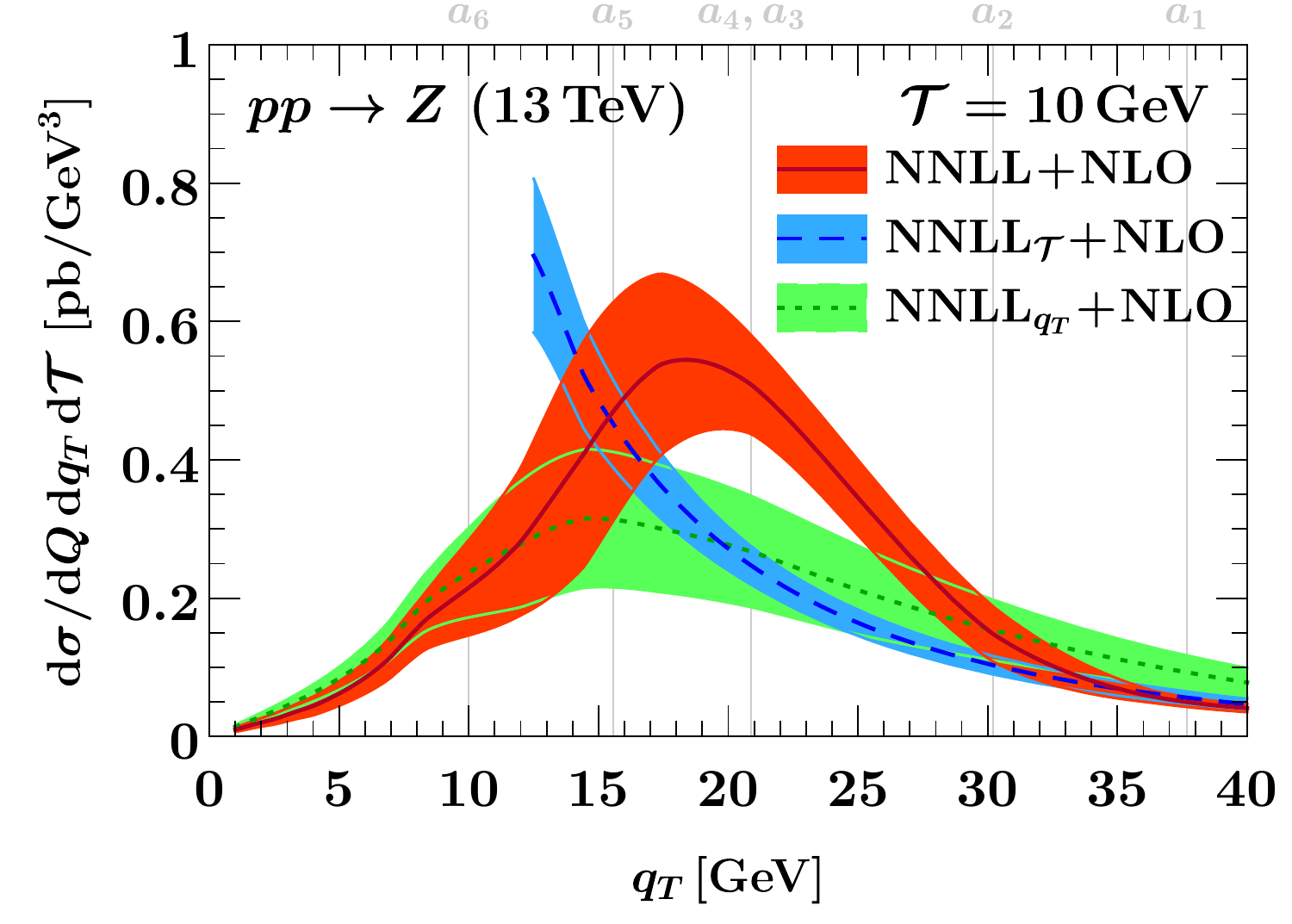}%
\caption{The double-differential Drell-Yan cross section for fixed $q_T$, as a function of $\Tau$ (left) and for fixed $\Tau$, as a function of $q_T$ (right).
The solid red curves are slices of the surface plots shown in the top left panel in \fig{3d_plot}, up to Jacobians.
The blue dashed (green dotted) curve corresponds to the middle (bottom) panel of \fig{3d_plot_fo_scet1_scet2_only}.
The thin vertical lines indicate the transition points $a_i$ described in \sec{profiles}.
The SCET$_\I$ prediction (dashed blue) has an unphysical edge at $\Tau = q_T$, see \fig{3d_plot_fo_scet1_scet2_only}, and is not shown beyond $\Tau = 0.8\,q_T$ to avoid distraction.
See the text for details on the uncertainty bands.}
\label{fig:slice_plots}
\end{figure*}

To further highlight the differences between our fully double-differential resummation
and the single-differential resummation at either NNLL$_{q_T}$ or NNLL$_\Tau$,
we take slices of the surface plots and overlay them in \fig{slice_plots},
keeping $q_T$ (left) or $\Tau$ (right) fixed.
The solid red curve corresponds to the matched and fully resummed cross section in \eq{matching},
with the uncertainty band given by the total perturbative uncertainty $\Delta_\total$, see \eq{Delta_total}.
The matched SCET$_\I$ (dashed blue) and SCET$_\II$ (dotted green) predictions
correspond to the middle and bottom panel of \fig{3d_plot_fo_scet1_scet2_only}, respectively.
Their uncertainty bands are given by $\Delta^\I_\total$ and $\Delta^\II_\total$,
which only probe a subset of higher-order terms as predicted by the respective RGE, see \eqs{Delta_I_total}{Delta_II_total}.
The SCET$_\I$ prediction features an unphysical sharp edge at $\Tau = q_T$, cf.\ the middle panel of \fig{3d_plot_fo_scet1_scet2_only},
and for this reason is cut off at $\Tau = 0.8\,q_T$.

All panels in \fig{slice_plots} show that our final prediction smoothly interpolates between the SCET$_\I$ and SCET$_\II$ boundary theories,
both for the central values and for the uncertainties.
Specifically, the matched prediction tends towards SCET$_\I$ (SCET$_\II$) for small (large) values of $\Tau$ and large (small) values of $q_T$.
In the left column one clearly sees that SCET$_\II$ only captures logarithms of $\Tau$ at fixed order,
leading to a diverging spectrum as $\Tau \to 0$, while the complete NNLL result features a physical Sudakov peak.
Conversely, the SCET$_\I$ result diverges as $q_T \to 0$, but is rendered physical by the additional $q_T$ resummation at NNLL.

\begin{figure*}
\centering
\includegraphics[width=\WidthTwoSubfigs]{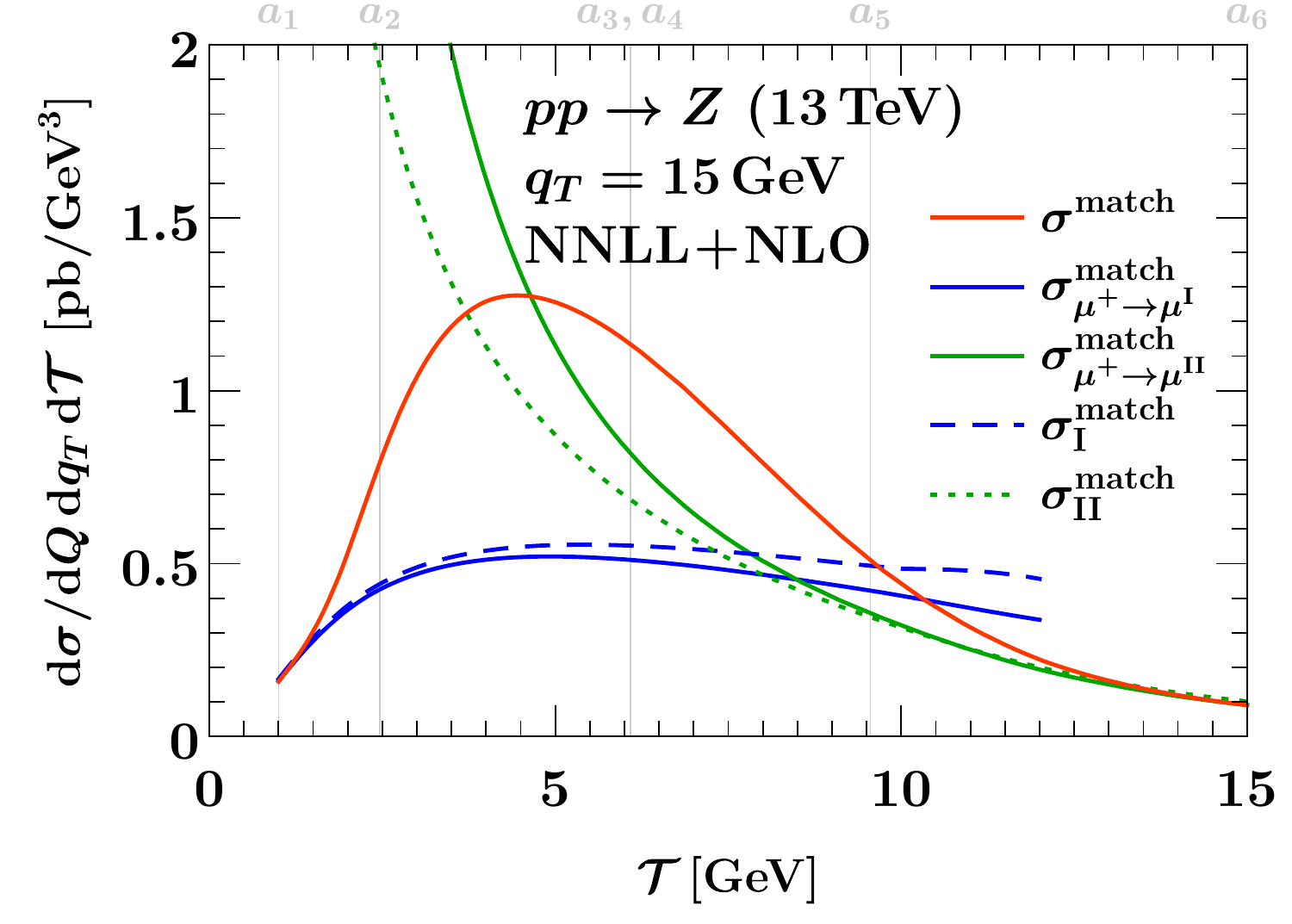}%
\hfill%
\includegraphics[width=\WidthTwoSubfigs]{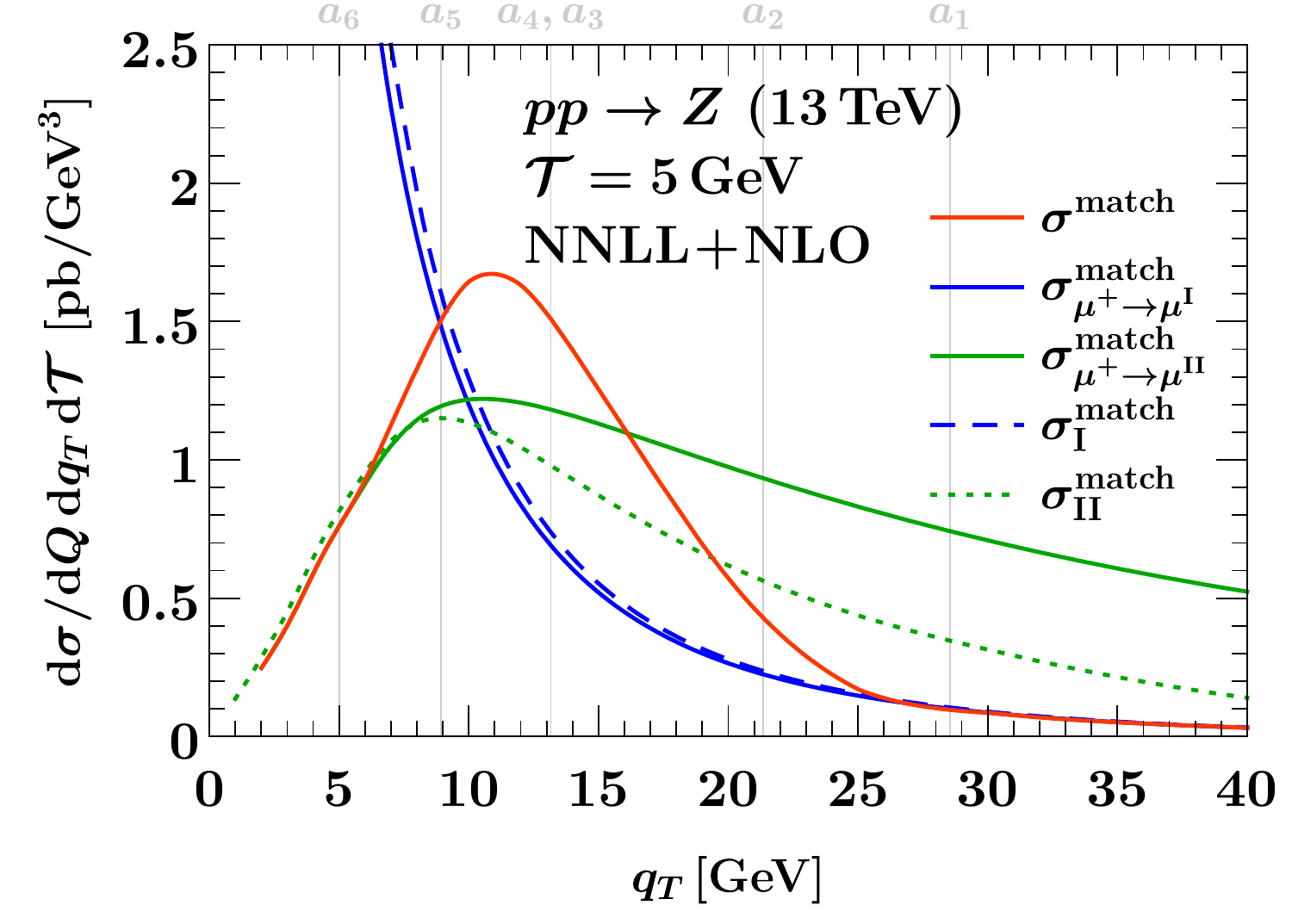}%
\caption{Slices of the double-differential Drell-Yan cross section at $q_T = 15 \GeV$ (left) and $\Tau = 5 \GeV$ (right).
The solid red, dashed blue, and dotted green curves are identical to the central results in \fig{slice_plots}.
The solid blue and green curves depict the SCET$_\I$ and SCET$_\II$ limits of our fully resummed result, given in \eqs{asymptote_scet_1}{asymptote_scet_2}.
The thin vertical lines indicate the transition points $a_i$ described in \sec{profiles}.}
\label{fig:asymp_plots}
\end{figure*}

We would like to stress that our fully resummed prediction does not exactly agree with either boundary theory,
even beyond the final transition points $a_1$ and $a_6$ where the intrinsic SCET$_+$ resummation is turned off.
The reason for this is that even in these limits, the matched cross section in \eq{matching}
improves over the matched SCET$_\I$ and SCET$_\II$ cross sections in \eq{matching_scet1} and \eq{matching_scet2}
by an additional resummation of power-suppressed terms, cf.\ \eqs{asymptote_scet_1}{asymptote_scet_2}.
To assess the size of this effect,
we again compare both single-differential resummations (dashed blue and dotted green) to our matched prediction (solid red) in \fig{asymp_plots},
but for reference include the case where $\sigma_+$ in the matched prediction is evaluated at $\mu^\I$ (solid blue) or $\mu^\II$ (solid green) directly.
One can easily verify from e.g.\ the right panel
that for $q_T$ above the right-most vertical line (where $a<a_1$),
the difference between the solid blue and the dashed blue curves indeed amounts to a small power-suppressed set of higher-order terms,
while our best prediction (solid red) recovers the solid blue curve as it must.
Similarly, for $q_T$ below the left-most vertical line (where $a > a_6$),
the difference between the solid green (and solid red) and dashed green curves
can be seen to be a small correction,
reflecting the size of power-suppressed higher-order terms predicted by the SCET$_\I$ RGE in this region.
The asymptotic limits are interchanged in the left panel, where $a < a_1$ towards the left and $a > a_6$ towards the very right of the plot.

%===============================================================================
\subsection{Single-differential spectra with a cut on the other variable}
\label{sec:results_spectrum_with_cut}
%===============================================================================

\begin{figure*}
\centering
\includegraphics[width=\WidthTwoSubfigs]{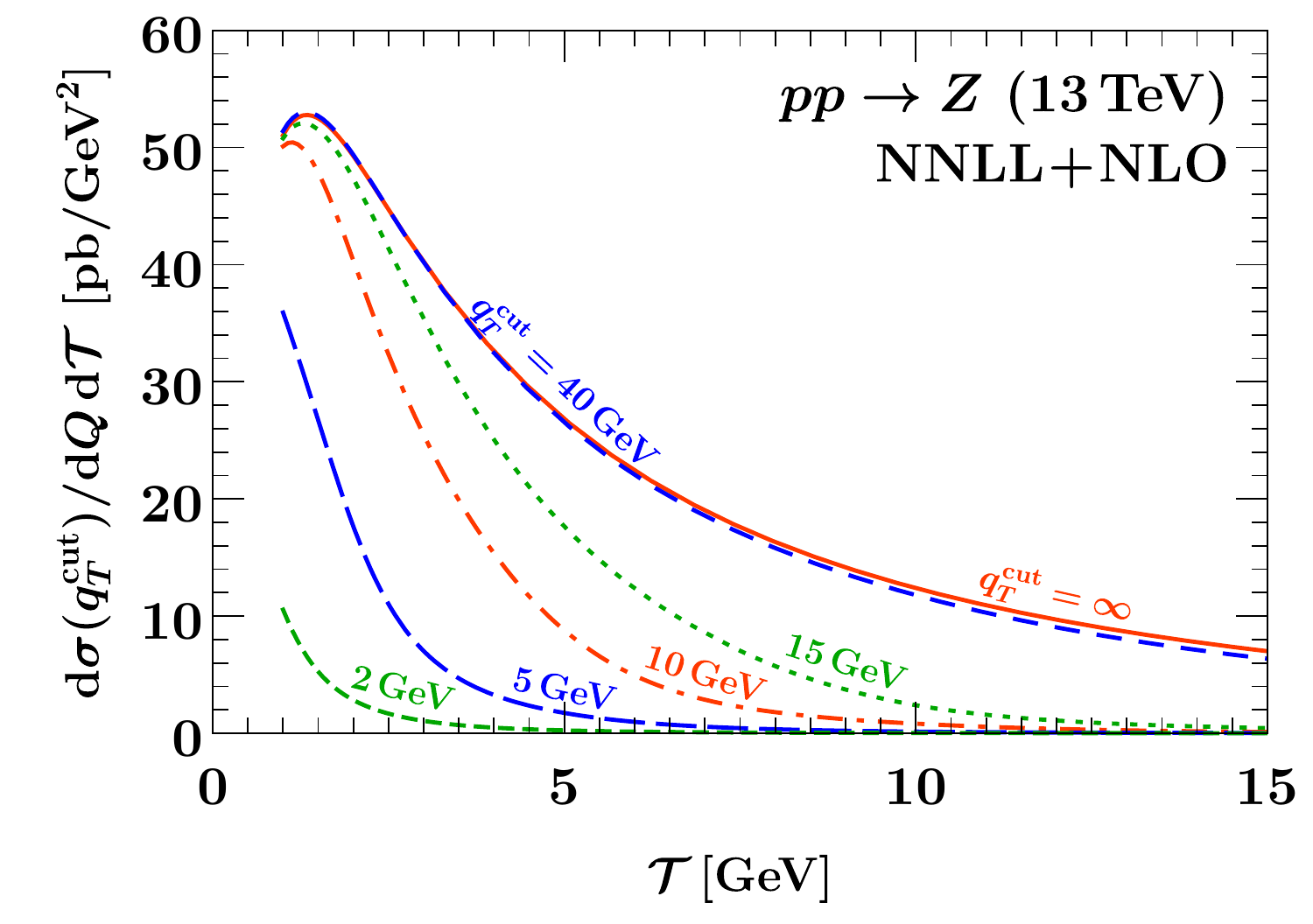}%
\hfill%
\includegraphics[width=\WidthTwoSubfigs]{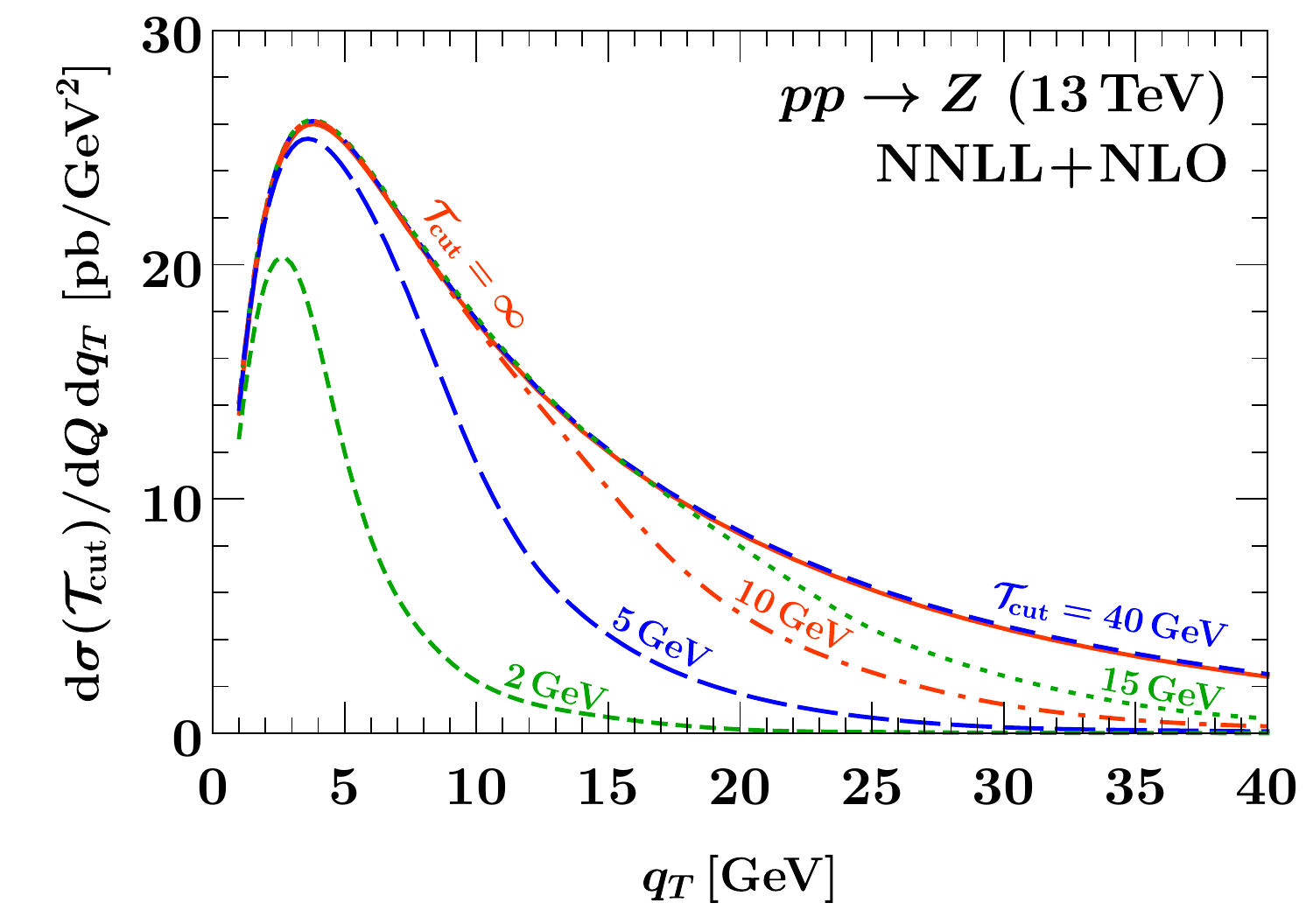}
\caption{The single-differential $\Tau$ (left) and $q_T$ (right) spectrum with a cut on the other variable at NNLL$+$NLO.
The different curves represent different values of the cut.
The solid orange lines correspond to the inclusive single-differential spectrum obtained by lifting the cut.}
\label{fig:recoverSingleDiff_plots}
\end{figure*}

So far we have turned our attention to the cross section differential in both $q_T$ and $\Tau$.
In addition to this double spectrum, our setup also predicts the fully matched and resummed double cumulant cross section,
and the single-differential $q_T$ (or $\Tau$) spectrum with a cut on the other variable;
selected results for these observables where already discussed in \sec{diff_cumul_scale_setting} from a more technical point of view.
In \fig{recoverSingleDiff_plots}, we show some more detailed results for the single-differential spectra with an additional cut,
where the left panel shows $\df \sigma(\qTcut)/\df\Tau$ as a function of $\Tau$ for various values of $\qTcut$,
and the right panel shows $\df \sigma(\TauCut)/\df q_T$ as a function of $q_T$ for various values of $\TauCut$.
By increasing the value of the cut, they can be seen to approach the inclusive single-differential spectra (orange solid),
with which they must agree when sending $\qTcut \to \infty$ or $\TauCut \to \infty$, respectively.
This is by construction because we employ cumulant scale setting as appropriate for this prediction, cf.\ \sec{diff_cumul_scale_setting}.
We observe that cuts on the other variable shape either spectrum in a very nontrivial way.
Tight cuts $\lesssim 10\GeV$ push the peak to lower values and suppress the tail,
where the $q_T$ spectrum is somewhat more resilient to cuts on $\Tau$ than vice versa.
Intermediate cuts $\sim 10-15 \GeV$ keep the peak and mostly lead to a suppression in the tail,
while the effect of cuts $\gtrsim 40 \GeV$ is almost negligible in the $q_T$ and $\Tau$ ranges of interest.

%%%%%%%%%%%%%%%%%%%%%%%%%%%%%%%%%%%%%%%%%%%%%%%%%%%%%%%%%%%%%%%%%%%%%%%%%%%%%%%%
\section{Conclusions}
\label{sec:conclusions}
%%%%%%%%%%%%%%%%%%%%%%%%%%%%%%%%%%%%%%%%%%%%%%%%%%%%%%%%%%%%%%%%%%%%%%%%%%%%%%%%

In this paper we calculated the Drell-Yan cross section double-differential in the transverse momentum $q_T$ of the lepton pair and the 0-jettiness $\Tau$.
Both $\Tau$ and $q_T$ probe the initial state radiation, leading to Sudakov
double logarithms of $\Tau/Q$ and $q_T/Q$ in the cross section.
We performed, for the first time, the simultaneous resummation of both kinds of logarithms,
achieving next-to-next-to-leading logarithmic accuracy and matching the result to next-to-leading fixed order.
We accomplish this resummation by using SCET$_\I$ and SCET$_\II$ to describe
the regions $\Tau \ll q_T \sim \sqrt{\Tau Q}$ and $\Tau \sim q_T \ll  \sqrt{\Tau Q}$, respectively,
and SCET$_+$ to describe the bulk of phase space in between these boundaries~\cite{Procura:2014cba}.

Obtaining reliable numerical predictions required several nontrivial steps:
(1)~Matching several factorized cross sections for the different regions of phase space,
for which we use a Venn-diagram method to avoid double counting.
(2)~Choosing appropriate profile scales for the various ingredients in the factorization formulas
that respect all relevant canonical scaling relations and at the same time
smoothly interpolate between the different regions of phase space,
and varying these scales to estimate perturbative uncertainties.
This is significantly more involved than in the usual single-differential case,
and is further complicated by the requirement to choose scales in impact parameter ($b_T$) space for SCET$_\II$.
For example, the rapidity scale for the collinear-soft function in SCET$_+$ has a canonical scaling
that does not coincide with any scale on the SCET$_\I$ and SCET$_\II$ boundaries.
(3)~Ensuring that scales and scale variations are still, to the extent possible, inherited from the single-differential resummation of $\Tau$ and $q_T$.
This makes our setup flexible enough to incorporate other procedures for estimating the uncertainty in the individual resummations.
(4)~To handle the transition between SCET$_\I$, SCET$_+$ and SCET$_\II$,
we introduced profile scales in terms of a regime parameter $a$,
designed such that $a=1$ for SCET$_\I$ and $a=2$ for SCET$_\II$.
The precise transition points in $a$ were chosen by comparing the various singular and nonsingular cross section,
and are varied as part of the uncertainty estimate.
(5)~We also introduced a new hybrid (i.e., $q_T$ and $b_T$ dependent) scale choice for $q_T$ resummation,
allowing the resummation to strictly take place in $b_T$ space,
while turning the resummation on and off using $q_T$.

We demonstrated that our simultaneous resummation of $\Tau$ and $q_T$ yields
the correct resummed single-differential cross sections after integrating over either $\Tau$ or $q_T$.
This requires choosing scales at the level of the differential or integrated (cumulative) cross section as appropriate,
which we discussed in detail.

While the predictions obtained here are of some direct phenomenological interest, as $\Tau$ has been measured in bins of $q_T$~\cite{Aad:2016ria}, our analysis is also an important step towards precise \emph{and} differential predictions for LHC cross sections in general. Specifically, the Monte Carlo event generator \textsc{Geneva}~\cite{Alioli:2012fc, Alioli:2015toa} is based on a NNLL$'$ resummed prediction for the cross section differential in $\Tau$, and would benefit from the simultaneous resummation of $q_T$. Indeed, our NNLL results clearly indicate that only resumming the logarithms of either $\Tau$ or $q_T$ gives a poor description of the double-differential cross section. Our methods apply at any order and for any color-singlet production process, allowing for a straightforward extension once the relevant perturbative ingredients become available. We hope that our analysis can pave the way for going
beyond single-differential resummations in many other contexts as well.

%%%%%%%%%%%%%%%%%%%%%%%%%%%%%%%%%%%%%%%%%%%%%%%%%%%%%%%%%%%%%%%%%%%%%%%%%%%%%%%%
\begin{acknowledgments}
We thank Goutam Das, Markus Diehl, and Markus Ebert for discussions.
We would also like to thank Markus Ebert for his contributions to \texttt{SCETlib}.
This work is supported by the ERC grant ERC-STG-2015-677323 and the D-ITP consortium, a program of the Netherlands Organization for Scientific Research (NWO) that is funded by the Dutch Ministry of Education, Culture and Science (OCW). FT and JM thank Nikhef for hospitality.
GL thanks DESY for hospitality and the DESY theory group thanks GL for an ample
supply of Dutch pepernoten.
\end{acknowledgments}
%%%%%%%%%%%%%%%%%%%%%%%%%%%%%%%%%%%%%%%%%%%%%%%%%%%%%%%%%%%%%%%%%%%%%%%%%%%%%%%%

%%%%%%%%%%%%%%%%%%%%%%%%%%%%%%%%%%%%%%%%%%%%%%%%%%%%%%%%%%%%%%%%%%%%%%%%%%%%%%%%
\appendix
\section{Plus distributions and Fourier transform}
\label{app:plus_distributions_and_fourier_transform}
%%%%%%%%%%%%%%%%%%%%%%%%%%%%%%%%%%%%%%%%%%%%%%%%%%%%%%%%%%%%%%%%%%%%%%%%%%%%%%%%

We use the following standard plus distributions with dimensionless arguments,
%%%
\begin{alignat}{3}
\cL_n(x)
&\equiv \biggl[ \frac{\theta(x) \ln^n x}{x}\biggr]_+
&&= \lim_{\beta \to 0} \biggl[
  \frac{\theta(x- \beta)\ln^n x}{x} +
  \delta(x- \beta) \, \frac{\ln^{n+1}\!\beta}{n+1} \biggr]\,,\\
\cL^a(x)
&\equiv \biggl[ \frac{\theta(x)}{x^{1-a}}\biggr]_+
&&= \lim_{\beta \to 0} \biggl[
  \frac{\theta(x-\beta)}{x^{1-a}} +
  \delta(x-\beta)\frac{x^a -1}{a} \biggr]
\,.\end{alignat}
%%%
In intermediate steps we need a two-dimensional plus distribution originally defined in \refcite{Procura:2014cba},
%%%
\begin{align}
\cL_\Delta(x_1,x_2) \equiv
\lim_{\beta \to 0} \frac{\df}{\df x_1}\, \frac{\df}{\df x_2}\,
\Big[&\theta(x_2 - x_1^2) \theta(x_1-\beta) \ln x_1\, (\ln x_2 - \ln x_1)
\nn \\ &
+ \frac14\,\theta(x_1^2 - x_2) \theta(x_2-\beta^2) \ln^2 x_2\Big]
\,.\end{align}
%%%
Our shorthands for distributions with dimensionful arguments
in one spatial dimension are
%%%
\begin{align}
\cL_n(k,\mu) \equiv \frac{1}{\mu} \cL_n\Bigl(\frac{k}{\mu}\Bigr)
\,, \qquad
\cL^a(k,\mu) \equiv \frac{1}{\mu} \cL^a\Bigl(\frac{k}{\mu}\Bigr)
\,.\end{align}
%%%
In terms of the second class of (power-like) plus distributions, we further define
%%%
\begin{align} \label{eq:def_flow_distribution}
\cV_a(x)
\equiv \frac{e^{-\gamma_E a}}{\Gamma(1+a)} \bigl[ a \cL^a(x) + \delta(x) \bigr]
\,, \qquad
\cV_a(k, \mu)
\equiv \frac{1}{\mu} \cV_a \Bigl( \frac{k}{\mu} \Bigr)
\,,\end{align}
%%%
which have a group property, assuming identical boundary condition $\mu$,
%%%
\begin{equation}
\int \! \df \ell \,
\mathcal{V}_a(\ell,\mu) \, \mathcal{V}_b(k - \ell,\mu) = \mathcal{V}_{a+b}(k,\mu)
\,, \qquad
\mathcal{V}_0(k, \mu) = \delta(k)
\,.\end{equation}
%%%
Shifting the boundary condition of $\cV_a(k, \mu)$ from $\mu$ to $\mu'$ is also straightforward,
%%%
\begin{equation}
\cV_a(k, \mu) = \Bigl( \frac{\mu'}{\mu} \Bigr)^a \, \cV_a(k, \mu')
\,.\end{equation}
%%%
We use the conventions from app.~C of \refcite{Ebert:2016gcn}
for logarithmic plus distributions in two integer spatial dimensions,
with $k_T^2 \equiv \vec{k}_T^2 \geq 0$,
%%%
\begin{align}
\delta(\vec{k}_T)
= \frac{1}{\pi} \delta(k_T^2)
\,, \qquad
\cL_n(\vec{k}_T,\mu)
&\equiv
\frac{1}{\pi\mu^2}
\biggl[\frac{\mu^2}{k_T^2}\ln^n\biggl(\frac{k_T^2}{\mu^2}\biggr)\biggr]_+^\mu
\equiv \frac{1}{\pi\mu^2} \cL_n\biggl(\frac{k_T^2}{\mu^2}\biggr)
\,.\end{align}
%%%

Our convention for the two-dimensional Fourier transform also follows \refcite{Ebert:2016gcn},
%%%
\begin{align} \label{eq:fourier_transform}
\frac{\df f}{\df \vec{p}_T}
= \int\! \frac{\df^2\vec{b}_T}{(2\pi)^2}\, e^{+i \vec{p}_T \cdot \vec{b}_T} \tilde f(\vec{b}_T)
\,, \qquad
\tilde f(\vec{b}_T)
= \int\! \df^2\vec{p}_T\, e^{-i \vec{p}_T \cdot \vec{b}_T} \frac{\df f}{\df \vec{p}_T}
\,.\end{align}
%%%
Here we make the mass dimension of $\df f/\df \vec{p}_T = \df f/(\df p_x \, \df p_y)$ explicit.
If $f$ is azimuthally symmetric, i.e., if for $p_T \equiv \abs{\vec{p}_T}, b_T \equiv \abs{\vec{b}_T}$,
%%%
\begin{equation}
\frac{\df f}{\df \vec{p}_T} = \frac{1}{2\pi p_T} \frac{\df f}{\df p_T}
\,, \qquad
\tf(\vec{b}_T) = \tf(b_T)
\,,\end{equation}
the azimuthal integral can be performed, leaving
%%%
\begin{align} \label{eq:fourier_transform_radial}
\frac{\df f}{\df p_T}
= p_T \int_0^\infty\! \df b_T\, b_T \, J_0(b_T p_T)\, \tilde f(b_T)
\,, \qquad
\tilde f(b_T)
= \int_0^\infty\! \df p_T \, J_0(b_T p_T) \frac{\df f}{\df p_T}
\,,\end{align}
%%%
where $J_0(x)$ is the zeroth-order Bessel function of the first kind.
Integrating the first expression in \eq{fourier_transform_radial} by parts,
the cumulant in $p_T$ is given by
%%%
\begin{align}
\int_0^{p_T^{\rm cut}}\df p_T\, \frac{\df f}{\df p_T}
= p_T^{\rm cut} \int_0^\infty\! \df b_T\, J_1(b_T p_T^{\rm cut})\, \tilde f(b_T)
\,,\end{align}
where $J_1(x)$ is the first-order Bessel function of the first kind. Fourier transforms of $\cL_n(\vec{p}_T,\mu)$ can be found in table~5 of \refcite{Ebert:2016gcn}, and are most conveniently expressed in terms of
%%%
\begin{equation} \label{eq:def_Lb}
L_b \equiv \ln \frac{b_T^2 \mu^2}{b_0^2}
\,, \qquad
b_0 \equiv 2 e^{-\gamma_E} \approx 1.12291\dots
\,.\end{equation}
%%%

%%%%%%%%%%%%%%%%%%%%%%%%%%%%%%%%%%%%%%%%%%%%%%%%%%%%%%%%%%%%%%%%%%%%%%%%%%%%%%%%
\section{Perturbative ingredients}
\label{app:perturbative_ingredients}
%%%%%%%%%%%%%%%%%%%%%%%%%%%%%%%%%%%%%%%%%%%%%%%%%%%%%%%%%%%%%%%%%%%%%%%%%%%%%%%%

%===============================================================================
\subsection{Anomalous dimensions}
\label{app:anom_dims}
%===============================================================================

We expand the $\beta$ function of QCD as
%%%
\begin{equation} \label{eq:expansion_beta}
\mu \frac{\df \as(\mu)}{\df \mu}
= \beta[\as(\mu)]
\,, \qquad
\beta(\as) = -2\as \sum_{n = 0}^\infty \beta_n \left( \frac{\as}{4\pi} \right)^{n+1}
\,.\end{equation}
%%%
The coefficients in the \MSbar scheme are, up to three loops~\cite{Tarasov:1980au, Larin:1993tp},
%%%
\begin{align} \label{eq:coeffs_beta}
\beta_0 &= \frac{11}{3}\,C_A -\frac{4}{3}\,T_F\,n_f
\,, \qquad
\beta_1 = \frac{34}{3}\,C_A^2  - \Bigl(\frac{20}{3}\,C_A\, + 4 C_F\Bigr)\, T_F\,n_f
\,, \\
\beta_2 &=
\frac{2857}{54}\,C_A^3 + \Bigl(C_F^2 - \frac{205}{18}\,C_F C_A
- \frac{1415}{54}\,C_A^2 \Bigr)\, 2T_F\,n_f
+ \Bigl(\frac{11}{9}\, C_F + \frac{79}{54}\, C_A \Bigr)\, 4T_F^2\,n_f^2
\nn\,.\end{align}
%%%
We work with $n_f = 5$ light flavors.
The cusp and noncusp anomalous dimensions are expanded as
%%%
\begin{equation} \label{eq:expansion_anom_dims}
\Gamma^i_\cusp(\as) = \sum_{n = 0}^\infty \Gamma^i_{n} \Bigl( \frac{\as}{4\pi} \Bigr)^{n+1}
\,,\qquad
\gamma(\as) = \sum_{n = 0}^\infty \gamma_n \Bigl( \frac{\as}{4\pi} \Bigr)^{n+1}
\,.\end{equation}
%%%
The coefficients of the $\overline{\mathrm{MS}}$ cusp anomalous dimension to three loops are~\cite{Korchemsky:1987wg, Moch:2004pa, Vogt:2004mw}
%%%
\begin{align} \label{eq:coeffs_cusp_anom_dim}
\Gamma_0^i &= 4C_i
\,,\nn\\
\Gamma_1^i
&= 4 C_i \Bigl[ C_A \Bigl( \frac{67}{9} - \frac{\pi^2}{3} \Bigr)  - \frac{20}{9}\,T_F\, n_f \Bigr]
= \frac{4}{3} C_i \bigl[ (4 - \pi^2) C_A + 5 \beta_0 \bigr]
\,,\nn\\
\Gamma_2^i
&= 4 C_i \Bigl[
   C_A^2 \Bigl(\frac{245}{6} -\frac{134 \pi^2}{27} + \frac{11 \pi ^4}{45} + \frac{22 \zeta_3}{3}\Bigr)
   +  C_A\, T_F\,n_f \Bigl(- \frac{418}{27} + \frac{40 \pi^2}{27}  - \frac{56 \zeta_3}{3} \Bigr)
   \nn \\ & \qquad
   +  C_F\, T_F\,n_f \Bigl(- \frac{55}{3} + 16 \zeta_3 \Bigr)
   - \frac{16}{27}\,T_F^2\, n_f^2
   \Bigr]
\,,\end{align}
%%%
where here and in the following $C_i = C_F \, (C_A)$ for $i = q\,(g)$.
The fixed-order boundary condition
of the resummed rapidity anomalous dimension \eq{resummed_rapidity_anom_dim} reads, through NNLL,
%%%
\begin{align} \label{eq:rapidity_anom_dim_fo_bundary_condition}
\tgamma^i_{\nu,\FO}(b_T, \mu)
&= \frac{\as(\mu)}{4\pi} \Bigl[ -2 \Gamma^i_0 L_b \Bigr]
+ \frac{\as^2(\mu)}{(4\pi)^2} \Bigl[
      -\Gamma^i_0 \beta_0 L_b^2 - 2 \Gamma^i_1 L_b  + \gamma^i_{\nu\,1}
   \Bigr]
   + \ord{\as^3}
\,,\end{align}
%%%
where we have already used that $\gamma^i_{\nu\,0} = 0$.
For our choice of regulator, the two-loop boundary condition
is given by~\cite{Luebbert:2016itl}
%%%
\begin{align} \label{eq:coeffs_rapidity_anom_dim}
\gamma^i_{\nu\,1}
= C_i \left[ -C_A \left( \frac{128}{9} - 56 \zeta_3\right)
- \beta_0 \frac{112}{9}\right]
\,.\end{align}
%%%

\paragraph{Hard function.}

The hard function $H_\kappa(Q, \mu)$ is proportional to the square of the hard matching coefficient [see \eq{hard_function_one_loop}].
The hard matching coefficient for $q\bar{q} \to Z/\gamma^\ast$ is renormalized as
%%%
\begin{align} \label{eq:rge_hard}
\mu \frac{\df}{\df \mu} C_{q\bq}^{V,A} (Q^2, \mu)
&= \gamma_{q\bq}^{V,A}(Q^2, \mu) \, C_{q\bq}^{V,A} (Q^2, \mu)
\,, \nn \\
\gamma_{q\bq}^{V,A}(Q^2, \mu)
&= \Gamma^q_\mathrm{cusp} \bigl[ \as(\mu) \bigr] \ln \frac{- Q^2-\img 0}{\mu^2}
   + 2 \gamma_C^q \bigl[ \as(\mu) \bigr]
\,.\end{align}
%%%
The coefficients of the quark noncusp anomalous dimension up to two loops are
%%%
\begin{align} \label{eq:coeffs_quark_anom_dim}
 \gamma^q_{C\,0} &= -3 C_F
\,,\nn\\
\gamma^q_{C\,1}
&= - C_F \biggl[
   C_A \Bigl(\frac{41}{9} - 26\zeta_3\Bigr)
   + C_F \Bigl(\frac{3}{2} - 2 \pi^2 + 24 \zeta_3\Bigr)
   + \beta_0 \Bigl(\frac{65}{18} + \frac{\pi^2}{2} \Bigr) \biggr]
\,.\end{align}
%%%

\paragraph{Beam functions and PDFs.}

In $b_T$ space, the TMD beam function is renormalized as
%%%
\begin{align} \label{eq:rge_scet2_beam}
\mu \frac{\df}{\df \mu} \tB_q(\omega, b_T, \mu, \nu)
&= \tgamma_B^q(\omega, \mu, \nu) \, \tB_q(\omega, b_T, \mu, \nu)
\,,\nn\\
\nu \frac{\df}{\df \nu} \tB_q(\omega, b_T, \mu, \nu)
&= -\frac{1}{2}\tgamma^q_\nu(b_T, \mu) \, \tB_q(\omega, b_T, \mu, \nu)
\,,\nn\\
\tgamma^q_B(\omega, \mu, \nu)
&= 2 \Gamma_\cusp^q[\as(\mu)] \ln \frac{\nu}{\omega} + \tgamma_B^q[\as(\mu)]
\,.\end{align}
%%%
We include a tilde to indicate that $\tgamma_B^q$ is related to the SCET$_\text{II}$ beam function,
even though it does not depend on $b_T$. Its coefficients through two loops are~\cite{Luebbert:2016itl}
%%%
\begin{align} \label{eq:coeffs_anom_dim_scet2_beam}
\tgamma_{B\,0}^q
&= 6C_F
\,, \nn \\
\tgamma_{B\,1}^q
&= C_F \Bigl[
(2 -24 \zeta_3) C_A
+ (3 - 4 \pi^2 + 48 \zeta_3) C_F
+ \Bigl(1 + \frac{4\pi^2}{3} \Bigr) \beta_0 \Bigr]
\,.\end{align}
%%%
The resummed rapidity anomalous dimension $\tgamma^i_\nu(b_T, \mu)$
is given in \eq{resummed_rapidity_anom_dim}.

The double-differential beam function satisfies the same RGE
as the inclusive SCET$_\text{I}$ beam function,%
\footnote{We note that \refcite{Procura:2014cba} incorrectly did not distinguish between $\gamma^i_B(\as)$ and $\tgamma^i_B(\as)$.
This lead to the noncusp contribution to the collinear-soft anomalous dimension being missing in their eq.~(3.26),
cf.\ our corrected \eq{rge_scetp_csoft}
and the nonvanishing two-loop noncusp coefficient in our \eq{coeffs_anom_dim_softs}.}
%%%
\begin{align} \label{eq:rge_scet1_beam}
\mu \frac{\df}{\df \mu} B_q(t, x, \vec{k}_T, \mu)
&= \int \! \df t' \,
\gamma_B^q(t - t', \mu) \, B_q(t', x, \vec{k}_T,\mu)
\,,\nn\\
\gamma_B^q(t, \mu)
&= -2 \Gamma^q_{\cusp}(\as)\,\cL_0(t, \mu^2) + \gamma_B^q[\as(\mu)]\,\delta(t)
\,.\end{align}
%%%
The coefficients of the SCET$_\text{I}$ quark beam anomalous dimension are~\cite{Stewart:2010qs, Gaunt:2014xga}
%%%
\begin{align} \label{eq:coeffs_anom_dim_scet1_beam}
\gamma_{B\,0}^q
&= 6C_F
\,, \nn \\
\gamma_{B\,1}^q
&= C_F \Bigl[ \Bigl(\frac{146}{9} - 80 \zeta_3\Bigr) C_A
+ (3 - 4 \pi^2 + 48 \zeta_3) C_F
+ \Bigl(\frac{121}{9} + \frac{2\pi^2}{3} \Bigr) \beta_0 \Bigr]
\,.\end{align}
%%%

We also require the one-loop coefficients of the PDF anomalous dimension,
%%%
\begin{align} \label{eq:DGLAP}
\mu \frac{\df}{\df \mu} f_i(x, \mu)
&= \sum_j \int_x^1 \! \frac{\df z}{z} \, 2P_{ij}[\as(\mu), z]\, f_j\Bigl( \frac{x}{z}, \mu \Bigr)
\,, \nn \\
P_{ij}(\as, z)
&= \sum_{n = 0}^\infty P^{(n)}_{ij}(z) \left( \frac{\as}{4\pi} \right)^{n + 1}
\,.\end{align}
%%%
Note that we expand $P_{ij}(\as, z)$ in $\as/(4\pi)$.
The one-loop coefficients are
%%%
\begin{align} \label{eq:DGLAP_coeff_lo}
P^{(0)}_{qq}(z) = 2C_F \, \theta(z) P_{qq}(z)
\,,\quad
P^{(0)}_{qg}(z) = 2T_F \,\theta(z) P_{qg}(z)
\,,\end{align}
%%%
in terms of the standard color-stripped one-loop QCD splitting functions
%%%
\begin{alignat}{3} \label{eq:p_ij}
P_{qq}(z) &= 2 \mathcal{L}_0(1-z)-\theta(1-z)(1+z) + \frac{3}{2}\delta(1-z)
&&= \Bigl[ \theta(1-z) \frac{1+z^2}{1-z} \Bigr]_+
\,, \nn \\
P_{qg}(z) &= \theta(1-z) \bigl[ 1-2z(1-z) \bigr]
   \,.\end{alignat}
%%%

\paragraph{Soft and collinear-soft functions.}

The RGE of the beam thrust soft function reads
%%%
\begin{align} \label{eq:rge_scet1_soft}
\mu\frac{\df}{\df\mu} S_i(k, \mu)
&= \int \! \df k' \,
\gamma_S^i(k - k', \mu) \, S_i(k', \mu)
\,, \nn \\
\gamma_S^i(k, \mu)
&= 4\Gamma_\cusp^i(\as)\, \cL_0(k, \mu) +
\gamma^i_S[\as(\mu)] \, \delta(k)
\,.\end{align}
%%%
For the double-differential soft function in $b_T$ space we have
%%%
\begin{align} \label{eq:rge_scet2_soft}
\mu \frac{\df}{\df \mu} \tS_i(k, b_T, \mu,\nu) &= \tgamma_S^i(\mu, \nu) \, \tS_i(k, b_T, \mu, \nu)
\,,\nn\\
\nu \frac{\df}{\df \nu} \tS_i(k, b_T, \mu,\nu) &= \tgamma^i_\nu(b_T, \mu) \, \tS_i(k, b_T, \mu, \nu)
\,,\nn\\
\tgamma^i_S(\mu, \nu)
&= 4 \Gamma_\cusp^i(\as)\, \ln \frac{\mu}{\nu} + \tgamma^i_S[\as(\mu)]
\,,\end{align}
%%%
where we again use a tilde on the $\mu$ anomalous dimension
to indicate that it relates to the SCET$_\text{II}$ soft function.
The RGE of the collinear-soft function in $b_T$ space reads
%%%
\begin{align} \label{eq:rge_scetp_csoft}
\mu\frac{\df}{\df\mu} \tcS_i(k, b_T, \mu, \nu)
&= \int \! \df k' \,
\gamma_\cS^i(k - k', \mu, \nu) \, \tcS_i(k', b_T, \mu, \nu)
\,, \nn \\
\nu \frac{\df}{\df \nu} \tcS_i(k, b_T, \mu, \nu)
&= \frac{1}{2} \tgamma^i_\nu(b_T, \mu) \, \tcS_i(k, b_T, \mu, \nu)
\,, \nn \\
\gamma_\cS^i(k, \mu, \nu)
&= -2 \Gamma_\cusp^i(\as)\, \cL_0\Big(k, \frac{\mu^2}{\nu}\Big) + \gamma^i_\cS[\as(\mu)] \delta(k)
\,.\end{align}
%%%
The soft and collinear-soft noncusp anomalous dimension coefficients are only nonzero starting at two loops
and can be inferred from consistency,
%%%
\begin{align} \label{eq:coeffs_anom_dim_softs}
\gamma^i_{S\,0} &= \gamma^i_{\cS\,0} = \tgamma^i_{S\,0} = 0
\,, \nn \\
-\gamma^i_{S\,1}
&= \gamma^i_{\cS\,1}
= \tgamma^i_{S\,1}
= C_i \left[
   C_A \left( \frac{128}{9} - 56 \zeta_3 \right)
 + \beta_0 \left( \frac{112}{9} - \frac{2\pi^2}{3} \right)
\right]
\,.\end{align}
%%%

%===============================================================================
\subsection{Fixed-order ingredients}
\label{app:fixedorder}
%===============================================================================

\paragraph{Hard process.}

The Born cross section for $q\bq \to Z/\gamma^\ast \to \ell^+ \ell^-$ is given by
%%%
\begin{equation} \label{eq:sigma_Born}
\frac{\df \sigma_B^q}{\df Q}
= \frac{8\pi \alpha_\text{em}^2}{3N_c\Ecm^2 Q} \biggl[Q_q^2 + \frac{(v_q^2 + a_q^2) (v_\ell^2+a_\ell^2) - 2 Q_q v_q v_\ell (1-m_Z^2/Q^2)}
{(1-m_Z^2/Q^2)^2 + m_Z^2 \Gamma_Z^2/Q^4}\biggr]
\,,\end{equation}
%%%
where $Q_q$ is the quark charge in units of $\abs{e}$, $v_{\ell,q}$ and $a_{\ell,q}$ are the standard vector and axial couplings of the leptons and quarks, and $m_Z$ and $\Gamma_Z$ are the mass and width of the $Z$ boson.
The one-loop Wilson coefficient $C_{q\bq}^{V,A}(Q^2,\mu)$ from matching the quark current in QCD onto SCET was computed in refs.~\cite{Manohar:2003vb, Bauer:2003di}. This leads to the following hard function~\cite{Stewart:2009yx},
%%%
\begin{align} \label{eq:hard_function_one_loop}
H_{ij}(Q, \mu)
&= \sum_q \frac{\df \sigma_B^q}{\df Q} \, \bigl(\delta_{iq}\delta_{j\bq} + \delta_{i\bq}\delta_{jq}\bigr) \, |C_{q\bq}^{V,A}(Q^2, \mu)|^2
\,, \nn \\
|C_{q\bq}^{V,A}(Q^2, \mu)|^2
&= 1 + 2\Re \, \biggl\{ \frac{\as(\mu)}{4\pi} \, C_F \Bigl[-\ln^2 \frac{Q^2}{\mu^2} + 3 \ln \frac{Q^2}{\mu^2} - 8 + \frac{7\pi^2}{6} \Bigr] \biggr\} + \ord{\as^2}
\,,\end{align}
%%%
where $\Re$ denotes the real part.

\paragraph{Beam functions.}

The one-loop matching coefficients for the single-differential SCET$_\text{II}$ beam function in $b_T$ space are given by~\cite{Ritzmann:2014mka,Luebbert:2016itl}
%%%
\begin{align}
&\tilde{\cI}_{qj}(\omega, b_T, z, \mu, \nu)
\\
&= \delta_{qj} \delta (1-z)
+ \frac{\as(\mu)}{4\pi} \biggl[ \delta_{qj} \delta(1-z) \Bigl(\Gamma^q_0 \ln \frac{\nu}{\omega} + \frac{\gamma^q_{B,0}}{2}\Bigr) L_b
- L_b P^{(0)}_{qj}(z)
+ \tilde{I}_{qj}^{(1)}(z) \biggr]
+ \ord{\as^2}
\,, \nn \end{align}
%%%
where $L_b$ was defined in \eq{def_Lb} and the boundary conditions at $L_b = 0, \nu = \omega$ are
%%%
\begin{align}
\tilde{I}^{(1)}_{qq}(z) &= C_F \, \, \theta(z)\theta(1-z) \, 2(1-z)
\,, \nn \\
\tilde{I}^{(1)}_{qg}(z) &= T_F \, \theta(z)\theta(1-z) \, 4z(1-z)
\,.\end{align}
%%%
As for its anomalous dimension, we use a tilde to indicate
that these boundary conditions are part of the SCET$_\text{II}$ beam function,
even though they do not depend on $b_T$.

It is convenient to decompose the matching coefficients for the double-differential SCET$_\text{I}$ quark beam function as
%%%
\begin{equation} \label{eq:def_delta_beam}
\cI_{qj}(t, z, \vec{k}_T, \mu)
= \delta(\vec{k}_T)\,\cI_{qj}(t, z, \mu)
+ \Delta \cI_{qj}(t, z, \vec{k}_T, \mu)
\,,\end{equation}
%%%
where $\cI_{qj}(t, z, \mu)$ is the matching coefficient for the inclusive quark beam function~\cite{Stewart:2009yx,Stewart:2010qs},
%%%
\begin{align} \label{eq:inclusive_beam_function_coefficients}
\cI_{qj} (t, z, \mu)
&= \delta_{qj} \, \delta(t) \delta(1-z)
\\ & \quad
+ \frac{\as}{4\pi} \Bigl[
\Gamma_0^q \delta_{qj} \, \cL_1(t, \mu^2) \, \delta(1-z)
+ \cL_0(t, \mu^2) \, \tilde{P}^{(0)}_{qj}(z)
+ \delta(t) \, I^{(1)}_{qj}(z) \Bigr]
+ \ord{\as^2}
\nn \,,\end{align}
%%%
with the finite terms in this case given by
%%%
\begin{align} \label{eq:inclusive_beam_function_finite_termes}
I^{(1)}_{qq}(z)
&= 2C_F \, \theta(z) \Bigl[ \cL_1(1-z)(1+z^2) - \frac{\pi^2}{6} \delta(1-z) + \theta(1-z)\Bigl(1 - z - \frac{1+z^2}{1-z}\ln z \Bigr) \Bigr]
\,, \nn \\
I^{(1)}_{qg}(z)
&= 2T_F \, \theta(z) \Bigl[ P_{qg}(z) \Bigl( \ln \frac{1-z}{z} - 1 \Bigr) + \theta(1-z)\Bigr]
\,,\end{align}
%%%
and using the shorthand
%%%
\begin{align}
\tilde{P}^{(0)}_{qj}(z)
\equiv
P^{(0)}_{qj}(z) - \delta_{qj} \delta(1-z) \frac{\gamma_{B\,0}^q}{2}
=
\begin{cases}
2C_F \cL_0(1-z)(1+z^2)
\,,& j = q
\,, \nn \\
2T_F  \Bigl[ (1-z)^2 + z^2 \Bigr]
\,,& j = g
\,.\end{cases}
\end{align}
%%%
The $\Delta \cI_{qj}$ piece can be interpreted as a correction
over the limit $t \ll k_T^2$,
where recoil from collinear radiation is power suppressed
and the double-differential beam function becomes proportional to $\delta(\vec{k}_T)$.
Specifically, it scales as
%%%
\begin{equation}
\Delta \cI_{qj}(t, z, \vec{k}_T, \mu)
\sim \biggl[ \frac{1}{t} \biggr]_+ \biggl[ \frac{1}{k_T^2} \biggr]_+ \times \ORd{\frac{t}{k_T^2}}
\quad \text{for} \quad t \ll k_T^2
\,,\end{equation}
%%%
and by construction satisfies
%%%
\begin{equation}
\int \, \df^2 \vec{k}_T \, \Delta \cI_{qj}(t, z, \vec{k}_T, \mu)
= 0
\,.\end{equation}
%%%
At one loop it can be extracted from the full calculation of $\cI_{qj}(t, z, \vec{k}_T, \mu)$~\cite{Mantry:2010mk,Jain:2011iu}
and has the compact form
%%%
\begin{align} \label{eq:delta_beam_one_loop}
\Delta \cI_{qj}(t, z, \vec{k}_T, \mu)
&= \frac{\as(\mu)}{4\pi} \Delta I_{qj}^{(1)}(t, z, \vec{k}_T)
+ \ord{\as^2}
\,, \nn \\
\Delta I_{qj}^{(1)}(t, z, \vec{k}_T)
&= \frac{\theta(t)}{t} \tilde{P}^{(0)}_{qj}(z)
\biggl[ \frac{1}{\pi} \delta\Bigl( k_T^2 - \frac{1-z}{z}t \Bigr) - \delta(\vec{k}_T)\biggr]
\,.\end{align}
%%%
The second line is regular in $t$
because the term in square brackets vanishes as $t \to 0$.
After accumulating over the transverse plane up to $q_T^\cut > 0$, we have
%%%
\begin{equation} \label{eq:delta_beam_qT_cumulant}
\int \, \df^2 \vec{k}_T \, \theta(q_T^\cut - \abs{\vec{k}_T}) \, \Delta I_{qj}^{(1)}(t, z, \vec{k}_T)
= - \frac{\theta(t)}{t} \tilde{P}^{(0)}_{qj}(z) \, \theta \Bigl[ (q_T^\cut)^2 < \frac{1-z}{z} t \Bigr]
\,.\end{equation}
%%%

\paragraph{Soft and collinear-soft functions.}

The (beam) thrust soft function is~\cite{Schwartz:2007ib, Fleming:2007xt,Stewart:2009yx}
%%%
\begin{align}
S_i(k,\mu) &= \delta(k) + \frac{\as(\mu)}{4\pi} \Bigl[
- 4\Gamma_0^i \, \cL_1(k, \mu) +
\frac{\pi^2}{3} C_i\, \delta(k) \Bigr] + \ord{\as^2}
\,.\end{align}
%%%
The one-loop collinear-soft function in $b_T$ space is~\cite{Procura:2014cba}
%%%
\begin{align}
\tcS_i(k, b_T, \mu, \nu)
= \delta(k)
&+ \frac{\as(\mu)}{4\pi} \biggl\{-\Gamma_0^i L_b \, \cL_0(k, \mu) + \Gamma_0^i \Bigl[-\frac{1}{2} L_b^2 -L_b \ln \frac{\nu}{\mu} \Bigr] \, \delta(k)
- \frac{\pi^2}{3} C_i \, \delta(k)  \biggr\}
\nn \\
&+ \ord{\as^2}
\,.\end{align}
%%%

It is again convenient to decompose the double-differential soft function computed in \refcite{Procura:2014cba}
into separate pieces with distinct power counting,
%%%
\begin{equation} \label{eq:def_delta_soft}
S_i(k, \vec{k}_T, \mu, \nu)
= \delta(k)\, S_i(\vec{k}_T, \mu, \nu) + \Delta S_i(k, \vec{k}_T, \mu, \nu)
\,.\end{equation}
Here $S_i(\vec{k}_T, \mu, \nu)$ is the standard single-differential $q_T$ soft function,
which in $b_T$ space at one loop is given by~\cite{Chiu:2012ir}
%%%
\begin{equation}
\tS_i(b_T, \mu, \nu)
= 1 + \frac{\as(\mu)}{4\pi} \Bigl[ -\frac{\Gamma^i_0}{2} L_b^2
   + 2 \Gamma^i_0 L_b \ln\frac{\mu}{\nu}
   - \frac{\pi^2}{3} C_i \Bigr]
\,.\end{equation}
%%%
The second term in \eq{def_delta_soft} can again be interpreted as a correction, in this case over the limit $k \gg \vec{k}_T$
where the contribution of soft radiation to the $\Tau = k$ measurement becomes power suppressed.
In momentum space this term satisfies
%%%
\begin{align} \label{eq:delta_soft_scaling_momentum_space}
\int \! \df k \, \Delta S_i(k, \vec{k}_T, \mu, \nu) &= 0
\,, \nn \\
\Delta S_i(k, \vec{k}_T, \mu, \nu) &\sim \biggl[ \frac{1}{k} \biggr]_+ \biggl[ \frac{1}{k_T^2} \biggr]_+ \times \ORd{\frac{k_T^2}{k^2}}
\quad \text{for} \quad k_T^2 \ll k^2
\,.\end{align}
%%%
Equivalently, in position space we have
%%%
\begin{align} \label{eq:delta_soft_scaling_position_space}
\int \! \df k \, \Delta \tS_i(k, b_T, \mu, \nu) &= 0
\,, \nn \\
\Delta \tS_i(k, b_T, \mu, \nu) &\sim \biggl[ \frac{1}{k} \biggr]_+ \times \ORd{\frac{1}{b_T^2 k^2}}
\quad \text{for} \quad \frac{1}{b_T^2} \ll k^2
\,.\end{align}
%%%
At one loop, $\Delta S_i$ is given by
%%%
\begin{align} \label{eq:delta_soft_one_loop}
\Delta S_i(k, \vec{k}_T, \mu, \nu) &= \frac{\as(\mu)}{4\pi} \Delta S_{i,1}(k, \vec{k}_T) + \ord{\as^2}
\,, \nn \\
\Delta S_{i,1}(k, \vec{k}_T) &= 4C_i \biggl[ \frac{2}{\pi \mu^3} \cL_\Delta \Bigl( \frac{k}{\mu}, \frac{k_T^2}{\mu^2} \Bigr) - \delta(k) \cL_1(\vec{k}_T, \mu) \biggr]
\,.\end{align}
%%%
The second line is not yet manifestly independent of $\mu$,
but can be simplified noting that
%%%
\begin{equation} \label{eq:delta_soft_trick}
\cL_\Delta(x_1, x_2) - \delta(x_1)\cL_1(x_2)
= \frac{\df}{\df x_1} \frac{\df}{\df x_2} \theta(x_2 - x_1^2) \Bigl[ -\frac{1}{2} \ln^2 \frac{x_1^2}{x_2} \Bigr]
\,.\end{equation}
%%%
It is straightforward to show this by writing all three distributions
in terms of $\theta(x_1 - \beta)$ and $\theta(x_2 - \beta^2)$ for infinitesimal $\beta$,
collecting terms, and noting that the result is finite as $\beta \to 0$.
From \eq{delta_soft_trick} we can immediately read off the fixed-order double cumulant
of $\Delta S_{i,1}$ for $\Tau_\cut > 0$, $q_T^\cut > 0$,
%%%
\begin{align} \label{eq:delta_soft_double_cumulant}
\int^{\Tau_\cut} \! \df k \, \int \! \df^2 \vec{k}_T \,
\theta(q_T^\cut - \abs{\vec{k}_T}) \,
\Delta S_{i,1}(k, \vec{k}_T)
=4 C_F \, \theta(q_T^\cut - \Tau_\cut) \Bigl[ -2 \ln^2 \frac{\Tau_\cut}{q_T^\cut} \Bigr]
\,,\end{align}
%%%
where the dependence on $\mu$ drops out as expected.
Inserting \eq{delta_soft_trick} and integrating by parts also yields
the cumulant up to $\Tau_\cut > 0$ in position space,
%%%
\begin{align} \label{eq:delta_soft_position_space_cumulant}
\int^{\Tau_\cut} \! \df k \, \Delta \tS_{i,1}(k, b_T)
= 4C_F \biggl[ \frac{1}{4} x^2 \phantom{}_3F_4\Bigl(1,1,1;2,2,2,2;-\frac{x^2}{4}\Bigr) - 2 \ln^2 \frac{xe^{\gamma_E}}{2}\biggr]
\,,\end{align}
%%%
where $x \equiv b_T \Tau_\cut$ and $\phantom{}_iF_j(x_1,\ldots,x_i; y_1,\ldots,y_j;z)$ is the generalized hypergeometric function.
The right hand side of \eq{delta_soft_position_space_cumulant}
asymptotes to $1 / x^2$ as $x \to \infty$, as required by the scaling law in \eq{delta_soft_scaling_position_space}.
We also need the spectrum of $\Delta \tS_{i,1}$ at $\Tau > 0$ in position space,
%%%
\begin{align} \label{eq:delta_soft_position_space_spectrum}
\Delta \tS_{i,1}(\Tau, b_T)
= 4C_F \, \frac{1}{\Tau} \biggl[ \frac{1}{2} x^2 \phantom{}_2F_3\Bigl(1,1;2,2,2;-\frac{x^2}{4}\Bigr) - 4 \ln \frac{xe^{\gamma_E}}{2}\biggr]
\,,\end{align}
%%%
where this time $x \equiv b_T \Tau$ and the term in square brackets again asymptotes to $1/x^2$ as $x \to \infty$.

%===============================================================================
\subsection{Renormalization-group evolution}
\label{app:rg_kernels}
%===============================================================================

\paragraph{\boldmath SCET$_\I$.}

The closed-form solution of \eq{rge_scet1_beam} is~\cite{Fleming:2007xt, Ligeti:2008ac}
%%%
\begin{align}
B_q(t, x, \vec{k}_T, \mu)
&= \exp \Bigl[ 4K^q_\Gamma(\mu_B, \mu) + K^q_{\gamma_B}(\mu_B, \mu) \Bigr]
\nn \\
&\quad \times
\int \! \df t' \,
\cV_{-2\eta^q_\Gamma(\mu_B,\,\mu)}(t-t', \mu_B^2) \, B_q(t', x, \vec{k}_T,\mu_B)
\,,\end{align}
%%%
where $\cV_\eta$ was defined in \eq{def_flow_distribution}.
Similarly, the solution of \eq{rge_scet1_soft} is
%%%
\begin{align}
S_i(k, \mu)
&= \exp \Bigl[ -4K^i_\Gamma(\mu_S, \mu) + K^i_{\gamma_S}(\mu_S, \mu) \Bigr]
\int \! \df k' \,
\cV_{4\eta^i_\Gamma(\mu_S,\,\mu)}(k-k', \mu_S) \, S_i(k', \mu_S)
\,.\end{align}
%%%
The definitions of $\eta^i_\Gamma$, $K^i_\Gamma$, and $K^i_\gamma$
and their approximate analytical form at NNLL are given for example
in app.~A.4 of \refcite{Ebert:2017uel}. The solution of the hard RGE in \eq{rge_hard}
in our notation is also found there.

\paragraph{\boldmath SCET$_\II$.}

Evolving first in $\nu$ and then in $\mu$ (from right to left), the solution of \eq{rge_scet2_beam} is
%%%
\begin{align}
\tB_q(\omega, b_T, \mu, \nu)
&= \exp \Bigl[ 2\eta^q_\Gamma(\mu_B, \mu)\ln \frac{\nu}{\omega} + K^q_{\tgamma_B}(\mu_B, \mu) \Bigr]
\nn \\
&\quad \times
\exp \Bigl[ -\frac{1}{2} \ln \frac{\nu}{\nu_B} \, \tgamma^q_\nu(b_T, \mu_B) \Bigr] \,
\tB_q(\omega, b_T, \mu_B, \nu_B)
\,,\end{align}
%%%
where the resummed rapidity anomalous dimension $\tgamma^q_\nu(b_T, \mu_B)$ is given by \eq{resummed_rapidity_anom_dim}.
For the double-differential soft function renormalized as in \eq{rge_scet2_soft}, we have
%%%
\begin{align}
\tS_i(k, b_T, \mu, \nu)
&= \exp \Bigl[ -4\eta^i_\Gamma(\mu_S, \mu)\ln \frac{\nu}{\mu_S} + 4K^i_\Gamma(\mu_S, \mu) + K^i_{\tgamma_S}(\mu_S, \mu) \Bigr]
\nn \\
&\quad \times
\exp \Bigl[ \ln \frac{\nu}{\nu_S} \, \tgamma^i_\nu(b_T, \mu_S) \Bigr] \,
\tS_i(k, b_T, \mu_S, \nu_S)
\,.\end{align}
%%%

\paragraph{\boldmath SCET$_+$.}

Again evolving first in $\nu$ and then in $\mu$,
the solution of the collinear-soft RGE in \eq{rge_scetp_csoft} in $b_T$ space is given by
%%%
\begin{align} \label{eq:rg_evolution_csoft}
\tcS_i(k, b_T, \mu, \nu)
&= \exp \Bigl[ 4K^i_\Gamma(\mu_\cS, \mu) + K^i_{\gamma_\cS}(\mu_\cS, \mu) \Bigr]
\int \! \df k' \,
\cV_{-2\eta^i_\Gamma(\mu_\cS,\,\mu)}\Bigl(k-k', \frac{\mu_\cS^2}{\nu}\Bigr)
\nn \\
&\quad \times
\exp \Bigl[ \frac{1}{2} \ln \frac{\nu}{\nu_\cS} \, \tgamma^i_\nu(b_T, \mu_\cS) \Bigr]\,
\tcS_i(k', b_T, \mu_\cS, \nu_\cS)
\,.\end{align}
%%%
The rapidity evolution factor on the second line does not depend on $k'$
and thus may be taken out of the convolution integral.

%===============================================================================
\subsection{Beam function convolutions with RG kernels}
\label{app:beam_function_convolutions_rg_kernels}
%===============================================================================

In SCET$_\I$ the combined beam and soft renormalization group running
in momentum space has the functional form $\cV_\eta(k, \mu)$ [see \eq{def_flow_distribution}],
where the convolution variable $k$ is ultimately fixed by the overall $\Tau$ measurement.
To evaluate the resummed SCET$_\I$ cross section as a function of $\Tau$ ($\Tau_\cut$)
we require convolutions of $\cV_\eta$ with the double-differential beam function
at finite $\Tau$ (integrated up to $\Tau_\cut$).
Convolutions of $\cV_\eta$ with the inclusive beam function reduce to a linear combination of
%%%
\begin{equation}
Q_i \! \int \! \df k' \,
\cV_\eta(k - k', \mu) \, \cL_n(Q_i k', \mu^2)
= \int \! \df k' \,
\cV_\eta(k - k', \mu) \, \cL_n \Bigl(k', \frac{\mu^2}{Q_i} \Bigr)
\,,\end{equation}
%%%
with $Q_i$ as in \eq{def_Tau_generic},
and are straightforward to evaluate using results in app.~B of \refcite{Ligeti:2008ac}.
The same is true for convolutions of the evolution kernel with the single-differential $\Tau$ soft function
and, in SCET$_+$, with the collinear-soft function.
(Cross terms between beam, csoft, or soft functions are absent at one loop.)
The only nontrivial ingredient are convolutions of $\cV_\eta(k, \mu)$
with the one-loop $\Delta I_{qj}^{(1)}$ piece defined in \eq{delta_beam_one_loop}.
Depending on the measurement, we distinguish the following four cases
for the resulting Mellin kernel:
\begin{enumerate}
\item
cumulant up to $\Tau_\cut > 0$, cumulant up to $q_T^\cut > 0$, $r \equiv (q_T^\cut)^2/(Q_i \Tau_\cut)$:
%%%
\begin{align}
&Q_i \! \int \! \df k \, \int \! \df^2 \vec{k}_T \,
\theta(\Tau_\cut - k) \, \theta(q_T^\cut - \abs{\vec{k}_T}) \,
\int \! \df k' \,
\cV_\eta(k-k', \TauCut) \, \Delta I_{qj}^{(1)}(Q_i k', z, \vec{k}_T)
\nn \\
&= \theta \Bigl( r > \frac{1-z}{z} \Bigr) P_{qj}^{(0)}(z) \,
\frac{e^{-\gamma_E \eta}}{\Gamma(1+\eta)}
\biggl[
- B_{1-r z/(1-z)}(1+\eta, 0)
\biggr]
\,,\end{align}
%%%
where $B_x(a,b)$ is the incomplete Beta function,
%%%
\begin{equation}
B_x(a,b) = \int_0^x \! \df t \, t^{a-1} (1-t)^{b-1}
\,.\end{equation}
\item
spectrum at $\Tau > 0$, cumulant up to $q_T^\cut > 0$, $r \equiv (q_T^\cut)^2/(Q_i \Tau)$:
%%%
\begin{align}
&Q_i \! \int \! \df k \, \int \! \df^2 \vec{k}_T \,
\delta(\Tau - k) \, \theta(q_T^\cut - \abs{\vec{k}_T}) \,
\int \! \df k' \,
\cV_\eta(k-k', \Tau) \, \Delta I_{qj}^{(1)}(Q_i k', z, \vec{k}_T)
\\
&= \frac{1}{\Tau} \, \theta \Bigl( r > \frac{1-z}{z} \Bigr) P_{qj}^{(0)}(z) \,
\frac{e^{-\gamma_E \eta}}{\Gamma(1+\eta)}
\biggl[ - \eta B_{1-r z/(1-z)}(1+\eta, 0)
- \Bigl( 1 - \frac{rz}{1-z} \Bigr)^{\eta} \biggr]
\nn \,.\end{align}
%%%
\item
cumulant up to $\Tau_\cut$, spectrum at $q_T > 0$, $r \equiv (q_T)^2/(Q_i \Tau_\cut)$:
%%%
\begin{align}
&Q_i \! \int \! \df k \, \int \! \df^2 \vec{k}_T \,
\theta(\Tau_\cut - k) \, \delta(q_T - \abs{\vec{k}_T}) \,
\int \! \df k' \,
\cV_\eta(k-k', \TauCut) \, \Delta I_{qj}^{(1)}(Q_i k', z, \vec{k}_T)
\nn \\
&= \frac{2}{q_T} \, \theta \Bigl( r > \frac{1-z}{z} \Bigr) P_{qj}^{(0)}(z) \,
\frac{e^{-\gamma_E \eta}}{\Gamma(1+\eta)}
\Bigl( 1 - \frac{rz}{1-z} \Bigr)^{\eta}
\,.\end{align}
\item
spectrum at $\Tau > 0$, spectrum at $q_T > 0$, $r \equiv (q_T)^2/(Q_i \Tau)$:
%%%
\begin{align}
&Q_i \! \int \! \df k \, \int \! \df^2 \vec{k}_T \,
\delta(\Tau - k) \, \delta(q_T - \abs{\vec{k}_T}) \,
\int \! \df k' \,
\cV_\eta(k-k', \Tau) \, \Delta I_{qj}^{(1)}(Q_i k', z, \vec{k}_T)
\nn \\
&= \frac{2}{q_T}\frac{1}{\Tau} \, P_{qj}^{(0)}(z) \,
\cV_\eta\Bigl( 1 - \frac{rz}{1-z} \Bigr)
\end{align}
%%%
\end{enumerate}
In the first three cases the overall $\theta$ function
cuts off the final PDF integral at
%%%
\begin{equation}
z < z_\cut \equiv \frac{1}{1 + r}
\,,\end{equation}
%%%
and the Mellin kernel is regular up to and including $z_\cut$.
In the last case we instead find a singularity at $z = z_\cut$,
i.e., the subtraction from $\cV_\eta$ now acts directly on the PDF integral.
In either case we have exploited that terms proportional to $\delta(1-z)$ are cut off since $r > 0$,
so we could replace $\tilde{P}_{qj}^{(0)}$ by $P_{qj}^{(0)}$ throughout.
We have also set $\mu = \Tau \, (\TauCut$) in the boundary condition of $\cV_\eta$ on the left-hand side,
which can always be achieved by a shift in $\mu$.
This ensures that the right-hand side depends only on the dimensionless parameters $r$ and $\eta$,
up to an overall dimensionful prefactor.
It is straightforward to check that for $\eta \to 0 $ (at fixed order),
the above results reduce to cumulants (spectra) of $\Delta I^{(1)}_{qj}$ itself.

\addcontentsline{toc}{section}{References}
\bibliographystyle{jhep}
\bibliography{qT-Tau0}

\end{document}